\documentclass[preprint]{aastex}

\newcommand\Stromgren{Str\"{o}mgren }
\newcommand\Teff{T_{\rm{eff}}} 
\newcommand\Fbol{F_{\rm{bol}}}
\newcommand\FeH{\rm{[Fe/H]}}
\newcommand\logg{\rm{log~g}}
\newcommand\EBV{\rm{E{\it (B-V)}}}
\newcommand\Eby{\rm{E{\it (b-y)}}}

\received{}
\accepted{}
\journalid{}{}
\articleid{}{}

\slugcomment{Astronomical Journal, submitted}

\shortauthors{Clem et al.} 
\shorttitle{$uvby$ Color-Temperature Relations}

\begin{document}

\title{EMPIRICALLY CONSTRAINED COLOR-TEMPERATURE \\ RELATIONS. II. 
$uvby^{1,\,2}$\footnotetext[1]{Based, in part, on observations made with the Nordic
Optical Telescope, operated jointly on the island of La Palma by Denmark, Finland,
Iceland, Norway, and Sweden, in the Spanish Observatorio del Roque de los Muchachos of
the Instituto de Astrofisica de Canarias.}\footnotetext[2]{Based, in part, on
observations obtained with the Danish 1.54m telescope at the European Southern
Observatory, La Silla, Chile.}}

\setcounter{footnote}{2}

\author{James L.~Clem and Don A.~VandenBerg}
\affil{Department of Physics \& Astronomy, University of Victoria,
       P.O.~Box 3055, Victoria, B.C. V8W~3P6, Canada}
\email{jclem@uvastro.phys.uvic.ca, davb@uvvm.uvic.ca}

\author{Frank Grundahl}
\affil{Department of Physics \& Astronomy, Aarhus University, Ny Munkegade, 8000 Aarhus C, 
       Denmark}
\email{fgj@phys.au.dk}
 
\and

\author{Roger A.~Bell}
\affil{Department of Astronomy, University of Maryland, College Park, MD 20742-2421}
\email{roger@astro.umd.edu}

\begin{abstract}
\label{sec:abstract}

A new grid of theoretical color indices for the \Stromgren $uvby$ photometric system
has been derived from MARCS model atmospheres and SSG synthetic spectra for cool dwarf
and giant stars having $-3.0\leq\FeH\leq+0.5$ and $3000\leq\Teff\leq8000\,$K.  At
warmer temperatures (i.e., $8000<\Teff\leq40000\,$K), this grid has been
supplemented with the synthetic $uvby$ colors from recent Kurucz atmospheric models
without overshooting (Castelli, Gratton, \& Kurucz, 1997, A\&A, 318, 841).  Our
transformations appear to reproduce the observed colors of extremely metal-poor
turnoff and giant stars:  the various $uvby$ color-magnitude diagrams (CMDs) for the
$\FeH\sim-2.2$ globular cluster M$\,$92 can be matched exceedingly well down to
$M_V\approx6$ by the same isochrone that provides a very good fit to published $BV$
data (see Paper I), on the assumption of the same distance and reddening.  Due to a
number of assumptions made in the synthetic color calculations, however, our
color--$T_{\rm eff}$ relations for cool stars fail to provide a suitable match to the
$uvby$ photometry of both cluster and field stars having $\FeH>-2.0$.  To overcome
this problem, the theoretical indices at intermediate and high metallicities have been
corrected using a set of color calibrations based on field stars having
well-determined distances from {\it Hipparcos}, accurate $\Teff$ estimates from the
infrared flux method, and spectroscopic $\FeH$ values.  In contrast with Paper I, star
clusters played only a minor role in this analysis in that they provided a
supplementary constraint on the color corrections for cool dwarf stars with
$\Teff\leq5500\,$K.  They were mainly used to test the color--$T_{\rm eff}$ relations
and, encouragingly, isochrones that employ the transformations derived in this study
are able to reproduce the observed CMDs (involving $u-v$, $v-b$, and $b-y$ colors) for
a number of open and globular clusters (including M$\,$67, the Hyades, and 47$\,$Tuc)
rather well.  Moreover, our interpretations of such data are very similar, if not
identical, with those given in Paper I from a consideration of $BV(RI)_C$ observations
for the same clusters --- which provides a compelling argument in support of the
color--$T_{\rm eff}$ relations that are reported in both studies.  In the present
investigation, we have also analyzed the observed \Stromgren photometry for the
classic Population~II subdwarfs, compared our ``final" $(b-y)$--$\Teff$ relationship
with those derived empirically in a number of recent studies, and examined in some
detail the dependence of the $m_1$ index on $\FeH$.\footnote{Tabular versions of our
\Stromgren color-temperature relations can be retrieved via {\it ftp} to the address
``uvphys.phys.uvic.ca" using the login I.D. ``star"  and ``vicmodel" as the password.  
A simple FORTRAN code (``uvby.for") is provided that interpolates within the low- and
high-temperature color tables (``uvbylo.data" and ``uvbyhi.data", respectively) to
yield $b-y$, $m_1$, $c_1$, and $BC_V$ indices for input values of $\Teff$, $\logg$,
and $\FeH$.}

\end{abstract} 

\keywords{photometry: $uvby$ --- stars: atmospheres --- stars: general --- stars:  
fundamental parameters --- color-magnitude diagrams (HR diagrams) --- globular
clusters: general --- globular clusters (M$\,$92, M$\,$3, 47$\,$Tucanae) --- open
clusters: general --- open clusters (M$\,$67, Hyades, NGC$\,$6791)}
 
\section{Introduction}
\label{sec:intro}

Among the wide variety of photometric systems available today, the $uvby$ system of
\citet{Stromgren1963} still remains one of the most valuable for the study of stellar
populations and Galactic structure.  Its usefulness stems from the fact that the four
intermediate-width filters are designed to isolate and measure certain key features in
a stellar spectrum which are highly sensitive to the underlying physical
characteristics of the star itself.  For example, \Stromgren $b-y$ provides an
accurate indicator of effective temperature similar to broadband $B-V$ or $V-I$ while
the two other \Stromgren indices, $m_1$ and $c_1$, yield photometric estimates of the
stellar metal abundance and surface gravity (or luminosity).  In this respect, the
ability of the $uvby$ system to provide precise estimates of these fundamental
parameters makes it much better suited than standard Johnson-Cousins $UBVRI$
photometry for the study of individual stars.

For years, \Stromgren photometry has provided a wealth of information on the chemical
and dynamical evolution of the Milky Way through its application to the field-star
populations in both the Galactic halo and the solar neighborhood \citep[see, for
example, the works of][]{CleggBell1973, SchusterNissen1988, SchusterNissen1989a,
SchusterNissen1989b, Olsen1984, Haywood2001}.  Moreover, it has also proven to be
extremely valuable in its application to star clusters.  While the narrowness of the
$uvby$ filters limited earlier investigations \citep{CrawfordBarnes1969,
CrawfordBarnes1970a, CrawfordPerry1966, CrawfordPerry1976} primarily to those stars in
nearby open clusters (e.g., the Hyades, NGC$\,$752, Praesepe, and the Pleiades) that
were bright and isolated enough to be readily observed using traditional
photomultipliers, the advent of modern CCD detectors meant that the \Stromgren system
could be extended to much fainter stars such as those found near the turnoff and main
sequence regions in more distant clusters \citep[see the pioneering studies
of][]{Anthony-Twarog1987a, Anthony-Twarog1987b, Anthony-TwarogTwarog1987}.  This
recent explosion in the amount of high-quality \Stromgren data for a large number of
open and globular clusters offers profound potential for refining our understanding of
these systems.  For instance, it has already been shown that both the $m_1$ and $c_1$
indices can reveal the existence of carbon and nitrogen abundance variations in
globular cluster RGB stars \citep{Hilker2000, Grundahl2002a}.  Moreover, the $c_1$
index has also been used to derive distance-independent cluster ages using techniques
akin to those developed by \citet{SchusterNissen1989b} for field stars
\citep{Grundahl2000a} while the $m_1$ index has provided photometric $\FeH$ estimates
for individual turnoff and giant stars \citep{Nissen1987, HughesWallerstein2000,
HilkerRichtler2000}.

Despite the availability of new high-quality $uvby$ photometry for a number of open
and globular clusters in the Galaxy, our ability to fully exploit these data using
stellar evolutionary models still remains somewhat difficult due to the lack of
accurate and reliable color--$\Teff$ relations and bolometric corrections for the
\Stromgren system that are needed to transform these theoretical models to the
observed cluster color-magnitude diagrams (CMDs).  While a number of empirical
calibrations and analytical formulae have been derived over the years that serve to
relate the \Stromgren $b-y$, $m_1$, and $c_1$ indices to the fundamental stellar
parameters of $\Teff$, $\FeH$, and $\logg$, respectively, they are often very specific
to certain types of stars that occupy limited regions of the H-R diagram and generally
rely on only a small number of stars in the solar neighborhood that have
well-determined properties from spectroscopic analysis.  As a result, these relations
cannot be trusted if they are applied to situations beyond the range in which they
were originally intended.  Alternatively, one may employ grids of theoretical colors
computed from model stellar atmospheres and synthetic spectra to interpret the
photometric data \citep{Bell1988}.  These synthetic colors are very useful for not
only confirming the empirical calibrations between various \Stromgren indices and
certain stellar properties, but also for characterizing the nature of stars in those
regimes not covered by these relations.  Furthermore, they are ideal for transforming
evolutionary models to the observational color-magnitude and color-color planes owing
to their broad coverage of stellar parameter space.  Their accuracy, however, is
ultimately limited by how well the models are able to reproduce the observed spectrum
of an actual star.  Some examples of these synthetic color computations for the $uvby$
system include the grid of MARCS colors for cool dwarfs and subgiants presented by
\citet{VandenBergBell1985} and those computed by Kurucz from his atmospheric models
\citep{Kurucz1993}.  While the $uvby$ colors of the latter cover a much larger range
in stellar parameter space than the \citeauthor{VandenBergBell1985} results, they are
known to have problems reproducing the observed colors of cooler dwarf and giant stars
\citep{Grundahl1998}.  Therefore, in order to accurately interpret and analyze cluster
$uvby$ photometry using current stellar models, we must first check if these
theoretical colors are in good agreement with the observed photometry for a collection
of stars with well-determined physical properties.

Recently, \citet[hereafter Paper I]{VandenBergClem2003} found that, by applying small
corrections to synthetic color transformations for the $BV(RI)_C$ system towards
cooler effective temperatures, it is possible to achieve good consistency with the
observational data for cool dwarf and giant stars in both the metal-poor and
metal-rich regimes.  Their so-called ``semi-empirical" approach resulted in a set of
$BV(RI)_C$ color--$\Teff$ relations and bolometric corrections that were able to
accurately and consistently interpret the observed $B-V$, $V-I$, and $V-R$ CMDs for
both a sample of clusters (such as the Hyades, M$\,$67, M$\,$92, 47$\,$Tuc, and
NGC$\,$6791) and {\it Hipparcos} field stars.  In addition, their predicted solar
metallicity $(B-V)$--$\Teff$ relationship and computed $(B-V)_{\odot}$ value agree
exceedingly well with those derived from the empirical analysis of
\citet{SekiguchiFukugita2000}.  In theory, these same methods could also be used to
overcome the problems with the theoretical colors computed for other photometric
systems, provided that a suitable amount of data, both for clusters and field stars,
are available in order to quantify what corrections are necessary to successfully
place the synthetic indices onto the observational systems.

The goal of the present investigation is to develop a set of accurate and reliable
semi-empirical color transformations for the \Stromgren $uvby$ system.  In contrast
with Paper I, which adopted previously published synthetic color grids as the initial
framework for cool stars, we choose instead to compute an entirely new grid of $uvby$
colors using MARCS model atmospheres and SSG synthetic spectra.  These new $uvby$
colors effectively supersede those reported by \citeauthor{VandenBergBell1985} since
they are computed from more recent versions of the MARCS/SSG programs and cover a
broader range in parameter space.  This new grid of $uvby$ colors is applicable to
both dwarfs and giants having $3000\leq\Teff\leq8000\,$K with metal abundances
extending from $\FeH=-3.0$ to +0.5.  At $\Teff>8000\,$K our new grid is
supplemented with the most recent Kurucz $uvby$ colors computed by \citet[hereafter
CGK97]{Castelli1997} from atmospheric models that neglect overshooting.  In the
analysis that follows we will explain how these purely theoretical colors can be
brought into better agreement with the observed $uvby$ data for a number of star
clusters by correcting them against a sample of {\it Hipparcos} field stars having
accurate $\Teff$ estimates.  Ultimately, the validity of our semi-empirical approach
will be demonstrated in much the same way as in Paper I for the $BV(RI)_C$
transformations by showing that they yield consistent fits of model isochrones to the
photometric data for a number of different clusters, regardless of which \Stromgren
color is considered.  More importantly, we will also show that the resultant
interpretations of these $uvby$ CMDs are virtually identical to those obtained in
Paper I, which analyzes $BV(RI)_C$ photometry of the same clusters, when reasonable
estimates for their distances, reddenings, and metallicities are assumed.

\section{Calculation of the Synthetic \Stromgren Colors} 
\label{sec:colorcalc}

The synthetic $uvby$ colors presented in this paper have been computed using the
latest versions of the MARCS model atmosphere and SSG spectral synthesis codes.  
Readers interested in the details of these programs are referred to
\citet{Houdashelt2000b} who provide extensive descriptions of the model calculations
as well as the recent improvements that have been implemented.  Below we give only a
brief overview of the underlying assumptions that are made in computing our synthetic
\Stromgren colors for cool stars, along with our choices for the various parameters
that must be defined in the model calculations.

The MARCS program \citep{Gustafsson1975} computes a flux-constant, chemically homogeneous,
plane-parallel stellar atmosphere assuming LTE and hydrostatic equilibrium.  Opacity
distribution functions (ODFs) are employed in the model calculations to represent atomic
and molecular opacities as a function of wavelength.  For all atmospheric models we assume
a value of $l/H_p=1.6$ for the mixing length parameter and solar abundance ratios given by
\citet{AndersGrevesse1989} with the small modifications to the carbon and nitrogen
abundances reported by \citet{Grevesse1990, Grevesse1991}.  Furthermore, we
consider enhancements to the $\alpha$ elements (O, Ne, Mg, Si, S, Ar, Ca, and Ti) by
+0.4~dex relative to the solar values for all models with $\FeH\leq-1.0$ and +0.25~dex for
those with $\FeH=-0.5$.  These enhancements reflect a growing body of spectroscopic
evidence \citep{ZhaoMagain1990, Kraft1998, Carney1996, Fulbright2000} which suggests that
most metal-poor halo and globular cluster stars exhibit an overabundance in the
$\alpha$-process elements of at least +0.3~dex relative to solar.

The SSG code \citep[hereafter BPT94]{Bell1994} uses a model atmosphere together with
an extensive absorption line list, Doppler broadening velocity, and the adopted
abundance table to create a synthetic stellar spectrum.  Our spectral models are
computed using an updated version of the Bell ``N" atomic and molecular line list.  
BPT94 has shown that this ``N" list yields the best fits between the synthetic and
observed solar spectra in the wavelength regions closely corresponding the locations
of the four \Stromgren filters.  In addition, for all models with $\Teff\leq4000\,$K,
a TiO line list is included in the computations to account for the increased strength
of TiO absorption features in M-type stars \citep{Houdashelt2000a}.  All spectra are
constructed at 0.1{\rm{\AA}} resolution covering an optical wavelength range of
$3000-8000${\rm{\AA}}, and we assume that the microturbulent velocity varies as a
function of surface gravity following the empirical relation $\xi=2.22-0.322\,\logg$
\citep{Gratton1996}.  It is important to note that our computed spectra do not allow
for variations in the carbon and nitrogen abundances and carbon isotope ratios that
are known to occur in field and cluster giant stars.

Finally, the \Stromgren colors are created by convolving each synthetic spectrum with
the $uvby$ transmission curves given by \citet{CrawfordBarnes1970b}, while accounting
for atmospheric extinction due to scattering by molecules and aerosols
\citep{HayesLatham1975}.  In order to give our colors the same zero point as the
standard system, we normalize the computed colors of a Vega model assuming
$\Teff=9650\,$K, $\logg=3.90$, and $\FeH=0.0$ \citep{DreilingBell1980} to its
corresponding observed indices, namely $b-y=0.004$, $m_1=0.157$, and $c_1=1.089$
\citep{CrawfordBarnes1970b}.  Importantly, it is assumed that we do not have to
transform the synthetic colors in any way to place them on the standard $uvby$ system.  
In other words, the $uvby$ transmission functions are assumed to be perfectly correct,
and the response of the 1P21 detector is the same as that given in the manufacturer's
literature.  In addition, when comparing our synthetic colors to observed $uvby$ data,
we must also assume that the data collected with current CCDs, which detect photons
and not flux, can be transformed to the original standard system with minimal error.

The final grid of synthetic colors are produced from spectra computed for $\FeH$ values
of $-$3, $-$2.5, $-$2, $-$1.5, $-$1, $-$0.5, 0.0, and +0.5 which, in each case, cover a
range in $\logg$ from $-$0.5 to 5.0 for $\Teff$'s between 3000 and 6000$\,$K and from
2.0 to 5.0 for $6000<\Teff\leq8000\,$K.  We mention that the computed colors for models
with $\Teff\leq4000\,$K and/or $\logg<0.5$ are highly uncertain because a detailed
comparison between the synthetic and observed spectra in the region of the $uvby$
filters for these types of stars has yet to be performed.  In addition, the extremely
low-gravity model atmospheres should incorporate spherical geometry rather than the
plane-parallel geometry we have assumed in our computations.  Nevertheless, the colors
that correspond to these models are included in this investigation since it is our goal
to present a set of $uvby$ color--$\Teff$ relations that cover the fullest extent of
stellar parameter space and are applicable to most of the H-R diagram.  In order to
accomplish this goal for early-type stars, we supplement our color grids with the
synthetic $uvby$ colors that were computed from the non-overshoot models of Kurucz by
CGK97 for hotter stars (i.e., $8000<\Teff\leq40000\,$K).\footnote{The CGK97 $uvby$
colors are currently available only on the homepage of R.~L.~Kurucz:  
http://kurucz.harvard.edu} While the latter colors remain largely untested against
photometric observations, some testament to their accuracy in reproducing the
observations of early-type stars is evident in the studies of \citet{RelyeaKurucz1978}
and \citet{Lester1986} who make use of $uvby$ colors computed from the older atmospheric
models of \citet{Kurucz1979}.  Although improvements have since been made to the Kurucz
models, particularly in the low-temperature atomic and molecular lines lists and in the
treatment of convection, it is unlikely that the colors for hotter models (i.e.,
$\Teff\gtrsim8000\,$K) would be significantly affected by such improvements.  
Therefore, we are reasonably confident that the raw synthetic \Stromgren colors of CGK97
are reliable for warmer stars, especially as Paper I has found no problems with the
BV(RI)$_C$ predictions from the same model atmospheres.

Finally, it is important to note that we have adopted the same bolometric corrections
which were reported in Paper I over those derived from the MARCS/SSG models for the
sake of consistency.  Indeed, the BC$_V$'s computed from the MARCS/SSG models tend to
show good agreement with those of Paper I in a systematic sense for $\Teff\geq4000\,$K
after applying a small zero-point shift to accommodate the different normalization
values adopted for the Sun.  Below 4000$\,$K, the MARCS/SSG BC$_V$'s tend to be
systematically larger (i.e., more positive).

\section{The Synthetic \Stromgren Colors: Tests and Calibrations} 
\label{sec:colorcalib}

The most obvious way to check the accuracy of our newly calculated \Stromgren colors
is to assess how well they reproduce the observed $uvby$ CMDs for a sample of stellar
clusters, both open and globular, which span a broad range in metallicity.  
Fortunately, the recent observational efforts of \citet{Grundahl1999} has resulted in
a large amount of high-quality CCD photometry on the $uvby$ system for a number of
metal-poor and metal-rich clusters, including M$\,$92, M$\,$3, 47$\,$Tuc, M$\,$67, and
NGC$\,$6791.  These data are ideal for our tests since they were obtained in all four
\Stromgren filters and subjected to the same reduction and calibration techniques
\citep[see][]{Grundahl1998, Grundahl2000a, Grundahl2002b}.  Moreover, these clusters
have metallicities that range from $\FeH\sim-2.2$ for the globular cluster M$\,$92
\citep{ZinnWest1984, CarrettaGratton1997} to $\sim+0.4$ for the metal-rich open
cluster NGC$\,$6791 \citep{PetersonGreen1998}.  As a result, their photometry can be
used to provide stringent constraints on the accuracy of the synthetic $uvby$
color--$\Teff$ relations over a wide range in stellar parameter space.

\begin{figure} 
\plotone{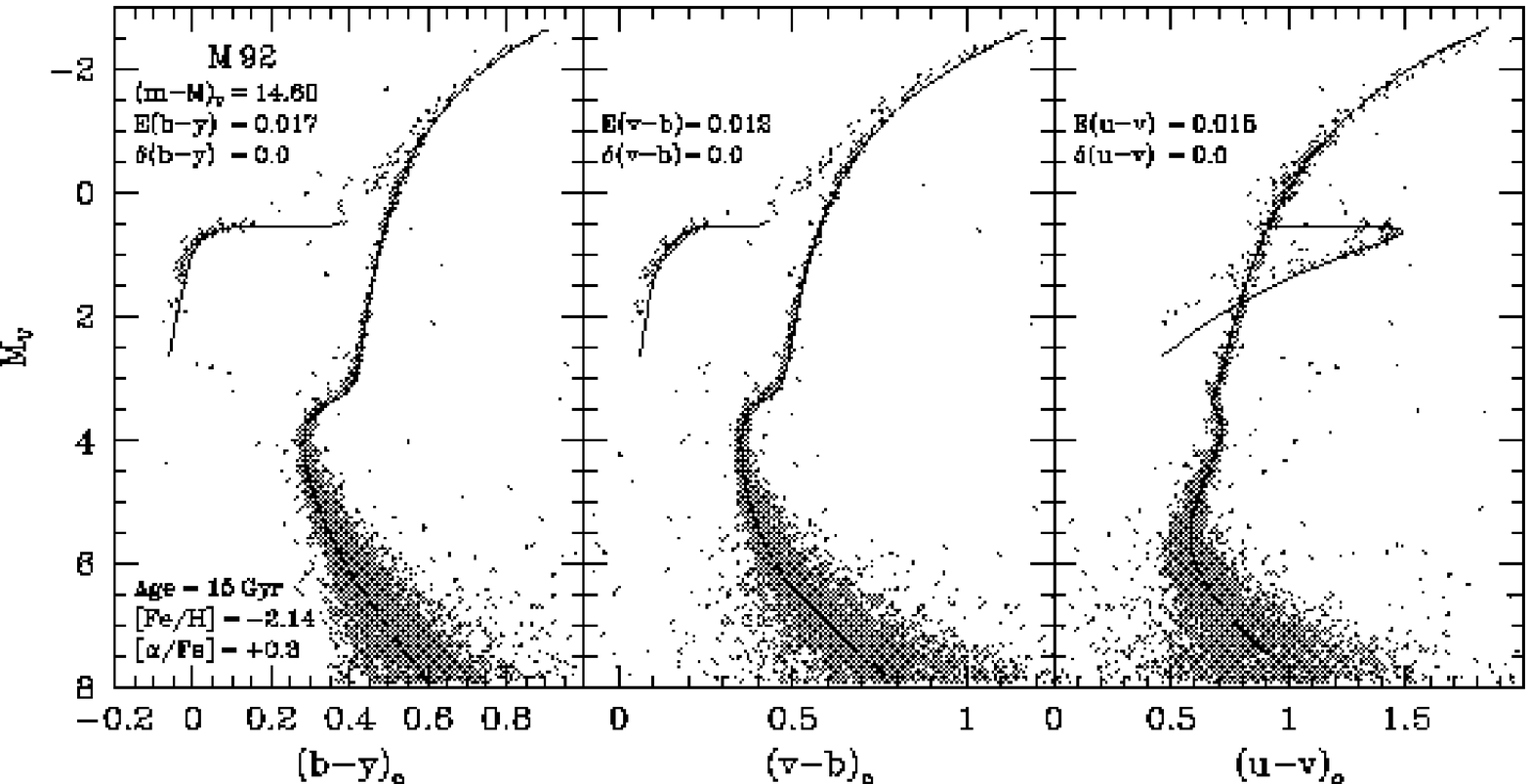} 
\caption{Various $uvby$ CMDs for the metal-poor globular cluster M$\,$92 overlaid with
a 15~Gyr, $\FeH=-2.14$ isochrone and ZAHB model \citep{BergbuschVandenBerg2001}. These
models are transposed to the observed planes using our synthetic \Stromgren colors and
the bolometric corrections from Paper I.  Reddening corrections \citep[corresponding 
to $\EBV=0.023$,][]{Schlegel1998} are as indicated, and the adopted apparent
distance modulus is based on the local Population~II subgiant HD~140283.  Only in the
case of the $u-v$ data was an additional color shift applied to the observations (see 
footnote 6), aside from the reddening.} 
\label{fig:m92} 
\end{figure}

To begin our assessment of the quality of the computed \Stromgren colors, we will
examine the fits of isochrones to the various $uvby$ CMDs of the globular cluster
M$\,$92.  Indeed, this same cluster played an important role in the testing of the
BV(RI)$_C$ transformations and bolometric corrections for extremely metal-poor stars
in Paper I.  In order to be consistent with Paper I, we will assume the same apparent
distance modulus [$(m-M)_V=14.6$, \citealt{Grundahl2000a}] and reddening value
[$\EBV=0.023$, \citealt{Schlegel1998}] for the purpose of our
analysis.\footnote{Assuming E$(b-y)=0.74\EBV$, E$(m_1)=-0.32\Eby$, and
E$(c_1)=0.20\Eby$ \citep{CrawfordMandwewala1976}, we find that E$(v-b)=0.50\EBV$ and
E$(u-v)=0.65\EBV$.} Figure \ref{fig:m92} presents the fit of a 15~Gyr isochrone and
zero-age horizontal branch (ZAHB) model \citep{BergbuschVandenBerg2001} for
$\FeH=-2.14$, which is within $\pm0.1$~dex of the values derived by
\citeauthor{ZinnWest1984} and \citeauthor{CarrettaGratton1997}, and
[$\alpha$/Fe]$\,=+0.3$ \citep{Carney1996} to the cluster data on three different
\Stromgren color-magnitude planes.  Upon initial inspection, it is quite obvious that
both the ZAHB model and the isochrone provide superb and consistent fits to the
photometric data on all three CMDs from the tip of the red giant branch, through the
turnoff region, and down to $M_V\approx6$.\footnote{According to
\citet{Grundahl2000a}, the $u$ photometry for stars in M$\,$92 could suffer from a
zero point problem in the sense they are $\sim0.04$~mag too faint. Therefore, we have
applied a $-0.04$~mag shift to the $u-v$ photometry in Figure \ref{fig:m92} to
compensate for this discrepancy.} Moreover, our interpretations of the cluster $uvby$
data is completely consistent with that obtained in Paper I where the same 15~Gyr
isochrone was found to provide the best fit to the $B-V$ data for M$\,$92 (see their
Fig.~1) reported by \citet{StetsonHarris1988}.  To be sure, the cluster reddening,
metallicity, and distance may not be exactly as we have assumed here, and the
isochrones may be deficient in some respects, but to within all of these
uncertainties, the isochrone fits to M$\,$92 presented in Figure \ref{fig:m92} seem to
indicate that our synthetic $uvby$ color for $\FeH\leq-2.0$ are able to reproduce the
observed photometry of old, metal-poor stars quite successfully.

\begin{figure} 
\plotone{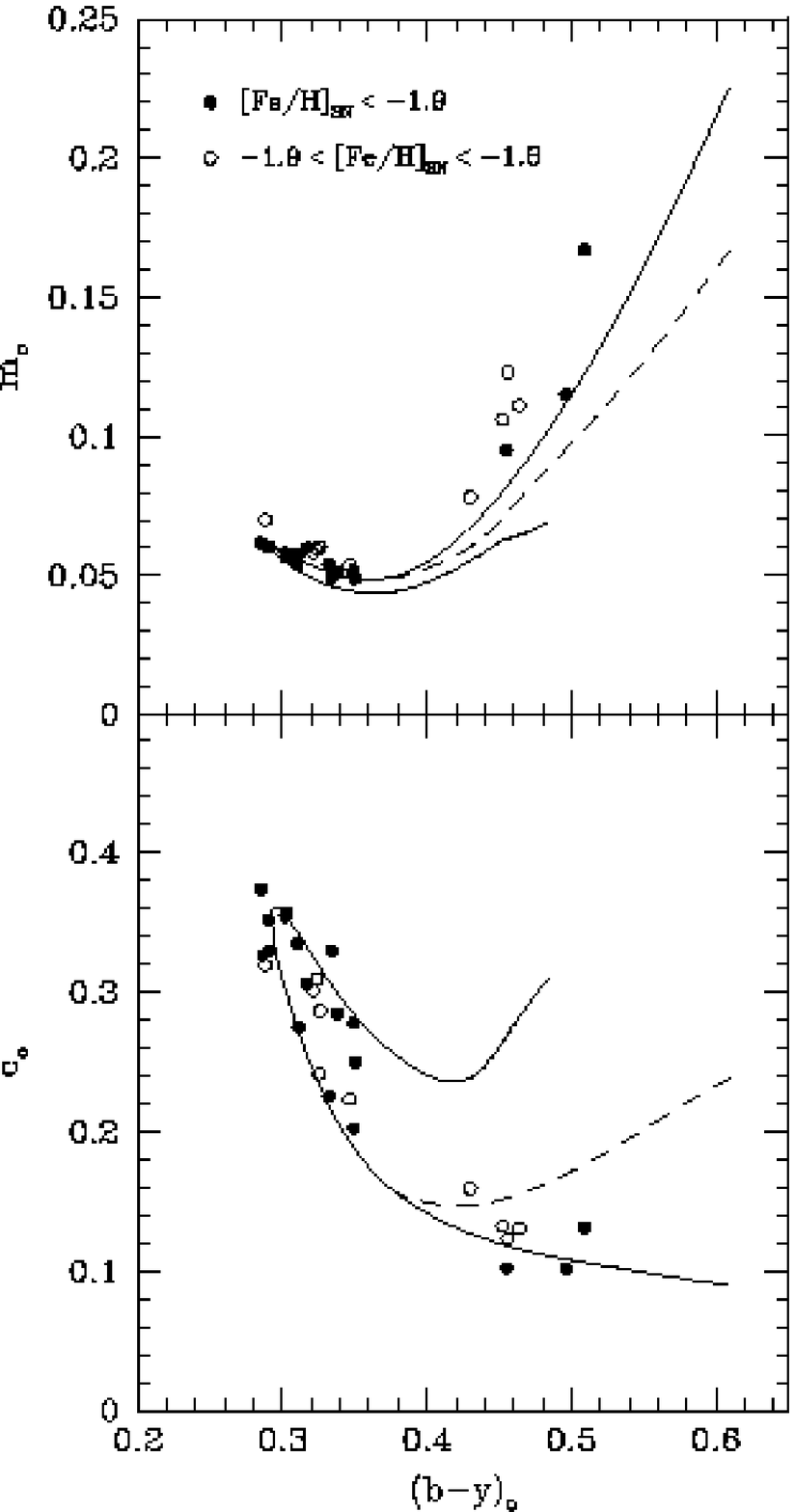} 
\caption{Color-color plots for a sample of metal-poor field stars with $uvby$
photometry from the survey of \citet{SchusterNissen1988}. The {\it filled} and {\it
open circles} denote stars having photometric metallicity estimates from
\citet{SchusterNissen1989b} within the indicated ranges.  The {\it dashed} and {\it
solid curves} represent, in turn, the same 15~Gyr, $\FeH=-2.14$ isochrone used in the
previous figure before and after corrections were applied to the synthetic $uvby$
color-temperature relations (see the text).}
\label{fig:snbym1c1} 
\end{figure}

Apart from testing our synthetic $uvby$ colors for $\FeH\leq-2.0$ using cluster stars,
we can also make use of a collection of field stars from the
\citet{SchusterNissen1988} study that have precise $uvby$ photometry and photometric
metallicity and reddening estimates derived from the calibrations of
\citet{SchusterNissen1989a}.  In Figure \ref{fig:snbym1c1} we compare, on two
dereddened color-color planes, the same 15~Gyr, $\FeH=-2.14$ isochrone used above with
the distribution of field stars having photometric metallicity estimates corresponding
to $\FeH<-1.8$. As indicated by the {\it dashed curves}, this isochrone matches the
warmer turnoff stars (those having $(b-y)_o\leq0.4$) quite well, but it deviates from
the loci defined by the few cooler dwarfs with $(b-y)_o\geq0.4$.  Based on these
distance-independent plots, we have adjusted the synthetic $v-b$ and $u-v$ colors at
$\Teff\leq5500\,$K and $\logg\geq3.5$ in order to alleviate these discrepancies.  
When the resultant empirically corrected transformations are employed, the 15 Gyr
isochrone is given by the {\it solid curves}, which clearly provide much improved fits
to the coolest field dwarfs.  [Because the adjustments to the $v-b$ and $u-v$ colors
are small in comparison with the breadth of the main-sequence photometry of M$\,$92 at
$(b-y)_o\sim 0.4$ (see Figure \ref{fig:m92}), they have no discernible impact on the
quality of the isochrone fits to the cluster CMD.  Note, as well, that the
displacement of the {\it open circles} to the left of the {\it solid curve} in the top
panel of Figure \ref{fig:snbym1c1} is consistent with them being $\approx0.25$~dex
more metal rich than the isochrone.]  We are thus led to conclude from Figure
\ref{fig:snbym1c1} that the synthetic colors corresponding to cool metal-poor dwarf
stars are in error and that, in order for our $uvby$ colors to accurately describe the
properties of such stars, some corrections to the transformations derived from model
atmospheres are necessary, even at metallicities slightly below [Fe/H] $= -2.0$. (The
corrections to the colors for metal-poor stars are discussed in more detail in Section
\ref{subsec:othercolcor}. Possible causes of the discrepancies are discussed below.)

\begin{figure} 
\plotone{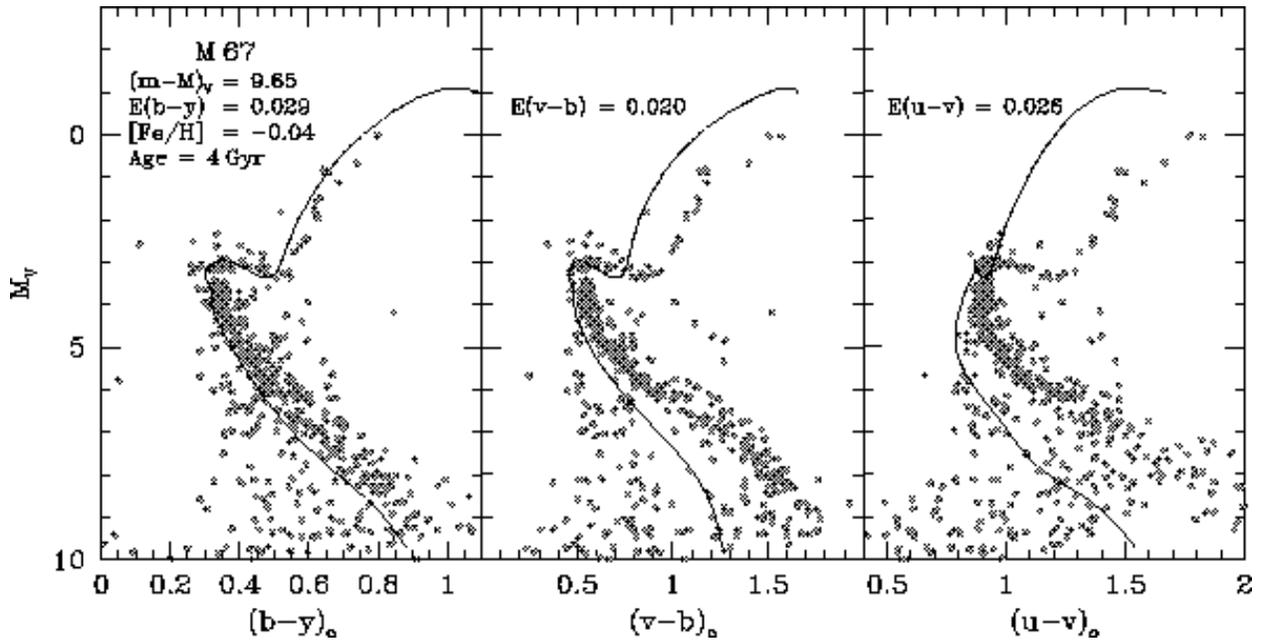} 
\caption{Same as Fig.~\ref{fig:m92} except for the open cluster M$\,$67.  The
$solid~line$ plotted in each panel represents the same 4~Gyr, $\FeH=-0.04$ isochrone
that provided the best fit to the broadband photometry of M$\,$67 in Paper I.  This
isochrone, when transformed to the respective CMDs using our purely theoretical
$uvby$ colors, obviously fails to reproduce the photometric data, particularly for
$v-b$ and $u-v$.}
\label{fig:m67badplot} 
\end{figure}

Turning now to solar metallicity stars, we present in Figure \ref{fig:m67badplot} an
analogous plot of $uvby$ CMDs, in this case for the open cluster M$\,$67.\footnote{It
is important to note that the $uvby$ photometry for M$\,$67 presented here is
unpublished, and the photometric zero points in the transformation from the
instrumental to the standard system are still preliminary.  Due to this fact, we have
performed a detailed comparison of our M$\,$67 CCD photometry to the photoelectric
photometry published by \citet{Nissen1987} and found differences of $b-y=0.013$,
$v-b=0.002$, and $u-v=0.025$ based on a total of 57 stars in common (our photometry
being redder in each case).  In order to be consistent with the data presented by
Nissen and his collaborators, we have applied these offsets to the photometry
presented in Figure \ref{fig:m67badplot}.  Unfortunately, due to the limited range in
color of the Nissen et al. data set, we are unable to determine whether or not there
is also a difference in the color scale between our CCD photometry and their
photoelectric photometry.  Owing to the rich line spectrum in the $u$ and $v$ bands,
even small differences between the filters used for our M$\,$67 observations and those
employed by Nissen et al.  could give rise to some significant differences in the
photometry.  While a detailed discussion of the calibration of our M$\,$67 photometry
is beyond the scope of this paper, we do make note of the fact that the open cluster
IC$\,$4651, which has photoelectric $uvby$ photometry available from
\citet{Nissen1988}, was also observed during the same run as M$\,$67 to specifically
check the accuracy of our transformations to the standard system.  Since this
IC$\,$4651 data set includes only 10 stars covering a small range in color located
near the cluster turnoff, we still cannot rule out possible trends as a function of
color.  It suffices to say, however, that {\it if} there are scale differences between
our CCD photometry and the photoelectric photometry for M$\,$67, then it is reasonable
to expect that stars lying at either the bluest or reddest colors would be affected
the most, whereas those near the turnoff would remain relatively unaffected except for
a possible zero-point offset.  To further investigate the quality of our photometry,
we have compared the locations of the M$\,$67 main sequence and red giant branch with
not only the standard sequences derived \citet{Olsen1984}, but also stars from the
$Hipparcos$ catalog on a variety of different color-color and color-magnitude planes.  
We find that there are no perceptible differences (i.e., different main sequence
slopes or locations of RGB stars) extending as far red as $b-y=0.8$, $v-b=1.4$, and
$u-v=1.3$.} Again, for the sake of consistency, we use the same values for the cluster
distance, metallicity, reddening as those adopted in Paper I, which compared
isochrones with the $B-V$, $V-R$, and $V-I$ CMDs of M$\,$67.  Unlike the previous
example for M$\,$92, however, the fits of our 4~Gyr, $\FeH=-0.04$ isochrone (from
VandenBerg, Bergbusch, \& Dowler, in preparation) to the cluster data exhibit some
large discrepancies between the computed and observed CMDs.  While the isochrone may
be brought into better agreement with the cluster turnoff if the colors are shifted
redward by small amounts depending on the index plotted, these shifts alone would
obviously not be able to reconcile the disagreement in the main-sequence and
giant-branch regions on the $v-b$ and $u-v$ planes.  The cause of this difference is
unlikely to be a problem with the temperature scale of the isochrone itself since
Paper I has already shown that the $(B-V)$--$\Teff$ relation predicted by the
isochrone is in very good agreement with empirical relationships (see their Fig.~10).  
In addition, the luminosities and temperatures of the M$\,$67 giant stars, as derived
from $V-K$ photometry, are consistent with the predictions of the same 4~Gyr isochrone
used here (see their Fig.~27).

There are a number of possible explanations for the differences seen in Figure
\ref{fig:m67badplot}.  First, it is possible that the atomic and molecular line list
used for calculating our synthetic spectra is not comprehensive enough to produce
reliable \Stromgren colors, particularly in the $u$ and $v$ pass bands where line
blanketing is especially strong.  Second, bands of the blue system of CN occur in the
$u$, $v$ and $b$ filters are certainly strong enough to be readily visible in spectra
of stars in the globular cluster 47 Tuc \citep[$\FeH\approx-0.8$,][]{Dickens1979}.  
For simplicity, our spectral calculations do not allow for differences in CNO
abundance.  Third, line blocking depends upon the value adopted for the microturbulent
velocity.  Finally, \citet{Bell2001} have found that, by incorporating bound-free
transitions of Fe~{\sc I} into the SSG models, it is possible to obtain a better fit
to the solar UV flux. The effects of this opacity source have not yet been included in
our stellar models.  Any combination of these factors could give rise to the mismatch
between our synthetic colors and the observed M$\,$67 data as well as the metal-poor
field dwarfs in Figure \ref{fig:snbym1c1}.

To compensate for the problems in the $uvby$ colors mentioned above, it is clear that
some corrections to our color--$\Teff$ relations are necessary to bring them into
better agreement with the observed data.  Indeed, similar problems with the $BV(RI)_C$
transformations were dealt with in Paper I by applying suitable adjustments to the
colors in order to satisfy empirical constraints imposed by both cluster and field
stars.  This semi-empirical approach will also be adopted here to correct our
synthetic $uvby$ colors.  In an effort to quantify the necessary corrections in the
simplest and most straightforward manner possible, we have chosen to follow the
methods of \citet[hereafter HBS2000]{Houdashelt2000b} who calibrated their synthetic
colors using a sample of field stars having precise $\Teff$ estimates determined using
the infrared flux method (IRFM).

Although HBS2000 developed their techniques as a means of semi-empirically
correcting their synthetic $UBVRIJHK$ colors, their methods can be easily adapted to
the present study provided that a large enough sample of field stars with $uvby$
photometry is available.  The HBS2000 investigation employed a total sample of 101
field dwarf and giant stars taken from the studies of \citet[hereafter
BG89]{BellGustafsson1989} and \citet[hereafter SH85]{SaxnerHammarback1985}.  We
note, however, that this sample is mainly limited to stars with metallicities near
solar and contains only two cool dwarfs with $\Teff<5000\,$K.  Given the significant
discrepancies between the observed and computed M$\,$67 CMDs in the vicinity of the
lower main sequence on the $v-b$ and $u-v$ planes, such a small sample of cool
dwarfs could pose a problem in deriving the correct calibrations for these indices
towards cooler $\Teff$'s.  Therefore, we have supplemented the HBS2000 list with a
much larger sample of stars with IRFM temperatures from the works of
\citet[hereafter AAM96 and AAM99, respectively]{Alonso1996a, Alonso1999a} that not
only cover a broader range in metallicity, but also contain more cool dwarf stars.  
By combining the field-star lists from all of these studies, the final sample used
here will not only be ideal for investigating the dependence of the color
calibrations on $\Teff$ and $\logg$, but, with the increased range in metallicity,
on $\FeH$ as well.
 
\subsection{The Field Star Sample}
\label{subsec:fieldstar}

While all of the studies mentioned above rely on the IRFM to determine $\Teff$, a
number of distinct differences exist between the methods and models employed by each.  
For example, both BG89 and SH85 use MARCS atmospheres to calibrate the ratio of
bolometric to infrared flux, while AAM96 and AAM99 rely on the stellar models of
\citet{Kurucz1993}.  Moreover, the techniques for deriving the bolometric flux
($\Fbol$) differ in the fact that both BG89 and SH85 compute this quantity from a
combination of 13-color, UV, and near-IR photometry, whereas AAM96/AAM99 rely solely
on integrated $UBVRIJHK$ photometry.  Therefore, some disagreement both in the
computed $\Fbol$'s and IRFM temperatures could arise from these different treatments.  
For this reason, we feel it is important to check that the IRFM temperatures from
these four separate studies are not only consistent with each other, but also that the
stellar angular diameters, predicted from the $\Fbol$ and $\Teff$ estimates, assuming
$\Teff\propto(\Fbol/\theta^2)^{1/4}$, are in good agreement with recent
interferometric estimates.

Table \ref{tab:comparisons} presents a comparison of the $\Fbol$ and $\Teff$ values
for a number of stars in common between the different studies.  The 34 dwarf and giant
stars examined by both AAM96/AAM99 and BG89 differ by $\approx6.2\%$ ($\pm2.5\%$) in
the mean $\Fbol$ value, or $\approx65\,$K ($\pm82$) in $<$$\Teff$$>$, in the sense
that the BG89 temperatures are hotter.  However, one dwarf star in common between the
two samples, HD~8086, deviates by more than 3$\sigma$ from the mean
temperature.\footnote{The HR~8085/8086 pair are the coolest dwarf stars that have IRFM
temperatures in both the AAM96/AAM99 and BG89 data sets and deserve a short discussion
regarding the rather large difference between their $\Teff$ (and $\Fbol$) estimates.  
The fact that HR~8085/8086 have spectral types of K5V and K7V, respectively, is
difficult to reconcile with the large difference of $\sim450\,$K in their $\Teff$'s
found by AAM96.  We would expect the temperatures of these two stars to differ by
$\lesssim250\,$K given their spectral types.  AAM96 attribute the difference between
their temperatures and those of BG89 for this pair to the unreliability of the models
atmospheres used to calibrate the ratio of $F_{IR}/F_{bol}$ due to the presence of
molecular absorption features in the infrared for cooler stars.  While the BG89
$\Teff$ estimates may be more realistic for these stars, BG89 does note that a
temperature of 4000$\,$K is possible for HR~8086 based on its photometry.  It is
important to note, however, that \citet{TomkinLambert1999} have derived spectroscopic
$\Teff$ values for HR~8085/8086 of 4450$\,$K and 4120$\,$K, respectively.  Since their
estimates are in better agreement with those of BG89 we have chosen to adopt the
temperatures derived by the latter for the subsequent analysis.} If this star is
rejected from the sample, then the average $\Teff$ difference is decreased to
$-56\pm60\,$K, with only a slight change in $\Fbol$ ($6.0\pm0.2\%$).  While this
offset in $\Fbol$ is likely associated with the different methods used to calculate
bolometric flux described above, the fact that the derived $\Teff$'s are in agreement
to within $\approx60\,$K is quite reassuring when one considers that the uncertainties
typically quoted for the IRFM range from 50 to 150$\,$K.  For AAM96/AAM99 and SH85 we
find slightly better agreement:  mean $\Fbol$ and $\Teff$ differences of $2.5\pm2.2\%$
and $8\pm43\,$K, respectively, if the anomalous star HR~2085 is omitted from the
consideration.  We conclude from this analysis that the $\Teff$'s computed by
AAM96/AAM99 are consistent (to within the uncertainties of the IRFM itself) with those
derived by SH85 and BG89.

Apart from confirming the consistency of the $\Teff$'s derived in different studies,
we must also ensure that the angular diameters computed from the IRFM temperatures and
the $\Fbol$ values listed in Table \ref{tab:comparisons} are in good agreement with
those obtained from more direct interferometric estimates.  To proceed, we make use of
the angular diameters recently compiled by \citet[hereafter N99 and N01,
respectively]{Nordgren1999, Nordgren2001} for giant stars using the Naval Prototype
Optical Interferometer.  Table \ref{tab:angdiameters} presents the comparison between
the N99/N01 measurements and those angular diameters inferred the results of BG89 and
AAM99.  While we opt to compare the angular diameters measurements, one can also
compare the stellar radii once the distance to the star is known.  For this reason, we
have included the $Hipparcos$ parallax estimates for these stars so the reader can
easily calculate and compare the stellar radii from the information given.  It is
important to note that the uniform-disk angular diameters determined from
interferometry must be corrected for limb-darkening before they may be compared with
those derived from the IRFM.  N99/N01 accomplish this by applying correction factors
between uniform disk and limb-darkened angular diameters from a set of coefficients
derived by \citet{Claret1995}.  The errors in $\theta$ quoted in Table
\ref{tab:angdiameters} come directly from the N99 and N01 studies, whereas those
computed from the IRFM temperatures were determined assuming a 5\% uncertainty in
$\Fbol$ and $\pm100\,$K in $\Teff$ for stars from both BG89 and AAM96/AAM99.  The mean
differences between the IRFM-derived angular diameters and those from interferometry
are only $0.054\pm0.108\,$mas and $-0.004\pm0.094\,$mas for BG89 and AAM96/AAM99,
respectively.  Therefore, we can be quite confident that the IRFM temperature scale is
correct.

In order to proceed with the calibrations of the synthetic \Stromgren colors, we must
first isolate stars from the lists of HBS2000 and AAM96/AAM99 that have both $uvby$
photometry and parallax estimates from the {\it Hipparcos} catalog.  Our primary
source of $uvby$ data for the color calibrations is the catalog of \citet[hereafter
HM98]{HauckMermilliod1998} which provides \Stromgren photometry for more than 60000
stars.  Although the HM98 catalog is ideal for our selection process, we are mindful
of the fact that their final tabulated photometry often represents the weighted mean
of several measurements compiled from different studies over the past four decades.  
Indeed, the data coming from such a large number of independent sources is sure to
exhibit some inhomogeneities due to the different observational equipment and/or
calibration techniques used by the various observers.  This is particularly true for
stars which lie in regions of the H-R diagram where the $uvby$ system is not well
defined (i.e., extremely red and blue stars) and differences between different data
sets can be as high as 0.1~mag for the $m_1$ and $c_1$ indices (see Olsen
\citeyear{Olsen1995} for a relevant discussion on this problem for late-type,
metal-deficient giants).  We do note, however, that the HM98 catalog is dominated by
the photometric samples collected by \citet{Olsen1983, Olsen1984, Olsen1993,
Olsen1994}, \citet{SchusterNissen1988}, and \citet{Schuster1993}.  The $uvby$ data
reported in these studies are particularly noteworthy since the authors generally used
the same instrumentation and reduction procedures to produce their final calibrated
photometry.

Our field star sample initially consisted of 559 stars that have both parallax and
$uvby$ data from the {\it Hipparcos} and HM98 catalogs.  Stars were subsequently
excluded from this list if their IRFM temperatures are higher than 8000$\,$K or below
4000$\,$K, if they are flagged for variability or multiplicity in the {\it Hipparcos}
catalog, or if their $uvby$ indices seem suspect when checked against their
temperatures or spectral types.  This culling process left us with 495 stars that were
then examined individually for possible inhomogeneities in their $uvby$ photometry
taken from HM98.  The photometry for approximately 75\% of these (365 stars) comes
predominantly from the studies mentioned in the previous paragraph, and we are
confident that their $uvby$ data are reliable enough for the color calibrations.  In
fact, no individual $b-y$, $m_1$, or $c_1$ measurement from these studies deviates by
more than 0.02~mag from the mean values listed in HM98 for any of these 365 stars.  
As far as the remaining 25\% of the sample is concerned, we have chosen to exclude
them entirely from our analysis if any of their individual $uvby$ indices, taken from
the various sources, differ by more than 0.05~mag from the HM98 means.  Furthermore, 
any star having only one set of $uvby$ measurements was excluded if its colors do not 
correspond well (i.e., to within 0.05~mag) with those of stars with similar 
temperatures, gravities, and metallicities that were retained in our sample.

\begin{figure} 
\plotone{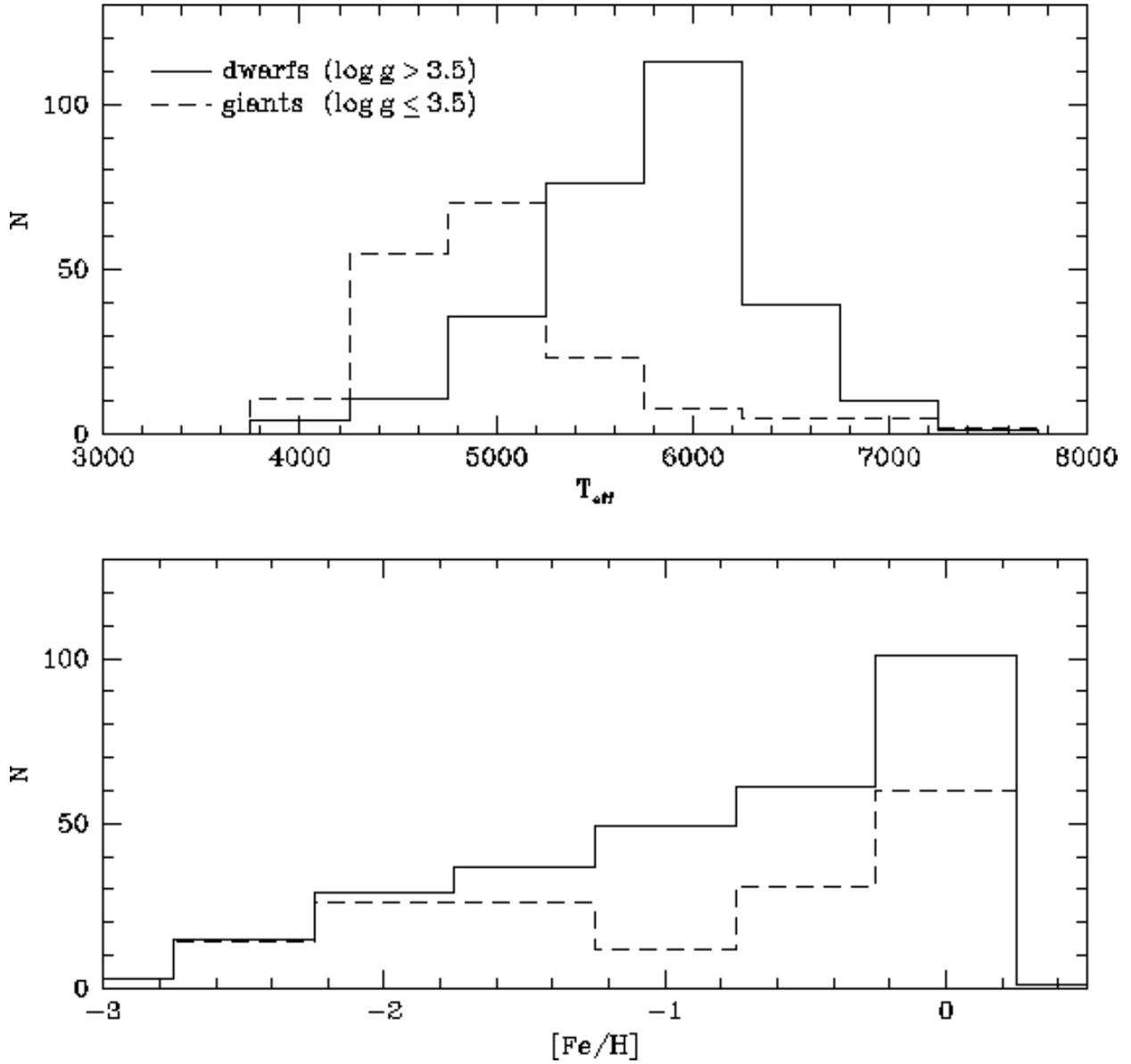} 
\caption{The distribution of our field star sample as a function of $\Teff$ and $\FeH$.  
Note that $\logg=3.5$ has been used to separate the dwarf- and giant-star
distributions.}
\label{fig:histogram}
\end{figure}

Our final sample consists of 478 field stars for which HBS2000 or AAM96/AAM99 provide
estimates for $\logg$ and $\FeH$.  In some cases, however, these values may not
necessarily be consistent with spectroscopic estimates.  This is especially true for
the stars listed by AAM96/AAM99, who assign only approximate values to the majority of
their sample for the reason that $\logg$ and $\FeH$ need only be accurate to within
0.5~dex and 0.3~dex, respectively, to obtain uncertainties in $\Teff$ of $\approx2\%$.  
Given the sensitivities of the \Stromgren $m_1$ and $c_1$ indices to metal abundance
and surface gravity, and the possible effects that uncertainties in these values may
have in the subsequent color calibrations, we have chosen to extract more precise
spectroscopic values from the $\FeH$ catalog of \citet{CayreldeStrobel2001}.  For
cases where the catalog provides more than one set of estimates for each star, we
adopt the median values for $\logg$ and $\FeH$.  Though the majority of these stars
are relatively nearby, some might be heavily reddened by local interstellar dust
clouds.  For this reason, we adopt the E$(B-V)$ values given by AAM96/AAM99, while
reddening estimates for stars from HBS2000 are derived from the extinction maps of
\citet{Schlegel1998} and corrected for distance using the expression
[1$-$exp($-$$|$$d$$\,$sin$\,$$b$$|$/$h$)], where $d$ is the star's distance (as
determined from the {\it Hipparcos} parallaxes), $b$ its galactic latitude, and $h$
the dust scale-height \citep[assumed to be 125~pc,][]{Bonifacio2000}.  The final
composite list of stars in our sample is given in Table \ref{tab:starlist}, and
histograms illustrating their distribution as a function of $\Teff$ and $\FeH$ are
shown in Figure \ref{fig:histogram}.

\subsection{Color Corrections at [Fe/H]$\,$=$\,$0.0}
\label{subsec:colorcalib0}

Given that a sizable fraction ($\sim35\%$) of the field stars in our sample have
metallicities within $\pm0.25$~dex of solar, they provide an excellent subset in which
to determine what corrections to the colors at $\FeH=0.0$ are needed to bring them into
better agreement with the observations.  In this section we aim to follow the methods of
HBS2000 by deriving corrections to synthetic colors based on simple polynomial fits to
the distribution of synthetic versus observed colors for a sample of field stars with
well-determined physical parameters.

\begin{figure}
\plotone{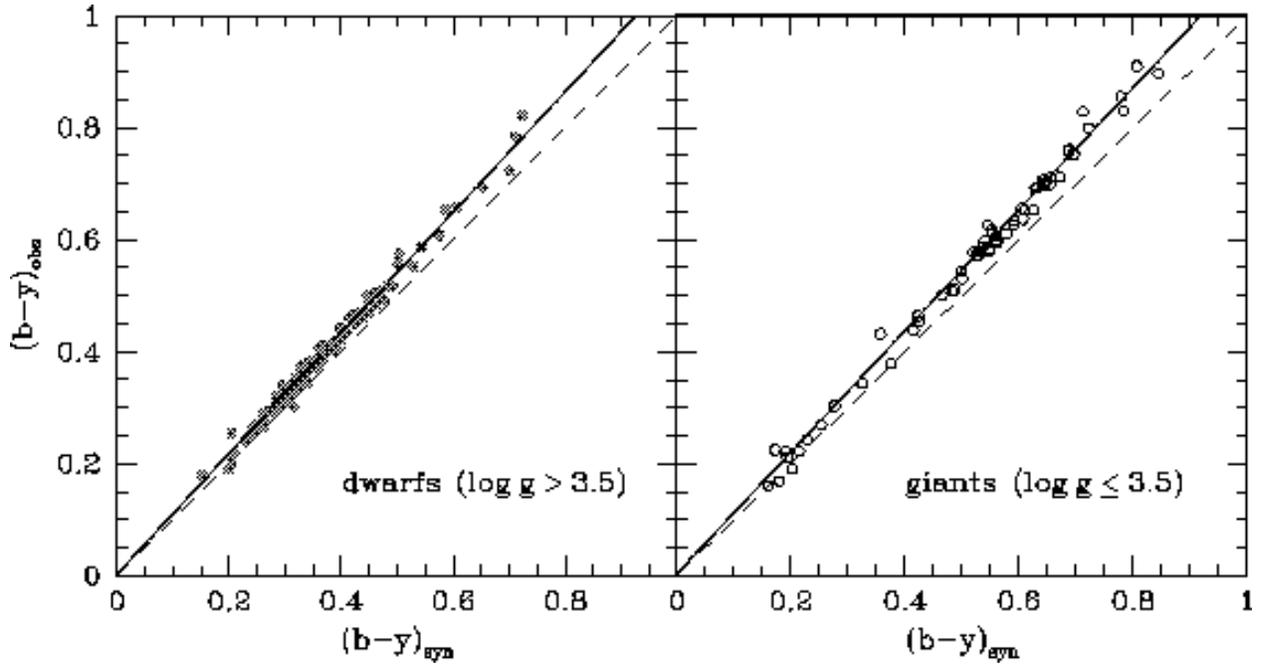}
\caption{The synthetic vs. observed $b-y$ photometry for field stars listed in Table
\ref{tab:starlist} with metallicities near solar (i.e., $-0.25\leq\FeH\leq+0.25$).  
The left-hand panel plots all dwarfs from the sample with $\logg>3.5$ whereas the
$open~circles$ in the right-hand panel indicate giants with $\logg\leq3.5$.  The
$solid~lines$ represent the linear, least-squares fits to the distribution of dwarfs
and giants in each panel while the {\it dashed lines} represent the lines of equality
between the observed and synthetic colors.  These fitted relations are used to correct
the synthetic $b-y$ colors to bring them into better agreement with the observed
photometry.}
\label{fig:bycalib}
\end{figure}

We begin with the calibration of the $b-y$ index.  In this case, as well as the
calibrations for the other \Stromgren colors that follow, a synthetic index for each star
is determined from direct interpolation within our color grid assuming the $\Teff$,
$\logg$, and $\FeH$ values listed in Table \ref{tab:starlist}.  This synthetic color is
then plotted against its observed, dereddened counterpart for all stars in the sample that
fall within $-0.25\leq\FeH\leq+0.25$ in order to establish the calibration of the model
colors at $\FeH=0.0$.  Figure \ref{fig:bycalib} presents such a plot for the $b-y$ index
with dwarfs and giant stars separated into different panels as a means of checking for
possible differences between stars of different gravity.  Inspection of the figure reveals
that the synthetic colors for both the dwarfs and giants exhibit noticeable systematic
deviations from equality ({\it dashed line}) towards cooler temperatures.  If
simple linear, least-squares fits are derived for each of the two sets separately, we
indeed find that the slopes are greater than unity (see Table \ref{tab:calibrations}).  
Furthermore, a single linear fit involving the dwarfs and the giants together show that
they follow very nearly the same trend as those obtained when they are treated separately
to within the errors of the fitted lines.  Based on this result, we conclude that our
synthetic $b-y$ colors at $\FeH=0.0$ can be suitably corrected to match the observed
field-star photometry using a single linear calibration.

\begin{figure} 
\plotone{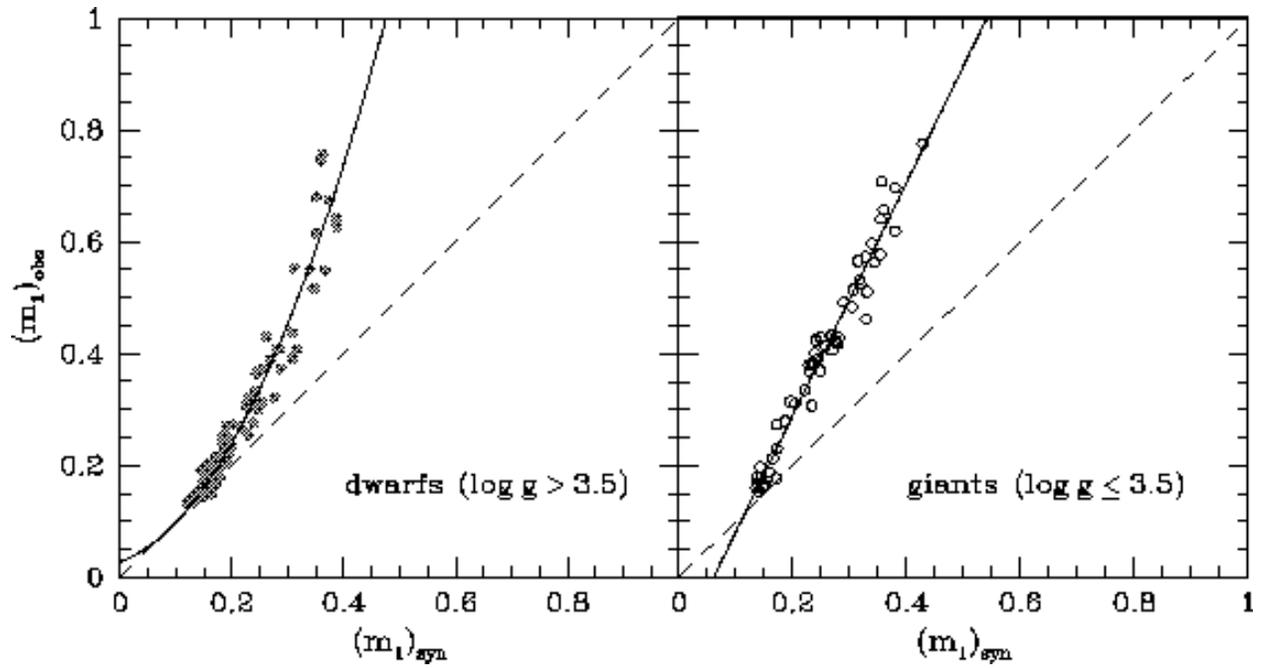} 
\caption{Same as Figure \ref{fig:bycalib} except for the $m_1$ index.  In this case,
the dwarf stars are best fitted using a second-order polynomial, whereas the giants
follow the indicated linear relation.}
\label{fig:m1calib}
\end{figure}

In Figure \ref{fig:m1calib} we present a plot comparing the synthetic versus observed
$m_1$ colors for the same subset of stars considered in Figure \ref{fig:bycalib}.  In
this case the synthetic colors exhibit substantial deviations from their observed
counterparts for both the dwarfs and the giants.  Unlike the $b-y$ index, however, the
dwarfs appear to follow a somewhat different trend than the giants, and a single
linear calibration would not be satisfactory.  Indeed, the interpretation of this
diagram is more complex that that of Figure \ref{fig:bycalib} in the sense that $m_1$
is sensitive to the abundances of some individual elements and isotopes (e.g., C, N,
and $^{12}$C/$^{13}$C) as well as the overall metal abundance, while $b-y$ depends
primarily on temperature.  However, since these deviations in Figure \ref{fig:m1calib}
appear to be consistent with the discrepancies between our 4~Gyr isochrone and the
$v-b$ CMD for M$\,$67 [recall that $m_1=(v-b)-(b-y)$] in Figure \ref{fig:m67badplot},
we proceed to correct the $m_1$ colors using the same procedure as that employed for
the $b-y$ index.  We fit the dwarf-star distribution using a second-order polynomial,
whereas a linear relation is derived for the giants.  The corresponding coefficients
of these fits are again given in Table \ref{tab:calibrations}.  This type of
calibration for the dwarf-star $m_1$ colors is not unreasonable.  For comparison, some
of the synthetic broadband colors computed by HBS2000, particularly the $B-V$ and
$V-I$ indices, exhibited large deviations among the coolest dwarfs in their sample.  
Their solution involved a separate cool-dwarf calibration that deviated from their
derived fit to the warmer dwarfs at a temperature of 5000$\,$K.  We have similarly
investigated if two separate linear calibrations, one for warm dwarfs and another for
cool dwarfs using 5000$\,$K as the dividing temperature, would adequately correct the
$m_1$ colors and found that the cool dwarfs appear to be ``over corrected" in a sense
that their calibrated colors extend too far to the red to adequately fit the $v-b$
(and $m_1$) photometry on the lower main sequence of M$\,$67.  Therefore, the original
second-order polynomial is used to correct the $m_1$ colors for $\logg>3.5$, and we
have taken great care to smoothly meld these calibrated dwarf-star colors to those for
the giants at $\logg=3.5$.

\begin{figure}
\plotone{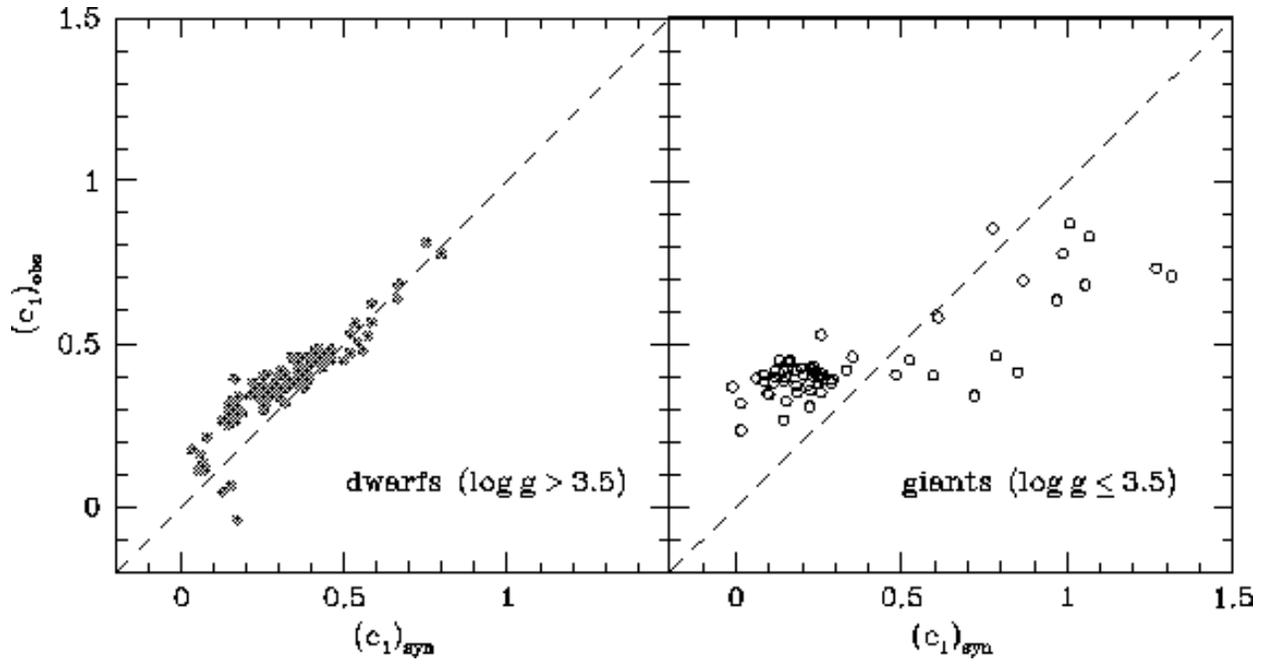}
\caption{Same as Figures \ref{fig:bycalib} and \ref{fig:m1calib} except for the $c_1$
index.  Note that both the dwarfs and giants do not appear to follow any specific
trend, either linear or polynomial, and as a result, it is not possible to derive a
satisfactory calibration for $c_1$ from such plots.}
\label{fig:c1calib}
\end{figure}

The final index left to calibrate is the \Stromgren $c_1$ index.  Upon inspection of
Figure \ref{fig:c1calib}, however, it would appear that this index poses even more of
a problem to calibrate.  While the dwarf stars are fairly well defined in the plot,
the giants show an appreciable scatter at large $c_1$ values and do not seem to follow
any specific trend.  As with the $m_1$ index, the correct interpretation of Figure
\ref{fig:c1calib} depends on the sensitivity of $c_1$ to the effects of surface
gravity, chemical abundance, $and$ temperature.  Moreover, the $u-v$ component of
$c_1$ is not a monotonic function of temperature.  Some additional factors that may
contribute to the problems with the $c_1$ index could be missing absorption lines
and/or the exclusion of the aforementioned Fe~{\sc I} opacity source in our SSG
spectra.  In addition, as mentioned in the previous section, the scatter at large
$c_1$ values for the giants could be associated with inhomogeneities in the HM98
catalog due to the fact that the \Stromgren system is not well established for these
types of stars.  However, we have been careful to exclude stars if their photometry
seems suspect, and we are confident that the scatter seen in the right-hand panel of
Figure \ref{fig:c1calib} is real.  To complicate matters further,
\citet{Grundahl2000b} first cited evidence that the $c_1$ colors of RGB stars in
globular clusters exhibit a rather large scatter that is much greater than the
photometric uncertainties.  This $c_1$ scatter has since been confirmed to be present
among RGB stars in {\it all} 21 globular clusters surveyed in the Grundahl program
(Grundahl et al., in preparation).  This effect has been interpreted as star-to-star
differences in the abundance of nitrogen \citep{Grundahl2002a}.  Since numerous NH
lines lie within the \Stromgren $u$ filter, the $c_1$ color of any star with an
abnormal abundance of nitrogen will be different from that of one with a normal
abundance.  As a result, our synthetic $c_1$ colors, which are derived from MARCS/SSG
models assuming scaled-solar abundances, cannot be expected to reproduce the observed
colors of field stars having abnormal abundances of nitrogen.

\begin{figure} 
\plotone{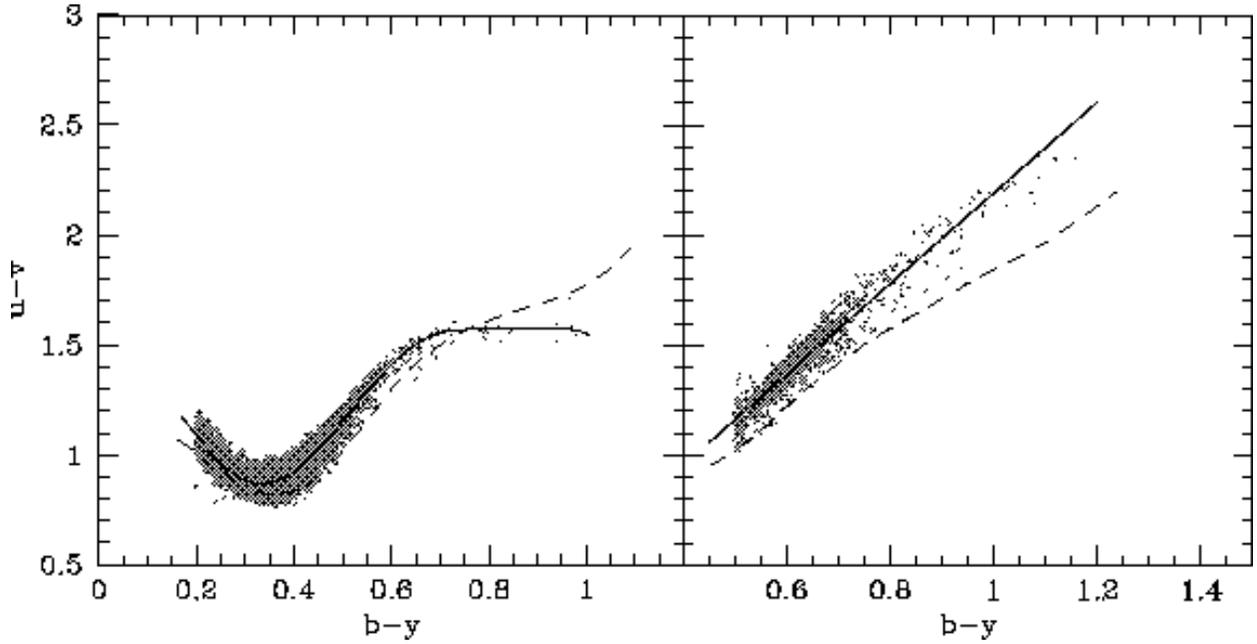} 
\caption{Two ($b-y$,~$u-v$) diagrams for a sample of dwarf and giant stars (left- and
right-hand panels, respectively) having parallax estimates from {\it Hipparcos} and 
$uvby$ data from the EHO catalog.  The
$solid~lines$ indicate the relationships derived to fit the distribution in the
photometric data and are used to correct our synthetic $u-v$ colors.  {\it Dashed
lines} represent the locus of a solar metallicity ZAMS model (for the dwarfs) and the
giant branch from our 4~Gyr isochrone {\it before} corrections to the $u-v$ colors are
applied.}
\label{fig:byuvcalib} 
\end{figure}

Given the obvious lack of agreement between the synthetic and observed indices in
Figure \ref{fig:c1calib}, it is clearly very difficult to derive any calibrations that
would adequately correct the dwarf and giant $c_1$ colors.  Consequently, we have
explored alternate techniques of correcting the $c_1$ colors for solar-metallicity
models, and found that the most straightforward way involved working with the
distribution of field dwarf and giant stars which have $uvby$ photometry and parallax
estimates from {\it Hipparcos} on the ($b-y$,~$u-v$) plane.  We choose to deal
predominantly with the synthetic $u-v$ colors rather than $c_1$ itself since the
latter includes a combination of both the $u-v$ and $v-b$ indices [recall that
$c_1=(u-v)-(v-b)$].  As the $v-b$ and $b-y$ indices have already been calibrated, we
only need to investigate what corrections are required to fix the synthetic $u-v$
colors.  In Figure \ref{fig:byuvcalib} we present the color-color plots for those
dwarfs and giants with accurate parallax estimates from {\it Hipparcos}.  Rather than
rely on the HM98 catalog as our source of photometry for this analysis, we have
instead chosen to extract the data from a catalog of accurate and homogeneous
$uvbyH\beta$ photometry recently compiled by E.~H.~Olsen (private communication) from
his published samples \citet{Olsen1983, Olsen1984, Olsen1993, Olsen1994}.  This
catalog, hereafter referred to as the EHO catalog, is comprised of almost 30000 stars
in the northern and southern hemispheres, all of which are reduced carefully to the
standard $uvby$ system.  Since these {\it Hipparcos} stars are relatively nearby, we
can safely neglect the effects of reddening, and assume they all have metallicities
near solar.  To ensure that the purest sample of dwarfs and giants are presented in
both panels, we impose cuts on the data based on the star's absolute magnitude and
$b-y$ color.  For instance, all cool dwarf stars plotted in the left-hand panel of
Figure \ref{fig:byuvcalib} have $M_V\geq10\;(b-y)$ and $b-y\geq0.2$ (corresponding to
$\Teff\lesssim7250\,$K), and we have isolated the giant stars to $M_V\leq4.5$ and
$b-y\geq0.5$ ($\Teff\lesssim5250\,$K).  In the case of the dwarfs a 6th order
polynomial, using $b-y$ as the independent variable, is used to fit the distribution
of data between $0.2\leq\,b-y\,\leq1.0$, while the giant stars are fit using a simple
linear relation for $0.5\leq\,b-y\,\leq1.2$.\footnote{While \citet{Caldwell1993} have
derived extensive color-color relations between field stars for the \Stromgren system,
their calibrations towards cooler temperatures are biased towards giant stars due to
paucity of extremely red dwarfs in their sample.  As a result, when their relations
are plotted on the data in Figure \ref{fig:byuvcalib}, we find that the warm dwarfs
are fit rather well, but the calibration shifts to giant stars around $b-y=0.6$.  
Therefore, we have chosen to derive our own calibrations rather than rely on theirs.}

These relations, which are indicated in each panel of Figure \ref{fig:byuvcalib} by a
{\it solid curve}, are subsequently used to correct our synthetic $u-v$ colors.  In
the case of the dwarf stars, the synthetic $u-v$ colors predicted from a
solar-metallicity ZAMS model ({\it dashed curve}) are forced into agreement with the
polynomial fit.  In general this meant applying redward shifts ranging approximately
from 0.01 to 0.1~mag in the $u-v$ colors for $0.3\leq\,b-y\,\leq0.7$, whereas a
combination of positive and negative corrections were required to match the
distribution of the cool dwarf stars at $b-y\gtrsim0.7$.  Similarly, the $u-v$ colors
corresponding to the giants in the right-hand panel are brought into agreement with
the derived linear relation by using the color predictions from the giant branch of
the 4~Gyr, $\FeH=-0.04$ isochrone ({\it dashed curve}).  These corrections for the
giants were generally much larger than for the dwarfs and ranged from +0.15 to +0.25
depending on $b-y$ color.

\begin{figure}
\plotone{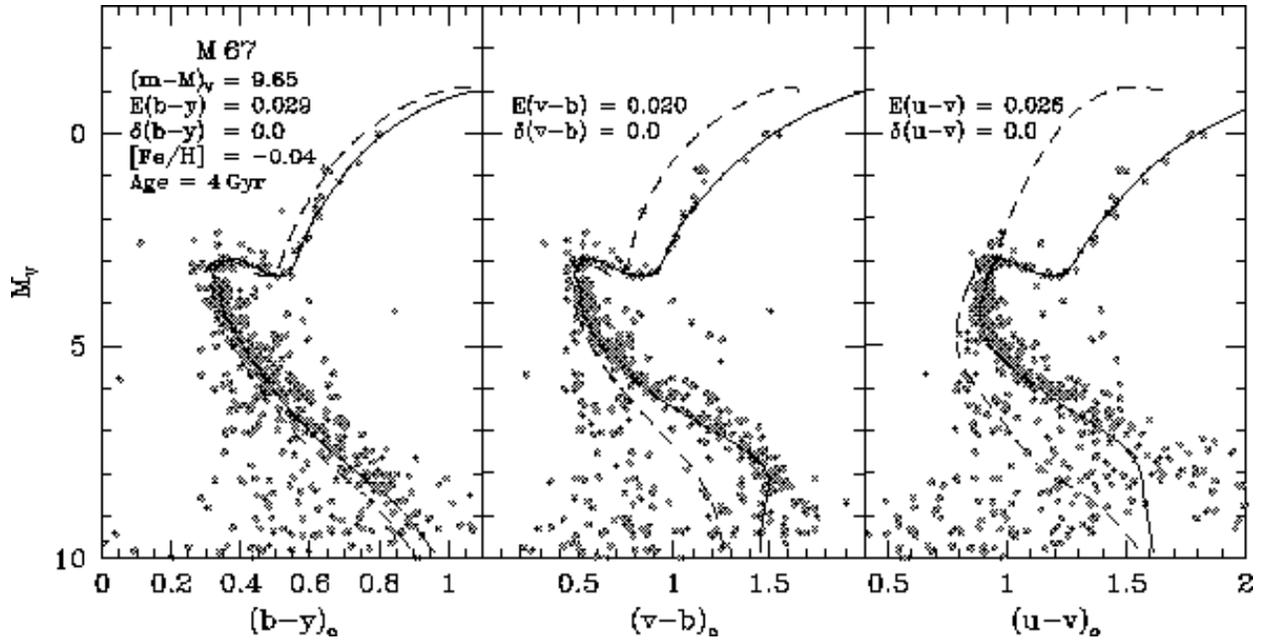}
\caption{Same as Figure \ref{fig:m67badplot} except the 4~Gyr isochrone ({\it solid lines}) 
has been transformed to the color-magnitude planes using our newly calibrated $uvby$ 
colors while the {\it dashed lines} represent the {\it uncalibrated} isochrone.} 
\label{fig:m67goodplot}
\end{figure}

With the synthetic \Stromgren colors at $\FeH=0.0$ now placed onto the observational
system as the result of our analysis of field stars, we can again assess how well we
can reproduce the various CMDs of M$\,$67.  Figure \ref{fig:m67goodplot} provides the
revised fits of the same 4~Gyr isochrone used in Figure \ref{fig:m67badplot}, except
that the transformation to the observed planes is accomplished using the calibrated
$uvby$ colors.  (Note that the same reddening and distance are adopted as in Figure
\ref{fig:m67badplot}.)  The uncalibrated and calibrated isochrones are shown as {\it
dashed} and {\it solid lines}, respectively.  Overall, the fits to the various M$\,$67
CMDs using the calibrated colors have been quite dramatically improved as compared
with those using the purely theoretical indices.  Importantly, the fits to all three
CMDs now show excellent consistency with each other as well as with the
interpretations of the $B-V$, $V-R$, and $V-I$ CMDs of M$\,$67 discussed in Paper I.

\begin{figure} 
\plotone{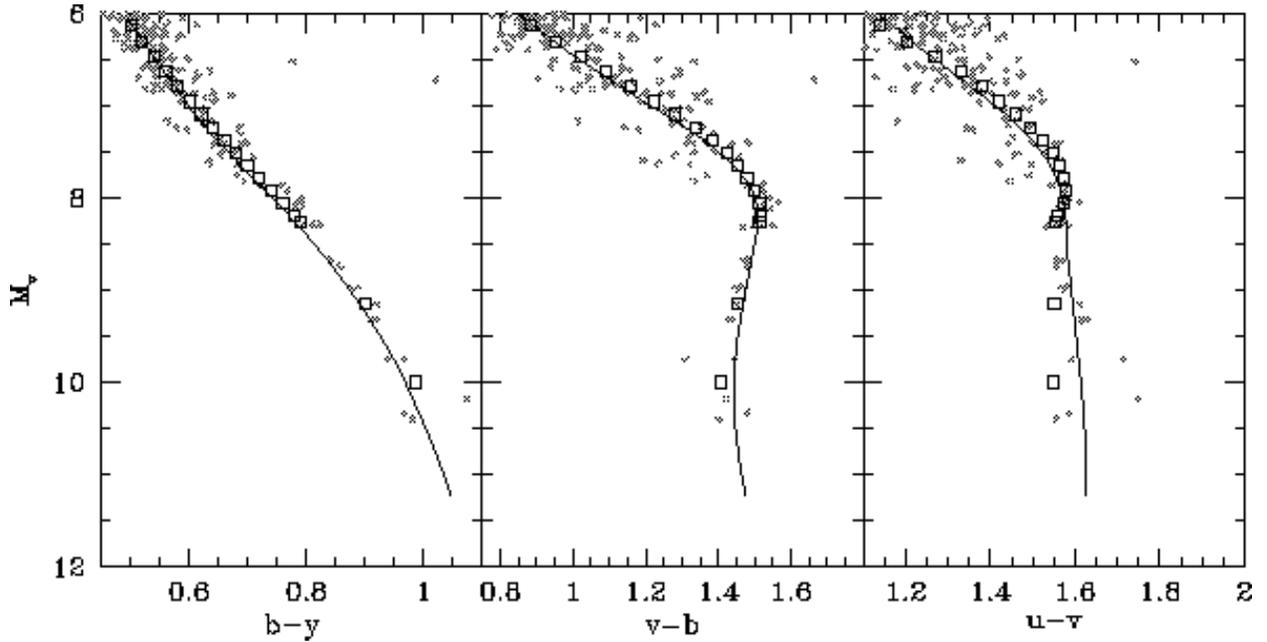} 
\caption{Various CMDs for K- and M-type dwarf stars in the solar neighborhood with
$M_V$ values derived from their {\it Hipparcos} parallaxes and $uvby$ photometry from
the EHO catalog.  Only those stars with $\sigma_{\pi}$/$\pi\leq0.1$ have been plotted.  
Our synthetic colors at $\Teff\lesssim4500$~K have been constrained by these data so
that our ZAMS for $\FeH=0.0$ is given by the $solid~curve$ in each panel.  As a check,
we compare our ZAMS fit to the empirical standard relation for late-type dwarf stars
({\it open squares}) as derived by \citet{Olsen1984}.}
\label{fig:ZAMSlower}
\end{figure}

It worth noting that the preceding calibrations of the synthetic colors are
technically valid for those dwarf and giant star models with $\FeH=0.0$ and
$4000\leq\Teff\leq8000\,$K since we have employed only solar-metallicity field stars
that fall within this temperature range.  While a detailed discussion of the
corrections made to models with metal abundances other than solar is deferred until
later, we make a few remarks here concerning the color corrections for $\Teff$'s
outside this range.  As mentioned in Section \ref{sec:colorcalc}, we have adopted the
synthetic $uvby$ colors of CGK97 for $\Teff>8000\,$K and have made small corrections
(typically less than 0.01$-$0.02~mag depending on the index) to our synthetic colors
at temperatures of 7500, 7750, and 8000$\,$K in order to meld our grid smoothly with
theirs.  At temperatures below 4000$\,$K, we apply corrections to the colors at
$\FeH=0.0$ in an effort to match the CMDs for a sample of extremely red field dwarf
stars from the EHO and {\it Hipparcos} catalogs.  In Figure \ref{fig:ZAMSlower} we
present the fits of a ZAMS model having $\FeH=0.0$ which has been transformed to the
indicated CMDs using the final corrected colors ({\it solid curve}) and overlayed on
the photometry for stars having extremely precise parallaxes (i.e.,
$\sigma_{\pi}/\pi\leq0.1$).  This technique is similar to that presented in Paper I,
which relied upon a large number of Gliese catalog stars to constrain the $BV(RI)_C$
color--$\Teff$ relations down to $M_V\sim13$ (see their Fig.~17).  However, very few
of these low-mass Gliese stars have observed $uvby$ data available in the EHO catalog
and we can only define our color transformations accurately down to $M_V\approx10.5$
($\Teff\approx3500\,$K).  Therefore, the corrections applied to the colors at 3000 and
3250$\,$K are somewhat more uncertain since we do not have any additional data for
extremely low mass stars that would help to better constrain them.  As an additional
check of our color corrections we plot the $uvby$ standard relation for late-type
dwarf stars derived by \citet{Olsen1984} as open squares in each panel of Figure
\ref{fig:ZAMSlower}.  Overall, the ZAMS and the standard relation agree quite well in
all three panels except at $M_V>9$ in the $u-v$ plot where the Olsen trend appears to
deviate from the field star distribution towards the blue.  (An implicit assumption
here is that the $T_{\rm eff}$ scale of the ZAMS models for very low mass stars is
accurate.  For some discussion of the reliability of this aspect of these models,
reference should be made to Paper I.)

\subsection{The Calibrated Colors and {\it Hipparcos} Field Stars}
\label{subsec:Hipparcos}

\begin{figure} 
\plotone{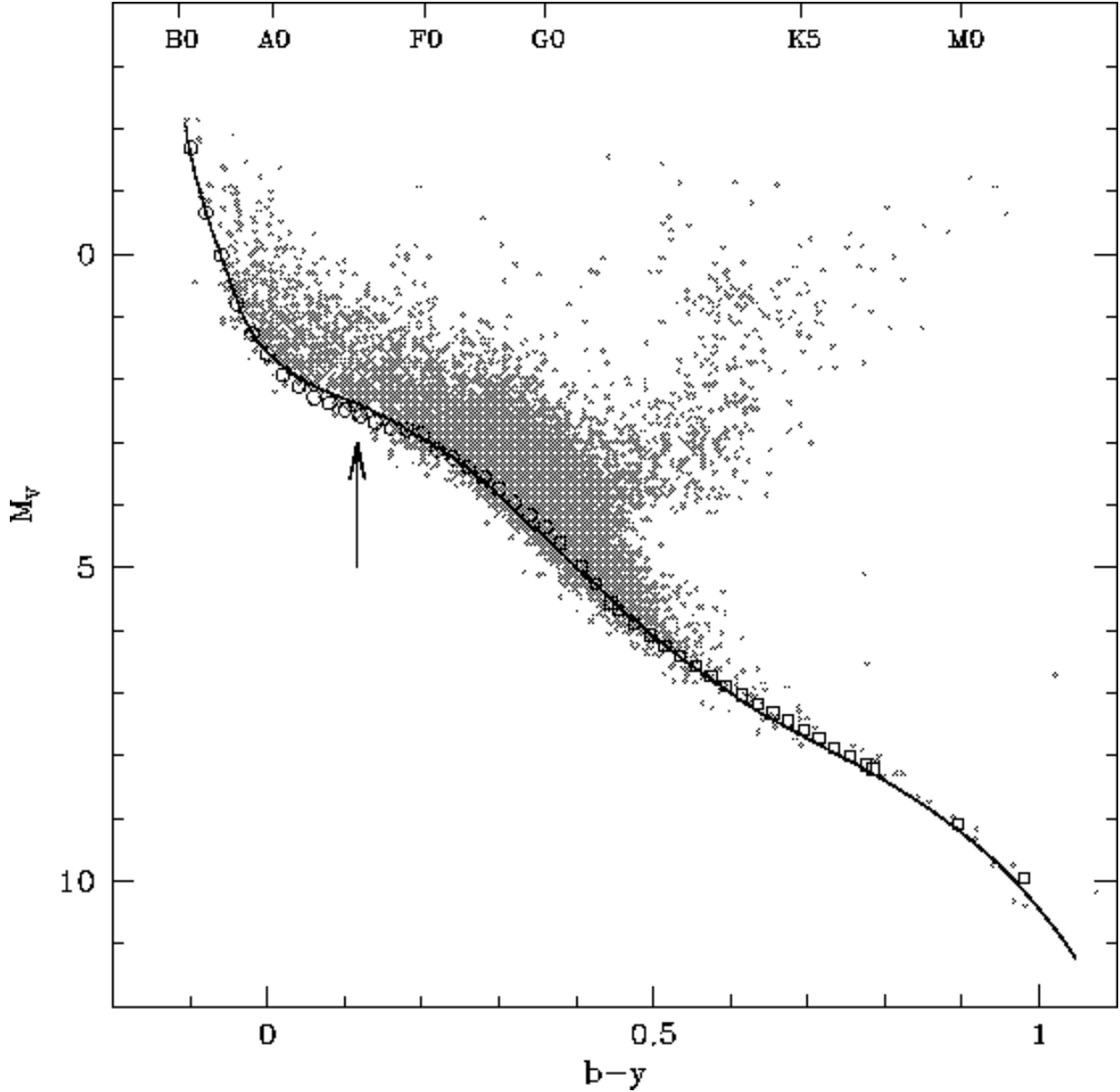} 
\caption{The ($b-y$,~$M_V$) diagram for field stars from the {\it Hipparcos} catalog
having accurate parallax estimates (i.e., $\sigma_{\pi}/\pi\leq0.1$) and available
$uvby$ photometry from EHO catalog.  Two empirical standard relations corresponding
to warm dwarfs \citep[$open~circles$]{PhilipEgret1980} and cool dwarfs
\citep[$open~squares$]{Olsen1984} are also plotted on the data to represent the
approximate location of solar metallicity dwarf stars in \Stromgren color-magnitude
space.  A ZAMS locus for $\FeH=0.0$ is transformed to the CMD using our newly
calibrated $b-y$ colors (and bolometric corrections from Paper I) and shown as a
{\it solid line}.  The vertical arrow located at $b-y\approx0.1$ denotes where our
cool star colors have been melded with those of CGK97 at a temperature of
8000$\,$K.}
\label{fig:ZAMSbyCMD} 
\end{figure}

To further illustrate the accuracy of our newly calibrated $uvby$ colors for
solar-metallicity stars, we demonstrate their ability to reproduce the observed
distribution of field stars on a variety of \Stromgren color-magnitude and
color-color planes.  For this investigation we again make use of the sample of
nearby {\it Hipparcos} stars described in the previous section.  Since this sample
is comprised primarily of stars with near-solar abundance lying close to the main
sequence, a ZAMS model for $\FeH=0.0$ is an appropriate locus to compare with the
$uvby$ data.  Figure \ref{fig:ZAMSbyCMD} presents the overlay of this ZAMS onto the
field-star photometry in the ($b-y$,~$M_V$) plane.  As mentioned earlier, the
\Stromgren photometry for each star was taken directly from the EHO catalog of
homogeneous $uvby$ data, while the broadband V magnitudes, which were used to derive
$M_V$, are from the original {\it Hipparcos} photometric catalog.  In addition to
the field-star data, we have plotted two empirical standard relations as defined by
\citet[$open~circles$]{PhilipEgret1980} for O$-$F-type main-sequence stars and by
\citet[$open~squares$]{Olsen1984} for G$-$M dwarfs.  The vertical arrow located at
$(b-y)\sim0.1$ indicates the region where our calibrated color-temperature relations
have been joined with those of CGK97 at a temperature corresponding to 8000$\,$K.  
Overall, the match to both the photometric data as well as the empirically defined
standard relations is quite good.

\begin{figure}
\plotone{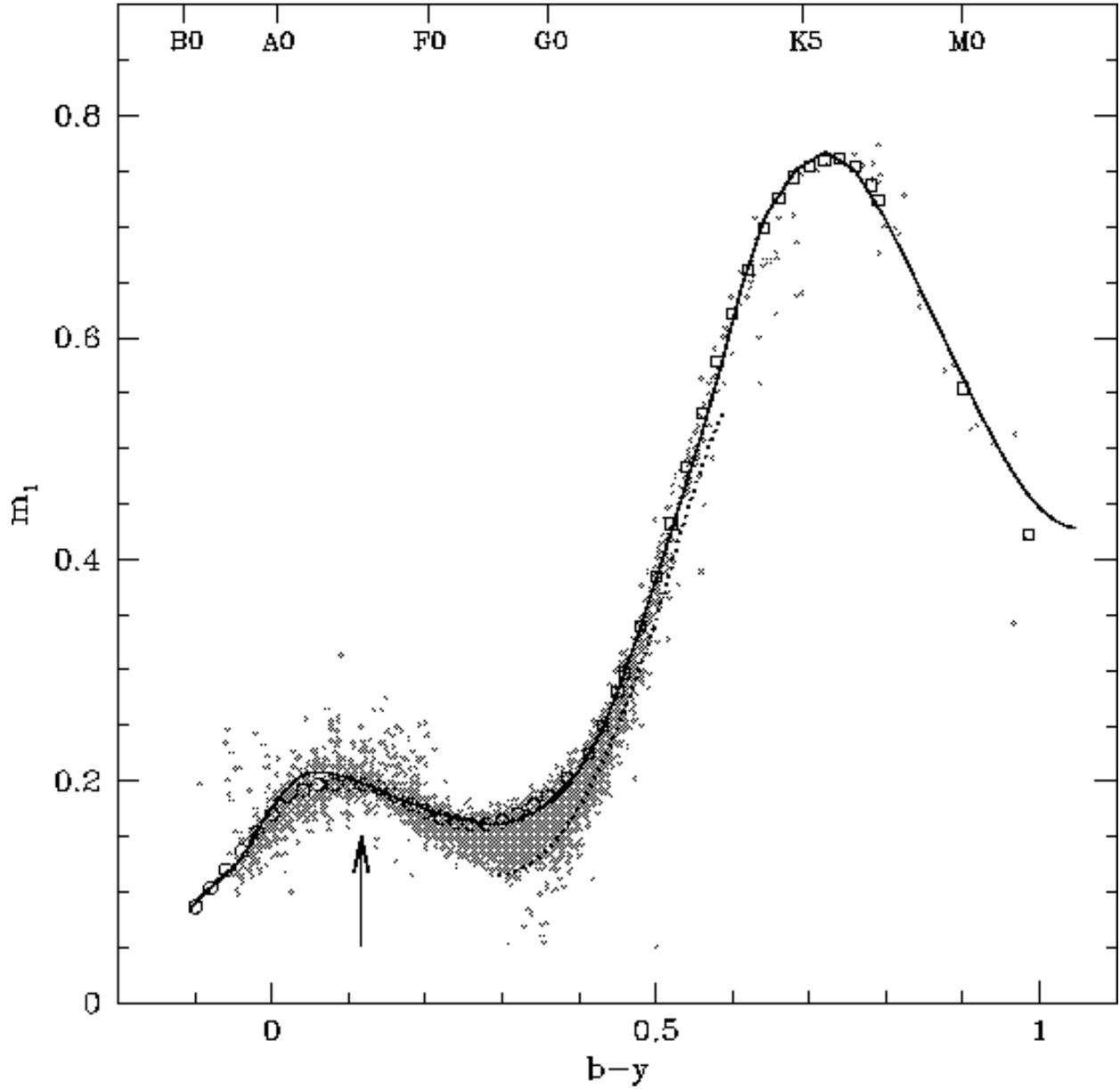}
\caption{As in Fig.~\ref{fig:ZAMSbyCMD}, except that the observations are plotted on
the ($b-y$,~$m_1$) diagram.  The $dotted~line$ below the solar metallicity ZAMS denotes
the location of an additional ZAMS model having $\FeH=-0.5$.}
\label{fig:ZAMSbym1} 
\end{figure}

\begin{figure}
\plotone{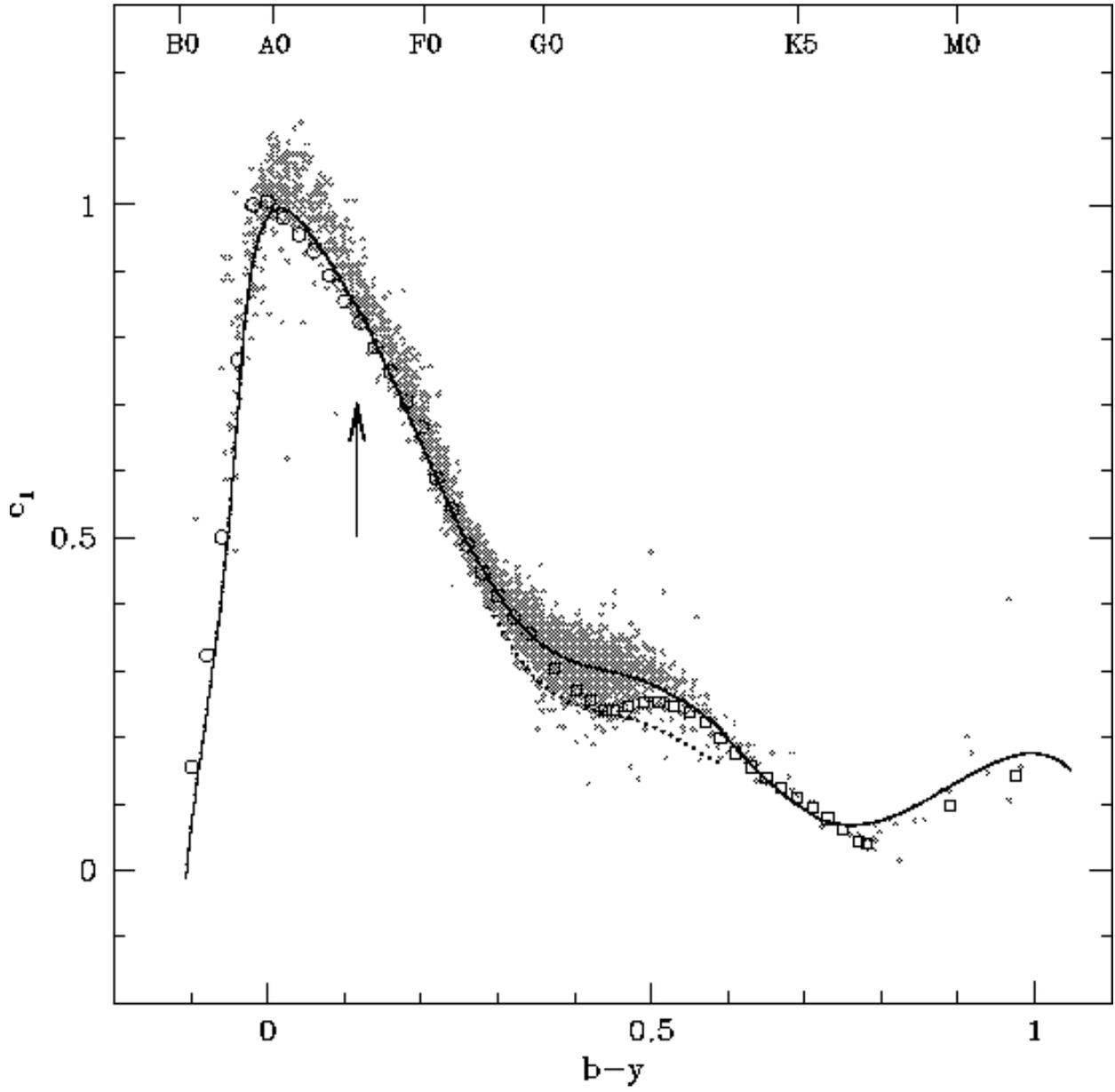}
\caption{As in Fig.~\ref{fig:ZAMSbym1}, except that the observations are plotted on
the ($b-y$,~$c_1$) plane.}  
\label{fig:ZAMSbyc1} 
\end{figure}

In Figures \ref{fig:ZAMSbym1} and \ref{fig:ZAMSbyc1} the same solar metallicity ZAMS
is transposed onto the ($b-y$,~$m_1$) and ($b-y$,~$c_1$) color planes to illustrate
how well it is able to reproduce the observed stellar distributions.  For the former
plot, the standard relation of \citeauthor{PhilipEgret1980} has been adjusted by
$-$0.01 in $m_1$ to better match the photometric means derived from main-sequence
spectral types \citep[see Fig.~2 of][]{PhilipEgret1983}.  Overall, the ZAMS locus
agrees quite well with the observations over a broad range in color.  We again stress
the fact that, for $b-y\lesssim0.1$, the colors are purely theoretical with no
corrections applied.  This type of diagram illustrates the unique sensitivity of the
$m_1$ index to chemical abundance in F- and G-type dwarf stars through a noticeable
spread in the $m_1$ colors at $0.2\leq\,b-y\,\leq0.5$.  While the location of our
solar-metallicity ZAMS locus corresponds well with the turnover in the standard
relations and the stellar data in this range, we have included an additional ZAMS
having $\FeH=-0.5$ ({\it dotted line}) to show that the majority of dwarfs with
slightly bluer $m_1$ colors have metallicities up to 0.5~dex less than solar.  
Indeed, our $\FeH=-0.5$ ZAMS follows the lower bound of the stellar distribution for
the F- and G-type dwarfs quite well in Figure \ref{fig:ZAMSbym1} with the few stars
having slightly bluer $m_1$ values at $b-y\approx0.35$ likely being even more metal
poor.

Upon inspection of the ($b-y$,~$c_1$) diagram in Figure \ref{fig:ZAMSbyc1}, there is a
difference between the \citeauthor{Olsen1984} standard relation and the ZAMS both at
extremely cool temperatures and in the color range corresponding to G-type stars.  
While the mismatch at the cool end of the main sequence is most likely due to the
small number of M dwarfs used to define the Olsen trend and has already been noted in
Figure \ref{fig:ZAMSlower}, the reason for the difference seen in the G dwarfs is not
immediately apparent.  Although the magnitude of this discrepancy is quite large
($\sim0.04$$-$$0.05\,$mag), we suggest that the explanation lies in the fact that the
$c_1$ index exhibits some sensitivity to star-to-star variations in metal abundance
within this temperature range --- which may explain the rather large spread in $c_1$
colors at $0.3\leq\,b-y\,\leq0.5$.  In support of this argument, the same $\FeH=-0.5$
ZAMS from the previous figure is again plotted to show that it defines the lower
distribution of stars very well for the F- and G-type dwarfs.  It would appear that
slightly more metal-poor stars are predicted to lie up to 0.07~mag below the trend
defined by the solar-metallicity dwarfs.

\begin{figure}
\plotone{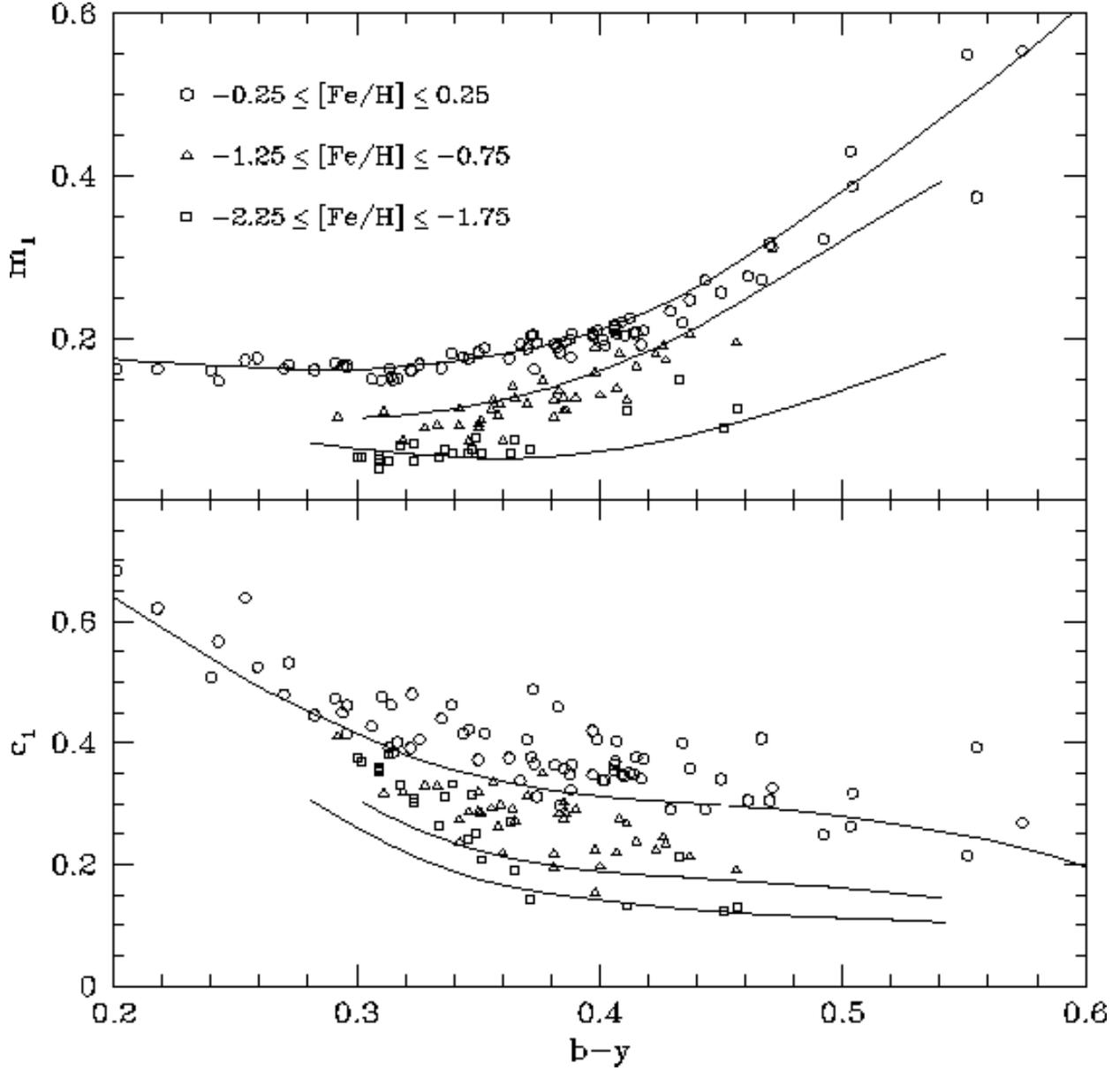}
\caption{Two color-color diagrams for the field star data from Table \ref{tab:starlist}
separated according to their metallicity values, as denoted by the different symbols.  
Three ZAMS models having $\FeH=0.0$, $-1.0$, $-2.0$ (in order of decreasing $m_1$ and
$c_1$) are represented by {\it solid lines}.}
\label{fig:ZAMSmetal}
\end{figure}

To further expand on this, Figure \ref{fig:ZAMSmetal} illustrates the metallicity
dependence of the \Stromgren $m_1$ and $c_1$ colors for F- and G-type dwarfs.  For
this purpose we have plotted only those field dwarfs stars in Table
\ref{tab:starlist} that have metallicity estimates.  The stars in each panel are
divided into separate metallicity bins as indicated by the different symbols and
overlaid with three ZAMS models for $\FeH=0.0$, $-1.0$, and $-2.0$ (in the order of
decreasing $m_1$ and $c_1$).  Recall that the colors employed for $\FeH=0.0$ ZAMS have
been calibrated as described in the previous section.  While we defer the discussion
of corrections to the colors made at other metallicities until the next section, it is
worth mentioning that the corrections to the $uvby$ colors towards cooler effective
temperatures (i.e., $\Teff\leq5500\,$K) at $\FeH=-1.0$ and $-2.0$ are primarily
constrained by the metal-poor field stars from the \citet{SchusterNissen1989b} sample
(see Figure \ref{fig:snbym1c1}) as well as the lower main sequences of the globular
clusters M$\,$3 and 47$\,$Tuc.  It is immediately obvious that stars with differing
chemical composition exhibit a rather large photometric spread both in $m_1$ and $c_1$
at $b-y$ colors between 0.3 and 0.5.  In the case of the bottom panel of Figure
\ref{fig:ZAMSmetal}, all three of our ZAMS models do an excellent job of reproducing
the lower bound to the distribution of dwarf stars in their respective metallicity
bins.  This is to be expected since a star that has evolved away from the main
sequence would have a larger $c_1$ index than another star of the same temperature and
metallicity but showing little evolution.  Given this evidence, it would seem that
Olsen's calibration may have been based on stars with slightly less than solar
abundances in this regime rather than actual solar-metallicity main-sequence stars.

\subsection{Color Corrections at [Fe/H]$\,\leq\,$--0.5 and [Fe/H]$\,$=$\,$+0.5}
\label{subsec:othercolcor}

Based on the analysis presented so far, we conclude that our calibrated $uvby$ colors
at $\FeH=0.0$ are able to provide both accurate and consistent interpretations of the
observed photometry for dwarf and giant stars having metallicities near the solar
value.  Moreover, our colors appear to do a reasonable job of reproducing the observed
photometry of metal-poor turnoff and giant stars (see Figure \ref{fig:m92}) and our
adopted color transformations in these temperature and gravity regimes for
$\FeH\leq-2.0$ remain purely theoretical.  However, from the evidence presented in
Figure \ref{fig:snbym1c1}, it seems clear that some adjustment to the cool dwarf-star
colors at extremely low metallicities is necessary in order to obtain consistency with
the field-star data.  To be more specific, we chose to keep all of the $b-y$
predictions at $\FeH\leq-2.0$ purely theoretical, but to apply some corrections to the
$v-b$ and $u-v$ colors at temperatures and gravities relevant to cool dwarfs (i.e., at
$\Teff\leq5500\,$K and $\logg\geq3.5$) to secure a better fit of the $\FeH=-2.14$
isochrone to the data.  In general, this meant iteratively forcing the $v-b$ colors
redder (i.e., making them more positive) and the $u-v$ colors bluer (i.e., more
negative) by increasing amounts towards cooler temperatures.  The justification for
this admittedly {\it ad hoc} procedure is simply that such adjustments are required to
satisfy the constraints imposed by the empirical data available to us at this time.  
Indeed, we are quite confident that our $uvby$ color transformations for
$\FeH\leq-2.0$ are able to reproduce the observed photometry for metal-poor stars
across a wide range in temperature and gravity.  Since it was necessary to make some
corrections to the synthetic colors for very metal-deficient stars and at $\FeH=0.0$,
it is to be expected that they will be necessary for essentially all [Fe/H] values.

In order to quantify what color corrections are necessary at intermediate
metallicities (i.e., $-1.5\leq\FeH\leq-0.5$), we have investigated if the same
techniques employed earlier for the correction of the colors at $\FeH=0.0$ might
continue to be applicable. However, the decrease in the number of stars from Table
\ref{tab:starlist} having lower metallicity values, combined with their limited ranges
in color (particularly for the dwarf stars), led us to conclude that there is not
enough information from the field-star sample to derive the necessary calibrations
adequately.  Therefore, we choose not to rely on our field-stars to calibrate the
colors for intermediate metallicities, but rather employ them later to test the
relevancy of the corrections to the $uvby$ colors we derive for $-1.5\leq\FeH\leq-0.5$
described below.

\begin{figure} 
\plotone{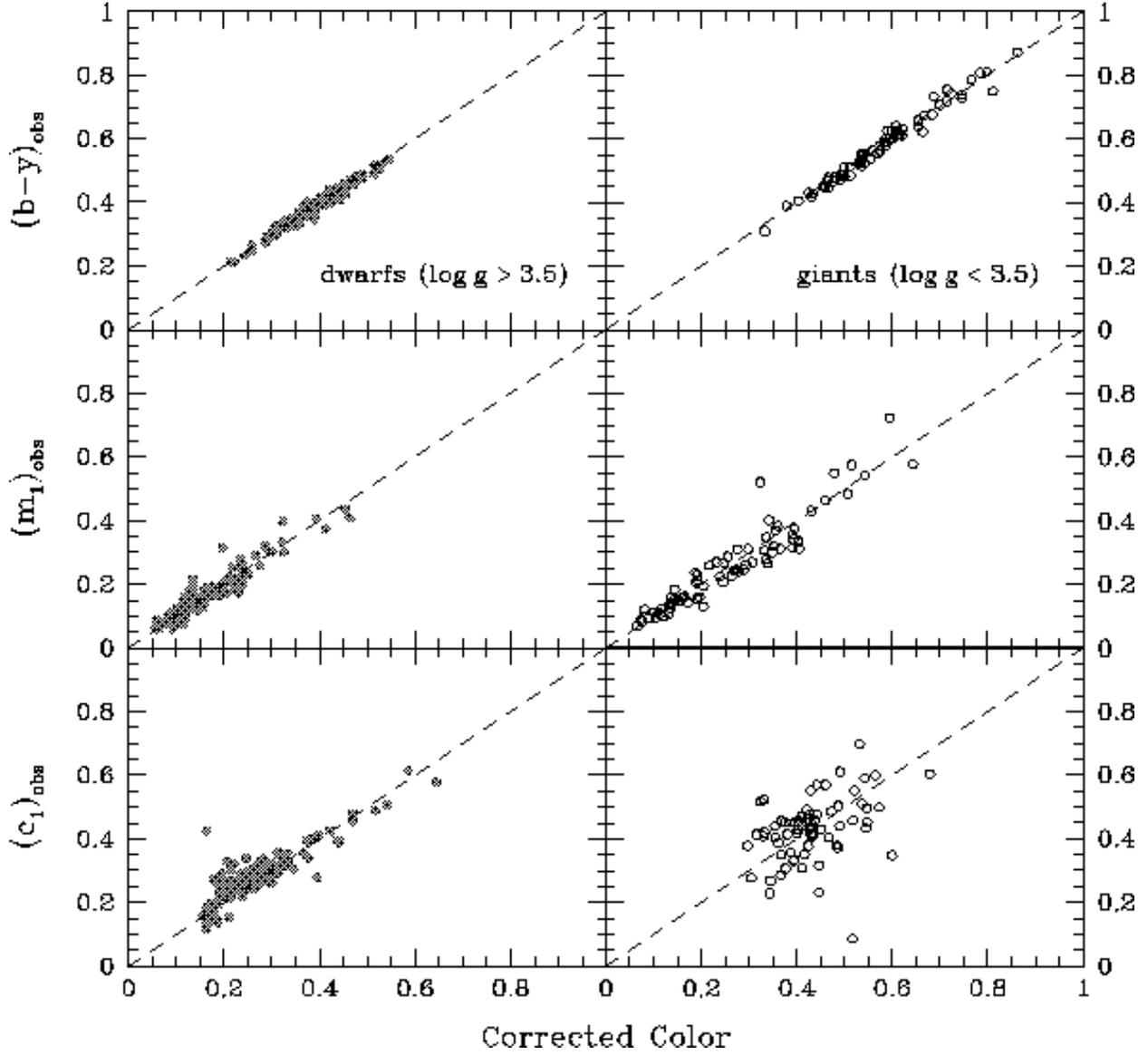} 
\caption{Plots of the {\it corrected} versus observed colors for dwarfs and giants 
with intermediate metallicities (i.e., $-1.75\leq\FeH\leq-0.25$) from our field star 
sample.  The {\it dashed line} indicates the line of equality.}
\label{fig:metalpoorcalib} 
\end{figure}

Since we have found that no corrections whatsoever are necessary for the synthetic
$uvby$ colors with $\FeH\leq-2.0$ towards warmer temperatures ($\Teff>5500\,$K) or
lower gravities ($\logg<3.5$), we have simply chosen to assume that the required
adjustments to the color transformations at intermediate metallicities in these same
temperature and gravity regimes are some fraction of those applied at $\FeH=0.0$.  In
particular, we assume that the size of this fraction scales linearly as a function of
$\FeH$.  For instance, the corrections applied to the synthetic colors at $\FeH=-0.5$
and $-1.0$ for a particular temperature and gravity correspond to $one$-$quarter$ and
$one$-$half$ of the corrections that are required at $\FeH=0.0$ for the same $\Teff$
and $\logg$.  For cool dwarf stars, however, it was necessary to correct the $v-b$ and
$u-v$ colors in a similar fashion as those for $\FeH\leq-2.0$; i.e., we have used the
$uvby$ data available to us from the metal-poor field-star sample of
\citeauthor{SchusterNissen1989b}, together with the precise $uvby$ photometry for the
globular clusters M$\,$3 and 47$\,$Tuc, to derive the transformations that yield the
best possible matches to the empirical data for cool cluster and field dwarfs having
$-1.5\leq\FeH\leq-0.5$.  Again, while our approach is {\it ad hoc}, we remark that
when the synthetic colors are corrected in this way, they seem to agree quite well
with observed colors for stars from our sample.  This is illustrated in Figure
\ref{fig:metalpoorcalib}, which plots the $calibrated$ versus observed colors of all
the dwarfs and giants in Table \ref{tab:starlist} having $-1.75\leq\FeH\leq-0.25$.  
There are clearly no systematic differences or inconsistencies between the corrected
and observed colors within this metallicity range, which lends considerable support to
our technique of scaling the corrections as a function of $\FeH$ as well as the
adjustments made for cooler dwarf stars.

\begin{figure}
\plotone{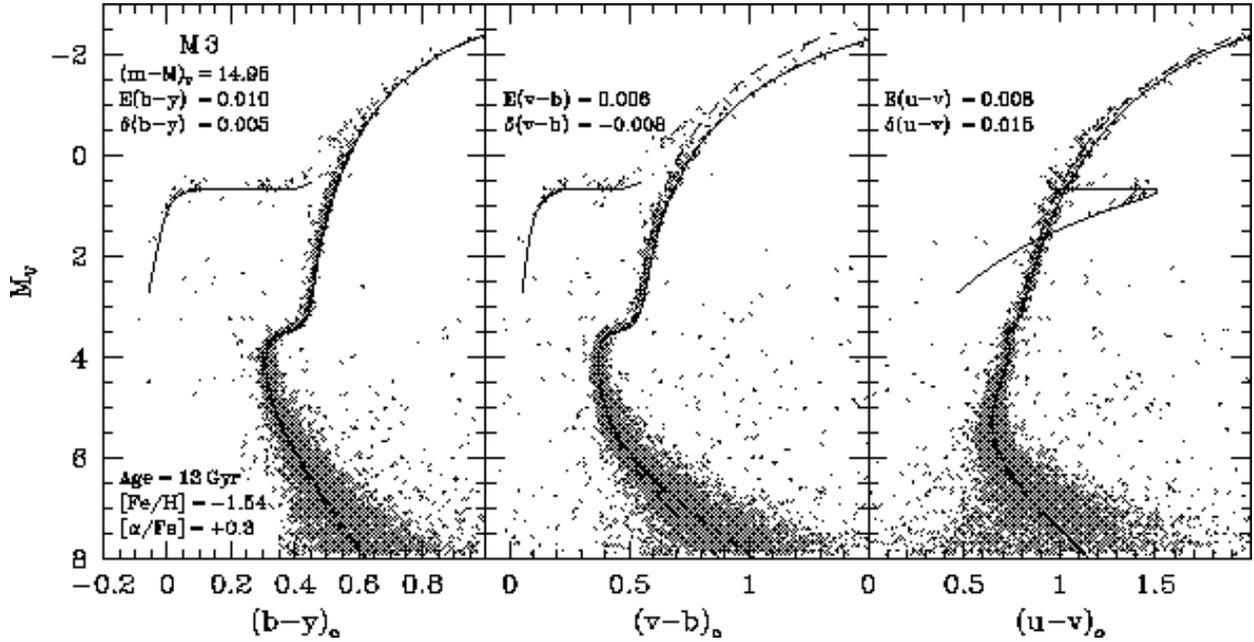}
\caption{Various $uvby$ CMDs for the globular cluster M$\,$3 overlaid with a 13~Gyr,
$\FeH=-1.54$ isochrone and consistent ZAHB model.  The indicated reddening is taken
from the dust maps of \citeauthor{Schlegel1998} while the apparent distance modulus 
is derived from the fit of the ZAHB model to the lower distribution of stars on the
horizontal branch.  The uncalibrated isochrone is represented by a $dashed~line$ in
each panel.}
\label{fig:m3}
\end{figure}

\begin{figure}
\plotone{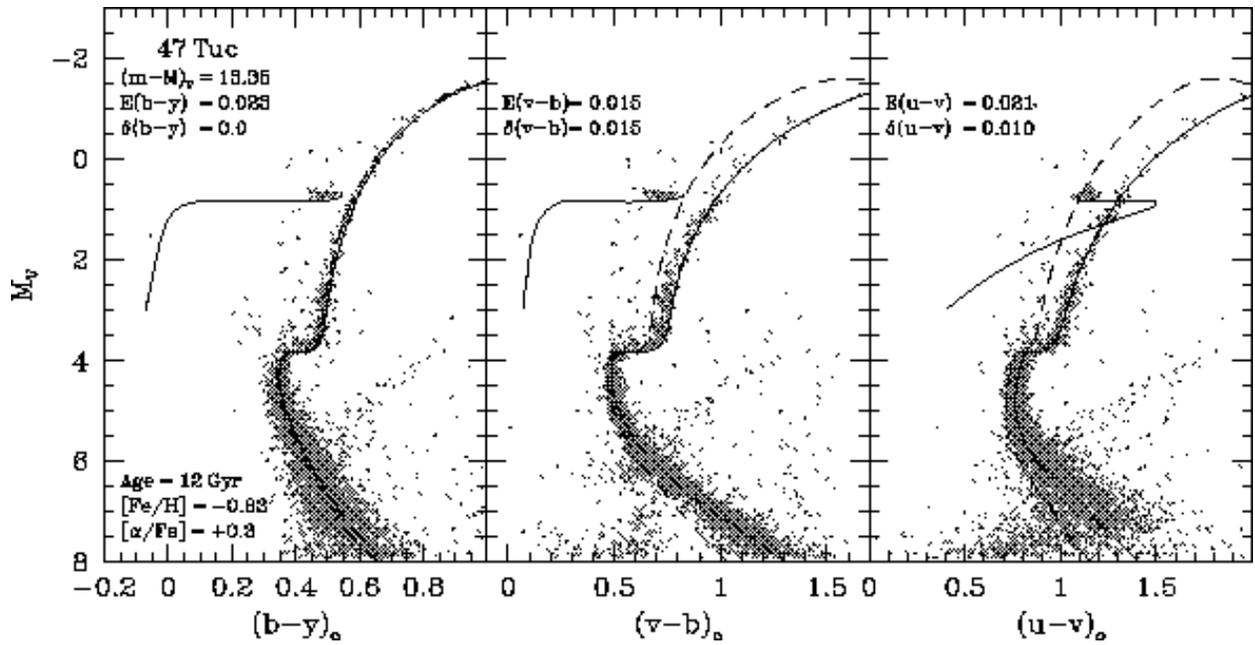}
\caption{As in Fig.~\ref{fig:m3}, except for the globular cluster 47$\,$Tuc (NGC$\,$104).  
Note that the apparent distance modulus and reddening values are identical to those
adopted in Paper I.  Our 12~Gyr, $\FeH=-0.83$ isochrone and ZAHB model are overlaid on
the cluster photometry.}
\label{fig:n104}
\end{figure}

Further illustrations of the accuracy of the corrected colors for intermediate
metallicities are given in Figure \ref{fig:m3} and \ref{fig:n104}, which present
comparisons of the CMDs for M$\,$3 and 47$\,$Tuc with relevant isochrones and ZAHB
models.  According to \citet{KraftIvans2003}, the iron abundance of M$\,$3 is between
$\FeH=-1.50$ and $-1.58$, which is within $\approx0.1$~dex of the \citet{ZinnWest1984}
estimate.  There seems to be general agreement that 47$\,$Tuc has $\FeH=-0.75\pm0.1$
\citep[see][]{KraftIvans2003, ZinnWest1984, CarrettaGratton1997}.  Isochrones for
metallicities within these ranges provide very good fits to the cluster data if the
foreground reddenings are taken from the \citeauthor{Schlegel1998} dust maps, and the
adopted distances are based on fits of ZAHB models to the lower bounds of the
respective distributions of horizontal-branch stars.  Of these two clusters, only
47$\,$Tuc was considered in Paper I, and the match reported therein of the same
isochrone used here to the $B-V$ fiducial derived by \citet{Hesser1987} is completely
consistent with those shown in Figure \ref{fig:n104}.  This is particularly
encouraging because cluster data has played no role whatsoever in our determination of
the corrections to the synthetic $uvby$ transformations at temperatures corresponding
to the turnoff stars in metal-poor globular clusters (i.e., $\Teff>5500\,$K), and yet
we find essentially the same interpretation of the M$\,$92 and 47$\,$Tuc CMDs as in
Paper I.  This consistency provides a strong argument that the color transformations
that have been derived in both investigations, as well as the model $\Teff$ scale, are
realistic.  It is also evident that the size of the color adjustments increases with
increasing $\FeH$ --- note the differences between the $dashed$ and $solid$ curves,
which represent the $uncalibrated$ and $calibrated$ isochrones, respectively.  The
same thing was found in Paper I.  Although we have applied small shifts to some of the
colors to obtain consistent fits to the turnoff data on the various color planes, it 
is not possible to say at this time whether they are due to small problems with the 
photometric zero-points, the adopted cluster parameters, the isochrones, or the 
color-temperature relations.

\begin{figure}
\plotone{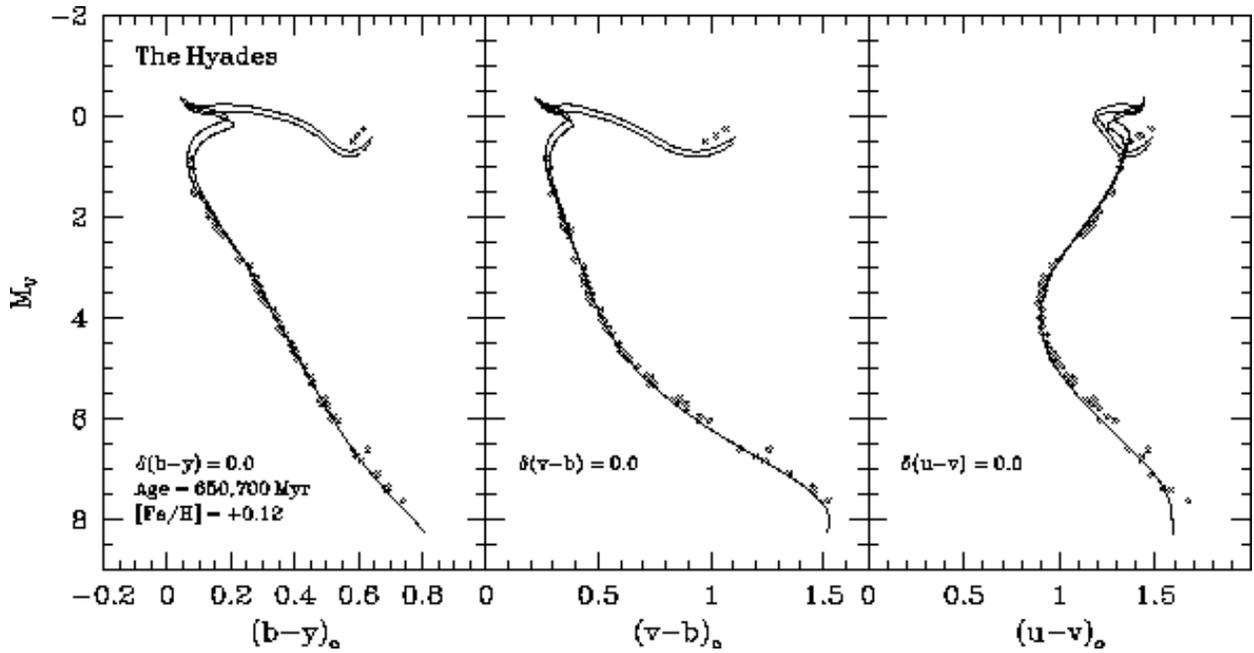}
\caption{\Stromgren CMDs for a sample of 77 ``high-fidelity'' single stars in the
Hyades open cluster.  Two isochrones having ages of 650 and 700~Myr with $\FeH=+0.12$
are overlaid on the photometric data.  Note that the absolute magnitude for each star
was derived from {\it Hipparcos} secular parallaxes by \citet{deBruijne2001}.}
\label{fig:hyades}
\end{figure}

\begin{figure}
\plotone{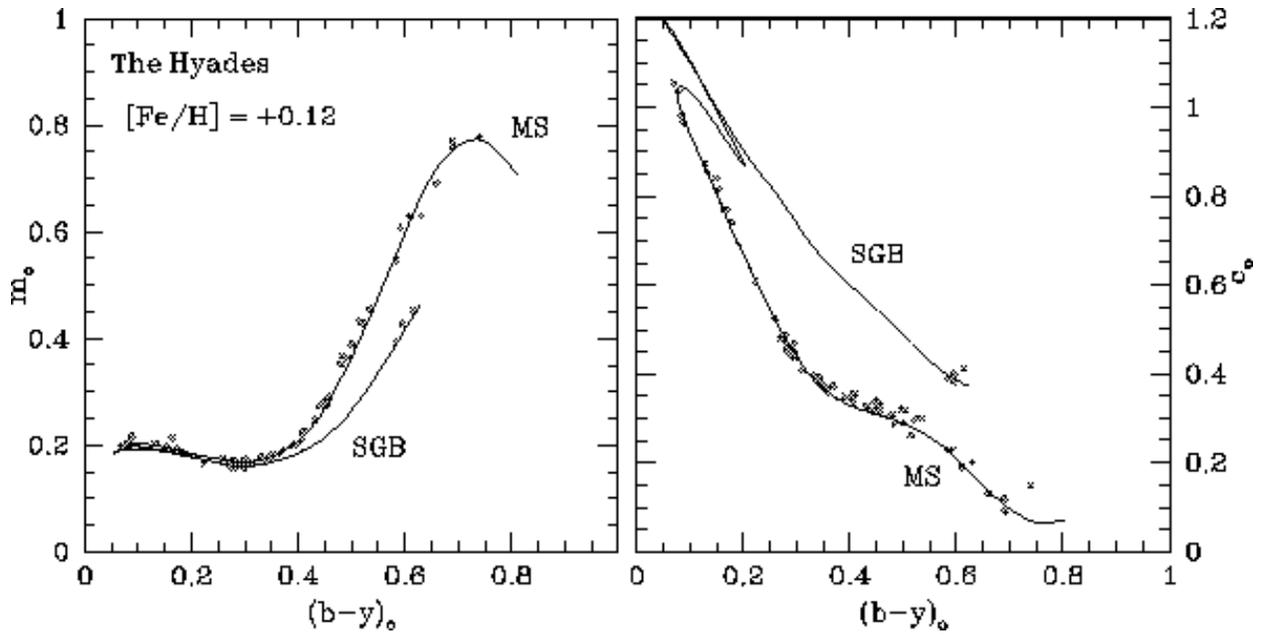}
\caption{\Stromgren color-color diagrams for the same sample of Hyades stars from the
previous figure overplotted with a 700~Myr, $\FeH=+0.12$ isochrone.  The 
main-sequence (MS) and subgiant (SGB) segments of the isochrone are labeled.}
\label{fig:hyadescc} 
\end{figure}

For the synthetic color corrections at $\FeH=+0.5$ we follow the same treatments as
mentioned above for the intermediate metallicity cases.  However, due to the fact that
there are only 2 stars from Table \ref{tab:starlist} which have $\FeH\geq+0.25$, we
cannot draw any meaningful conclusions as to the accuracy of our corrected colors at
$\FeH=+0.5$ from plots such as Figure \ref{fig:metalpoorcalib}.  Alternatively, we can
rely upon the observed $uvby$ photometry of the Hyades, which has $\FeH=+0.12\pm0.02$
\citep{Cayrel1985, BoesgaardFriel1990}, to test the colors at the metal-rich end. We
present the various \Stromgren CMDs for the Hyades in Figure \ref{fig:hyades}.  To
better constrain the models, we have selected stars from the ``high fidelity" list of
\citet{deBruijne2001}, who used secular parallaxes from {\it Hipparcos} to derive
individual $M_V$ values, and thereby produce exceptionally well-defined CMDs.  The
majority of $uvby$ photometry for this sample is taken from \citet{CrawfordPerry1966}
and \citet{Olsen1993}.  For the remaining stars not included in either of these
references we adopt the mean photometry from HM98.  The Hyades data presented in
Figure \ref{fig:hyades} have been overlaid with isochrones having $\FeH=+0.12$,
$Y=0.262$, and $Z=0.025$ and corresponding to ages of 650 and 700~Myr.  This chemical
mixture is justified in Paper I as giving the best fit to the mass--$M_V$ relationship
as defined from a sample of Hyades binaries (see their Fig.~21).  The superb quality
of the isochrone fits to the data is a testament to the quality of calibrated colors
at metallicities just above solar.  To further demonstrate this fact, we have plotted
the 700~Myr isochrone on the Hyades ($b-y$,~$m_1$) and ($b-y$,~$c_1$) diagrams in
Figure \ref{fig:hyadescc}.

\begin{figure} 
\plotone{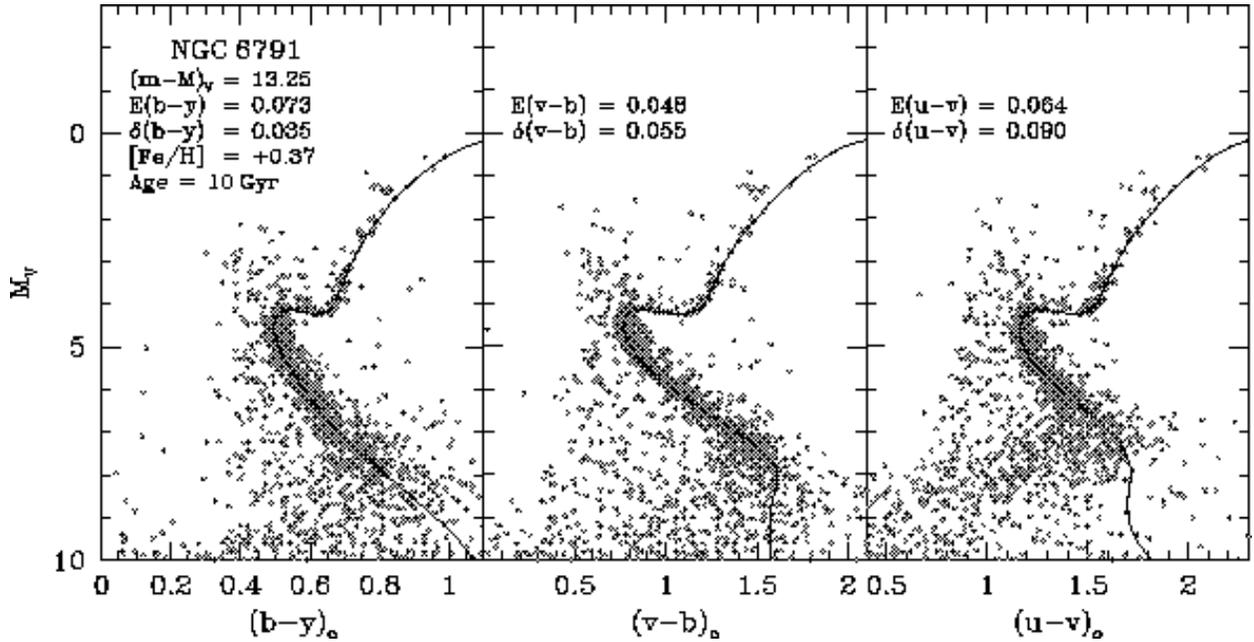} 
\caption{\Stromgren CMDs for the metal rich open cluster NGC~6791 fitted with a
10~Gyr, $\FeH=+0.37$ isochrone.  The indicated values of reddening, distance, and age
are the same as those adopted in Paper I from fits to cluster data on the $B-V$ and
$V-I$ CMDs.  Note the rather large color offsets required to properly fit the cluster
turnoff in each panel can be most likely attributed to uncertainties in the 
zero-points of our $uvby$ photometry (see the text).}
\label{fig:n6791} 
\end{figure}

As a final test of colors at $\FeH>0.0$ we present $uvby$ CMDs of the metal-rich open
cluster NGC$\,$6791 overlaid with our best fit 10~Gyr, $\FeH=+0.37$ isochrone.  Since
the metallicity of this cluster lies much closer to our set of colors at $\FeH=+0.5$
than the Hyades, its photometry can be used as a somewhat more stringent test of the
color-temperature relations at such high metal abundances.  The estimates for the
cluster distance and reddening indicated in Figure \ref{fig:n6791} are the same as the
values assumed in Paper I from the fits of the same 10~Gyr isochrone to the $B-V$ and
$V-I$ CMDs.  As mentioned in Paper I, these estimates may not necessarily be the
correct ones given that other authors have quoted somewhat lower $\FeH$ and age values
by about 0.2~dex and 2~Gyr, respectively.  Unfortunately, our interpretation of the
data is consistent with Paper I only if rather large color shifts are applied to the
isochrone in all three CMDs.  In this regard we note that our $uvby$ data for
NGC$\,$6791 is somewhat preliminary, and they appear to suffer from uncertainties in
the zero-points for the calibrated photometry.  Indeed, we have found that there is a
0.04~mag difference between our \Stromgren $y$ magnitudes and the Johnson $V$
magnitudes published by \citet{Stetson2003}.  Since their broadband photometry has
been standardized with extreme care and exhibits good consistency with other data
sets, we are inclined to conclude that the $uvby$ data presented here are in error, at
least with regards to the photometric zero-points.  However, we are unable to say if
there are also zero-point errors in the other three \Stromgren filters, and therefore,
we do not know to what extent the colors are affected by such errors.  More
observations are needed to shed light on this problem and to check the reliability of
our color transformations for stars having higher metallicities than that of the
Hyades.

\subsection{Population~II Subdwarfs}
\label{subsec:pop2subdwarfs}

To further assess the accuracy of the calibrated colors for sub-solar metallicities, we
have selected a number of Population~II subdwarfs from Table \ref{tab:starlist} that
are among the most well-studied metal-deficient field stars in the literature.  The goal
of this particular analysis is to check whether we can correctly reproduce the observed
\Stromgren colors for these subdwarfs provided accurate estimates of their parameters
are available.

Since the $b-y$ index is highly sensitive to $\Teff$, we first ensure that the IRFM
temperatures for our sample of subdwarfs are consistent with those from other studies.  
Table \ref{tab:subdwarftemp} presents such a comparison for temperatures extracted
from a number of different sources.  The second column lists the mean of those
temperatures quoted in the studies of Gratton et al.~\citeyearpar{Gratton1996,
Gratton2000} and/or \citet{Clementini1999}.  Each of these studies rely upon either
empirical or theoretical color-temperature relationships to derive $\Teff$.  The
effective temperatures presented by \citet{Axer1994} and \citet{Fuhrmann1998} were
computed by fitting theoretical spectra to Balmer line profiles, and
\citet{AllendePrietoLambert2000} derived $\Teff$ by analyzing the flux distribution in
the near-UV continuum.

Although all of these studies rely upon different methods of deriving $\Teff$, they
appear to yield quite consistent results.  In general, most of the IRFM temperatures
show good agreement with those taken from the indicated studies to within $\pm100\,$K,
the most notable case being HD~19445, for which the temperature estimates lie within
25$\,$K of each other.  The two subdwarfs HD~134439 and HD~134440, however, both have
IRFM temperatures that are $\approx$100$-$150$\,$K cooler than those derived from
the other studies.  While the reasons for the differences in $\Teff$ are not
immediately apparent, we will assess the implications for the $uvby$ colors of adopting
slightly warmer temperatures for these two stars.

In order to calculate the \Stromgren colors for our subdwarfs, accurate estimates of
the surface gravities and metallicities must supplement the IRFM temperatures for
these stars.  As mentioned earlier, the mean spectroscopic values for $\logg$ and
$\FeH$ from the catalog of Cayrel de Strobel et al.~\citeyearpar{CayreldeStrobel2001}
are favored over those included in the original AAM96 list from which these subdwarfs
were extracted.  In Table \ref{tab:subdwarfparam} we present the adopted stellar
parameters together with the dereddened $uvby$ and $B-V$ photometry for the sample of
subdwarfs.  Furthermore, since all of these subdwarfs have very accurate parallaxes
from {\it Hipparcos}, the $\logg$ estimates derived from isochrones of
\citet{BergbuschVandenBerg2001}, assuming the spectroscopic $\FeH$ values, have also
been included for comparison.  Table \ref{tab:subdwarfparam} also lists, again for
comparison, the photometric $\FeH$ estimates derived from \Stromgren metallicity
calibrations by \citet{SchusterNissen1989b}.  Finally, the \Stromgren photometry for
these subdwarfs are taken from the study of \citet{SchusterNissen1988} and corrected
for reddening using the $\EBV$ values from \citet{Carretta2000}.  It is important to
note that the subdwarf photometry is on the original \Stromgren system, and so the
comparisons which follow do not suffer from possible uncertainties in the
transformation from the CCD system to the original system.

In Table \ref{tab:subdwarfcalc} we list the results of numerous calculations carried out
in an attempt to match the observed colors of the subdwarfs.  The \Stromgren indices are
calculated from direct interpolation in the grid of calibrated colors, while the $B-V$
colors are derived from the broadband color transformations of Paper I.  For all of the
subdwarfs listed in the table, the first set of colors (Model A) is based on the stellar
parameters presented in Table \ref{tab:subdwarfparam} (i.e., the IRFM temperatures,
together with the spectroscopic values of $\logg$ and $\FeH$).  In addition, for a few
selected subdwarfs (HD~19445, HD~103095, HD~140283, and HD~201891) we investigate the
effects that uncertainties in the stellar parameters have on the computed photometry.

The majority of subdwarfs in Table \ref{tab:subdwarfcalc} show excellent agreement
between the observed and computed indices for the first set of parameters, considering
that the observed colors have uncertainties of $\sim0.01$~mag due to errors in the 
observations and transformation to the standard system.  The other calculations that are 
listed confirm that the $b-y$ colors is indeed most sensitive to uncertainties in 
$\Teff$, while the $m_1$ and $c_1$ indices are largely dependent on the accuracy of the 
adopted $\FeH$ and $\logg$ values, respectively.

However, a few halo stars show some disagreement between their observed and computed
colors. The most notable case is HD~25329, which has a difference of almost 0.06~mag
in the $m_1$ index.  It seems highly unlikely that errors in the adopted $\Teff$ for
this star could cause such a mismatch given the consistency of both the $\Teff$
estimates (see Table \ref{tab:subdwarftemp}) and the calculated and observed $b-y$ and
$B-V$ colors.  It is also unlikely that the star could have a surface gravity or
metallicity that deviates significantly from the parameters listed in Table
\ref{tab:subdwarfparam}.  Moreover, even if the reddening is non-zero, as we have
assumed, this would only serve to $increase$ the dereddened value of $m_1$ [since
$m_o=m_1+0.24\EBV$].  Finally, we also note that \citet{Olsen1993} obtains $uvby$
colors for HD~25329 ($b-y=0.525$, $m_1=0.305$, and $c_1=0.130$), which are in
excellent agreement with those presented in Table \ref{tab:subdwarfcalc}.

Thus, we are left to conclude that the discrepancies are due to chemical abundance
anomalies in the star's atmosphere.  It is known that HD~25329 exhibits unusually
strong CN absorption features for its classification as a metal-poor halo star
\citep{Spiesman1992}, and a few recent studies have shown that variations in the
abundances of carbon and nitrogen can affect the \Stromgren $m_1$ and $c_1$ indices
\citep{Grundahl2002a, Hilker2000}.  Specifically, the \Stromgren $v$ filter is
centered almost exactly on the CN band located at 4215{\rm{\AA}} while the NH
molecular band sits within the $u$ filter at 3360{\rm{\AA}}.  Therefore, we should
expect any star with abnormal abundances of carbon and nitrogen to have somewhat
different $m_1$ and $c_1$ values than one with ``normal" abundances.  This motivates
us to try to find a CN-enhanced model that is better able to reproduce the observed
indices of HD~25329, on the assumption of the same stellar parameters as before
(4842/4.66/$-$1.65).  The first such model (Model B) assumes that
[C/Fe]$\,$=$\,$[N/Fe]$\,$=$\,$+0.4, while models C and D assume that carbon and
nitrogen have been enhanced, in turn, by this amount. It appears that when both carbon
and nitrogen are enhanced by +0.4~dex, the resulting colors show the best agreement
with their observed counterparts.  In support of this result, we note that
\citet{Carbon1987} derived abundances of [C/Fe]$\,$=$\,$+0.44 and [N/Fe]$\,$=$\,$+0.45
for HD~25329 based on high-resolution spectroscopy.

For the few other subdwarfs whose calculated and observed colors differ, we include
additional models in which the values for $\Teff$ and/or $\FeH$ have been slightly
altered to produce better agreement.  For example, the second model for HD~140283
adopts the $\Teff$ listed in column 3 of Table \ref{tab:subdwarftemp}, which is
$\sim120\,$K higher than that obtained from the IRFM.  Clearly, this particular model
yields much better agreement for the $b-y$ and $B-V$ indices.  A somewhat higher
temperature was also obtained by \citet{Gratton1996} and \citet{Fuhrmann1998}.  Such
an increase in temperature is justified by the fact that HD~140283 likely has a
non-negligible reddening \citep{Grundahl2000a}, which was not taken into account by
AAM96.  Finally, for the subdwarf pair HD~134439 and HD~134440, Model B adopts
temperatures listed in column 2 of Table \ref{tab:subdwarftemp}.  While these somewhat
higher temperatures improve the agreement for the $b-y$ index, an additional model
(Model C) that assumes the photometric metallicities listed in column 5 of Table
\ref{tab:subdwarfparam} yields the best overall agreement in all three \Stromgren
colors.  In support of these new models, we note that \citet{Clementini1999} derived
$\FeH=-1.30$ for HD~134439 and $\FeH=-1.28$ for HD~134440.

In conclusion, our calibrated \Stromgren colors appear to provide a satisfactory match
to the observed photometry for most of the ``classical" subdwarfs.  Although a few of
the stars exhibit some discrepancies between their calculated and observed colors, we
have shown that these can be largely explained by slightly altering their basic
parameters within justifiable limits.  

\section{Previous \Stromgren Color--T$_{\rm{eff}}$ Relations and Calibrations}
\label{sec:comparisons}

Since our semi-empirically corrected $uvby$ transformations appear to be reliable in the
variety of tests presented so far, we now compare them with other $uvby$ color--$\Teff$
relations and calibrations that are available in the literature.  In addition, we
investigate if our colors can reproduce the loci of constant $\FeH$ in ($b-y$,~$m_1$)
space that are predicted by the \Stromgren metallicity calibrations of 
\citet{SchusterNissen1989a} for dwarfs and \citet{Hilker2000} for giants.
 
\subsection{Comparisons with Other Synthetic \Stromgren Color--T$_{\rm{eff}}$ Relations}
\label{subsec:synCTcomp}

The grids of synthetic \Stromgren colors considered for our comparison are those
derived from the previous MARCS/SSG models of \citet[hereafter
VB85]{VandenBergBell1985} as well as latest set from CGK97 computed from Kurucz model
atmospheres without overshooting.  It is important to note, however, that both VB85
and CGK97 adopted the same $uvby$ filter transmission functions
\citep{CrawfordBarnes1970b} and use Vega as their zero-point
standard.\footnote{Although the choice of Vega as a zero-point standard is common
between for the MARCS/SSG and ATLAS9 colors, the input stellar parameters for the
synthetic Vega models differ slightly.  Our calculations as well as those of VB85
adopt the \citet{DreilingBell1980} parameters of (9650/3.90/0.0) for the Vega model
while CGK97 use the \citet{CastelliKurucz1994} values of (9550/3.95/$-$0.5).  Despite
this fact, the actual difference in the derived colors corresponding to these two
separate Vega models is less than 0.007~mag for all three \Stromgren indices.}
Therefore, any differences between the synthetic grids from these studies can largely
be attributed to differences in the MARCS/SSG and ATLAS9 codes.

\begin{figure} 
\plotone{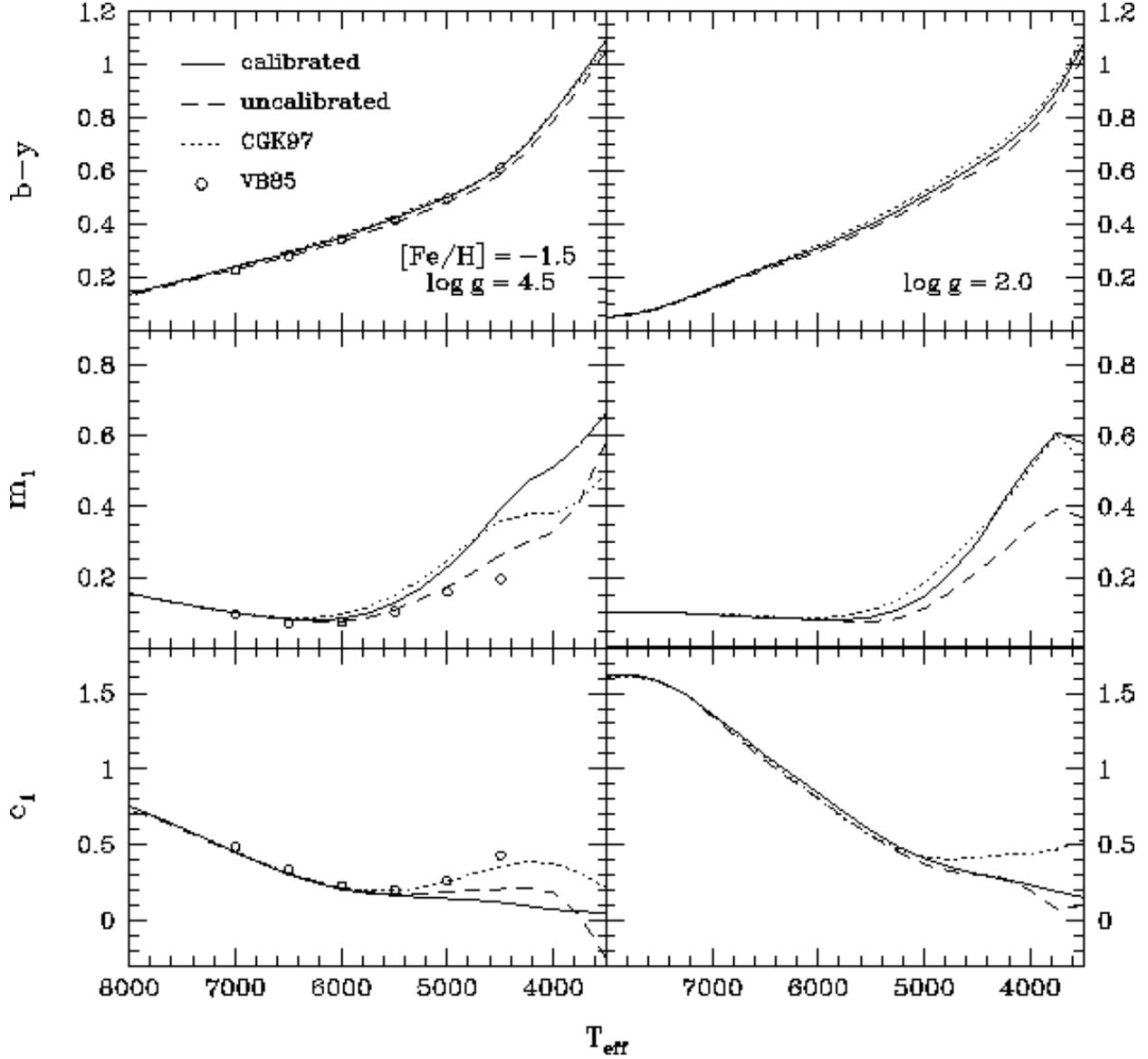} 
\caption{A comparison of synthetic color-temperature relations for $\logg$ values
corresponding to dwarfs and giants with $\FeH=-1.5$.  The {\it dashed lines} indicate
the trends from our purely synthetic colors, whereas the {\it solid lines} represent
the corrected colors.  Also plotted are the previous MARCS/SSG colors from
\citet[$open~circles$]{VandenBergBell1985} and those derived from the non-overshoot
Kurucz model atmospheres ({\it dotted lines}) as computed by \citet{Castelli1997}.}
\label{fig:compCTm15_6plot} 
\end{figure}

\begin{figure}
\plotone{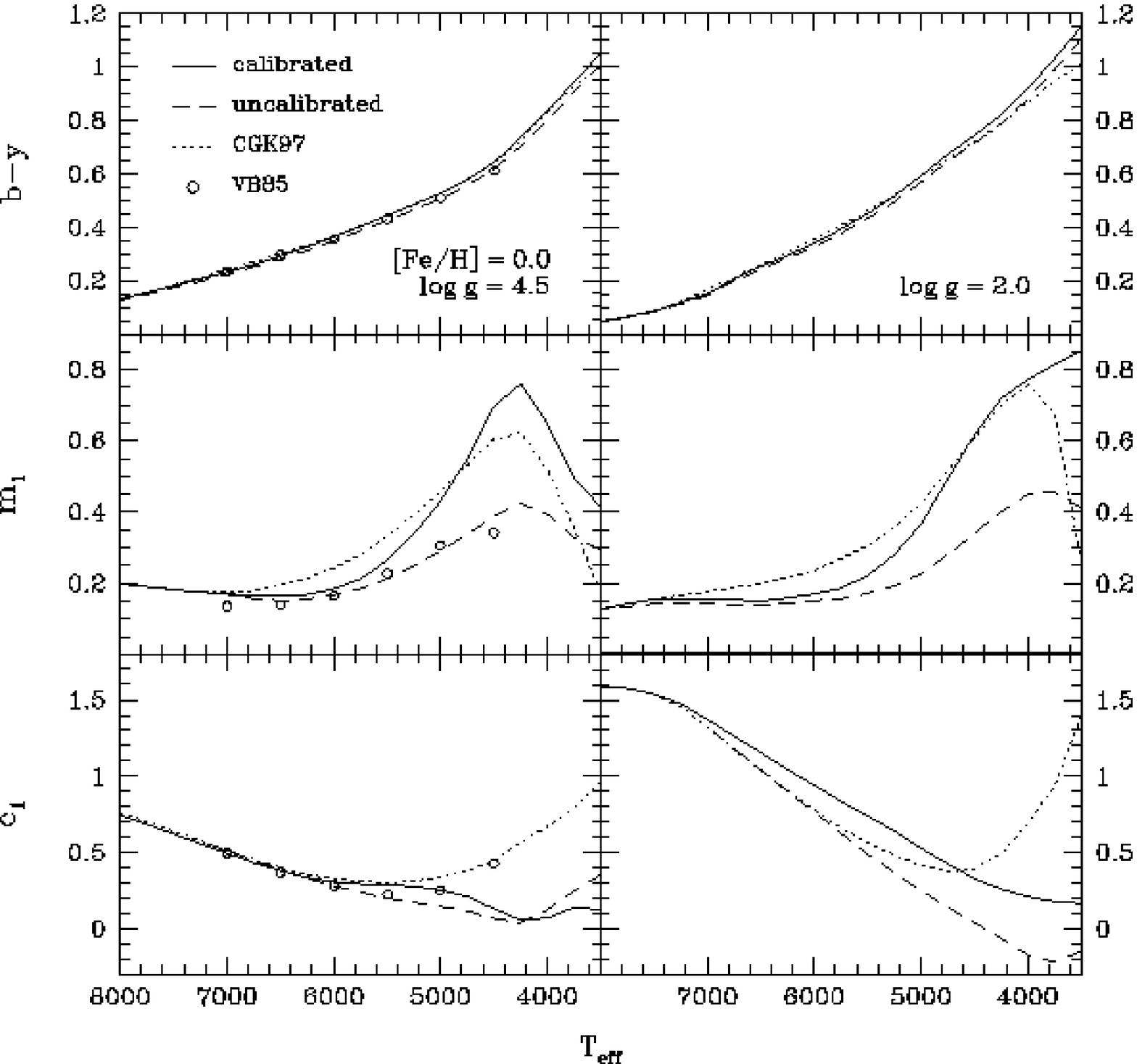}
\caption{Same as Fig.~\ref{fig:compCTm15_6plot} but for $\FeH=0.0$.}
\label{fig:compCTp00_6plot}
\end{figure}

The CGK97 study provides the optimal set of colors against which we will compare our
calibrated (and uncalibrated) color--$\Teff$ relations due to fact that their coverage
of parameter space for cool stars (i.e., $\Teff<8000\,$K) is comparable to our own.  
While comparisons between our uncalibrated MARCS/SSG colors and those of VB85 are
useful in investigating improvements in these models over the years, the colors
computed by the latter cover a much more limited range in temperature and gravity.  In
Figures \ref{fig:compCTm15_6plot} and \ref{fig:compCTp00_6plot} we compare our
calibrated and uncalibrated colors to those of CGK97 and VB85 for two representative
metallicities of $\FeH=-1.5$ and 0.0 and gravities of $\logg=4.5$ (dwarfs) and 2.0
(giants).  At first glance, there are only very small differences between the VB85 and
$uncalibrated$ dwarf $b-y$ colors in both metallicity cases.  In addition, there is
decent correspondence between the dwarf $m_1$ and $c_1$ indices of VB85 and our purely
synthetic ones for temperatures in the range of 5500$-$7000$\,$K.  For temperatures below
5500$\,$K, however, these VB85 indices start to deviate systematically from their
$uncalibrated$ counterparts.  These differences are largely due to advancements in the
MARCS/SSG modeling routines over the years, such as the inclusion of more detailed
atomic and molecular line lists and improved low-temperatures opacities since the VB85
colors were calculated.

Turning to the ATLAS9 color grids, the $b-y$ indices of CGK97 tend to show better
agreement with the $calibrated$ colors for both metallicity cases.  However, there are
some significant discrepancies in $m_1$ at temperatures between 4500 and 7000$\,$K ---
the $m_1$ indices of CGK97 tend to be systematically redder than our calibrated colors by
up to 0.1~mag. within the temperature range encompassing late-F through early-K type
dwarfs and giants.  For models cooler than $\sim4500\,$K, the CGK97 colors shift to
being systematically too blue for the dwarfs.  Moreover, fairly large discrepancies
between our $c_1$ colors and those of CGK97 are also quite obvious in the metallicity and
gravity regimes considered here.

These large discrepancies between our colors and those of CGK97 are most likely due to
differences in the absorption line lists that are used to compute the synthetic
spectra.  BPT94 noted that the lines lists of Kurucz include a number of predicted
absorption lines in the spectral regions of the \Stromgren $v$ and $u$ filters that
are not observed in the solar spectrum.  Such an overestimate in line absorption would
ultimately lead to fainter magnitudes in these passbands (hence larger values for
$m_1$ and $c_1$) than what are actually observed, especially for cooler and/or more
metal-rich models owing to the increased strength of metallic features towards later
spectral types and higher metallicities.

\begin{figure}
\plotone{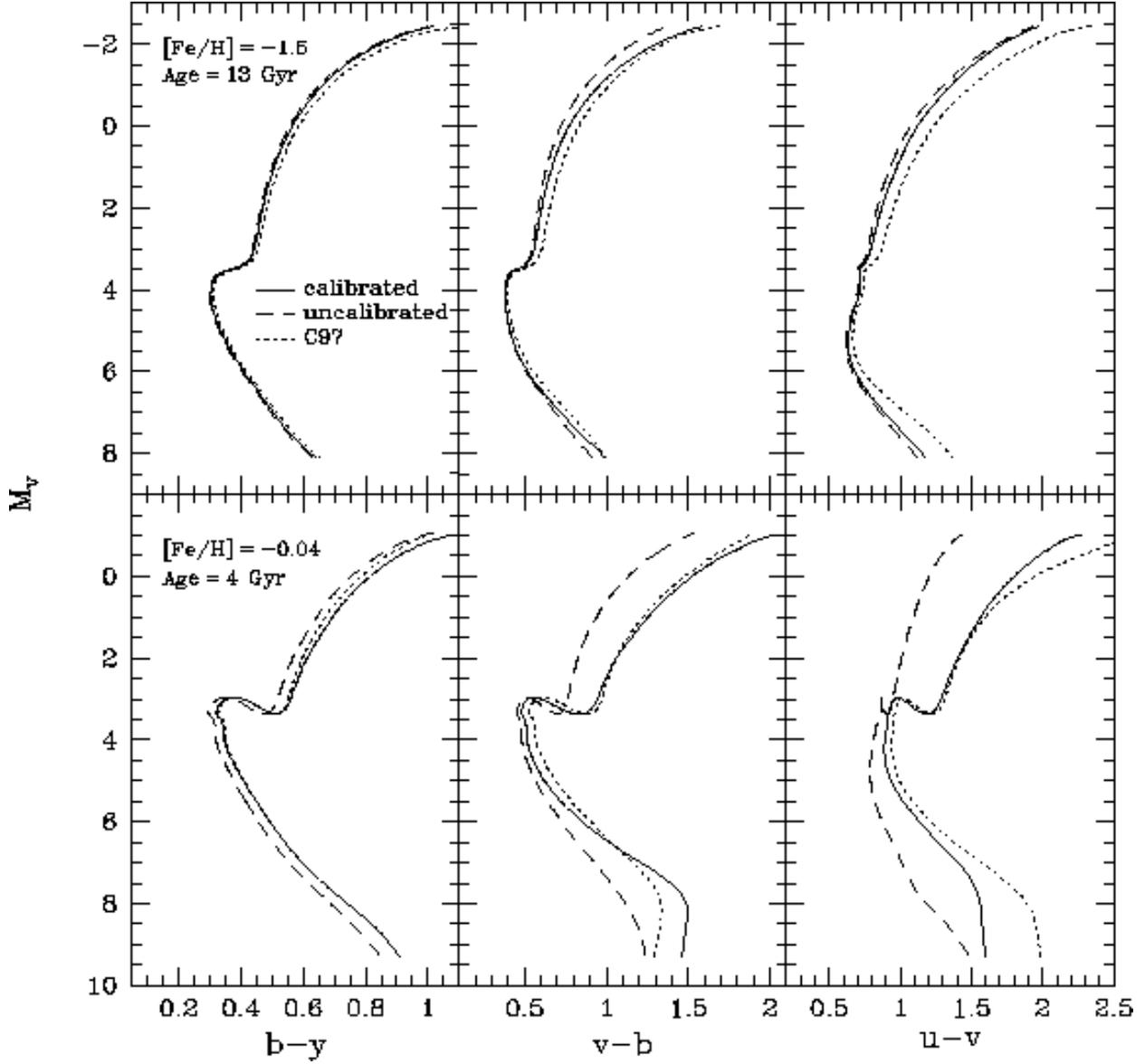}
\caption{A comparison of two isochrones having $\FeH\approx-1.5$ and 0.0 for ages of 13
and 4~Gyr, respectively, which have been transformed to the various $uvby$ CMDs using
the theoretical \Stromgren colors from our MARCS/SSG models and those of Kurucz
ATLAS9 non-overshoot models as reported by \citet{Castelli1997}.  The isochrones
corresponding to our calibrated $uvby$ colors are plotted with {\it solid lines}.}
\label{fig:compiso}
\end{figure}

To further exemplify the consequences of differences in the theoretical $uvby$
transformations considered here, Figure \ref{fig:compiso} plots two different
isochrones corresponding to the indicated ages and metallicities that have been
transformed to the various \Stromgren color-magnitude planes using the current color
transformations (both uncalibrated and calibrated) as well as those of CGK97.  We note
that identical bolometric corrections (namely those from Paper I) have been used to
derive the $M_V$ values in each CMD to ensure that any differences between the
isochrones are due solely to the color transformations.  While it is encouraging that
the calibrated and CGK97 $b-y$ isochrones agree quite well, there are fairly large
differences between the $v-b$ and $u-v$ isochrones.  As found in Paper I concerning
the $(B-V)$--$\Teff$ relations, the $uvby$ transformations of CGK97 will not provide
as good a match of the 4~Gyr isochrone to the M$\,$67 CMDs as those presented in this
study.

\subsection{Comparisons with Empirical ({\it b}$\,$--$\,${\it y})--T$_{\rm{eff}}$ Relations}
\label{subsec:empCTcomp}

There are several empirical calibrations of $b-y$ versus $\Teff$ that have been widely 
used in recent years:

\begin{itemize}

\item{\citet[hereafter SH85]{SaxnerHammarback1985} presented one of the first
empirical calibrations of $b-y$ vs. $\Teff$ based on stellar temperatures derived
using the IRFM.  Their examination was based on a total of 30 dwarf stars ranging in
temperature between 5800 and 7000 K whose colors fall within the range of
$0.20<b-y<0.40$.  While the stars used in their study covered a limited range in
metallicity, an abundance term was included in the final calibration to account for
the dependence on $\FeH$.}

\item{The IRFM temperatures for $\sim75$ stars used in the \citet[hereafter
GCC96]{Gratton1996} calibration were collected from the lists of BG89 and corrected to
the same temperature scale as \citet{BlackwellLynasGray1994}.  This large sample of
stars covered a much wider range in temperature than the aforementioned SH85 study,
and the final GCC96 calibration can be reasonably applied to solar-metallicity dwarf
stars with $0.06\leq\,b-y\,\leq0.95$.}

\item{The $(b-y)$--$\Teff$ calibrations of \citet[hereafter AAM96b]{Alonso1996b} for
dwarfs makes use of the same IRFM temperatures that we have used to calibrate the
synthetic $uvby$ colors in the present study.  This sample is by far the largest to
date with well-determined $\Teff$'s, and it includes stars with spectral types between
F0 and K5 covering a wide range in metallicity ($-3.0\leq\FeH\leq+0.5$).  To account
for the effects of differing metal abundances and gravity, their final dwarf-star
calibration is given as a function of both $\FeH$ and $c_1$.}

\end{itemize}

\begin{figure} 
\plotone{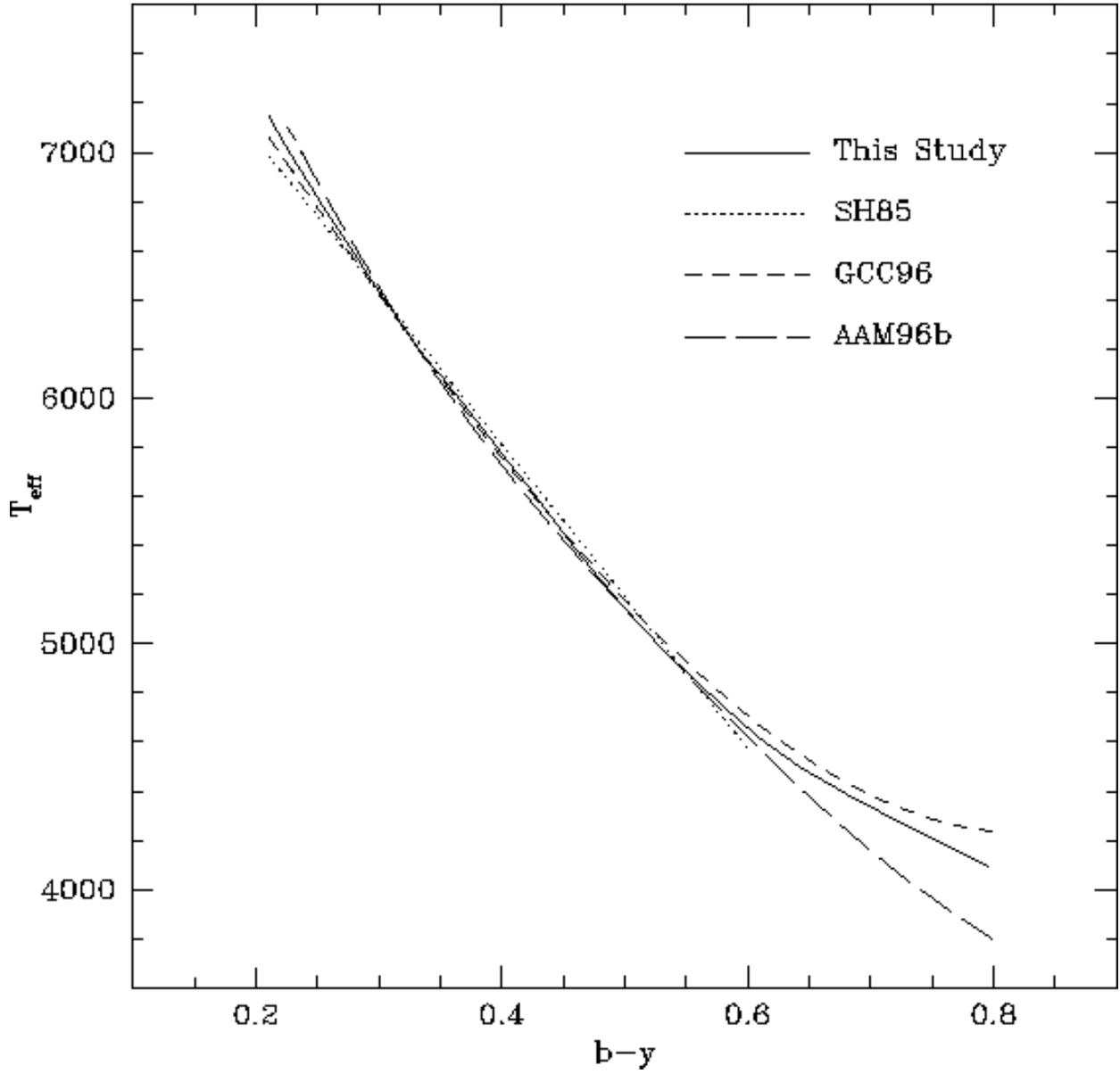} 
\caption{Empirically derived $(b-y)$--$\Teff$ relations from the indicated sources for
dwarf stars.  The predictions from our calibrated $b-y$ colors corresponding to the
temperatures of a solar metallicity ZAMS is indicated by a {\it solid line}.  Note 
that the calibration of SH85 has been extended beyond the intended limit of $b-y=0.4$ 
to demonstrate its validity for somewhat cooler dwarfs.}
\label{fig:compempCT} 
\end{figure}

Figure \ref{fig:compempCT} compares these three empirical $(b-y)$--$\Teff$
calibrations together with the relationship predicted by our solar metallicity ZAMS
model.  For those calibrations that include a metallicity term (SH85 and AAM96b),
$\FeH=0.0$ is assumed.  Also since the AAM96b calibration includes a $c_1$ term, we
opt to use the $c_1$ predictions from our ZAMS to plot their trend between $b-y$ and
$\Teff$.  It is quite reassuring that all three empirical calibrations agree rather
well in the range $0.2<b-y<0.6$, which illustrates the good consistency among the
several studies that have used the IRFM to infer stellar temperatures.  In addition,
Figure \ref{fig:compempCT} demonstrates that our calibrated \Stromgren colors can
accurately relate $b-y$ to temperature.  Indeed, our ZAMS is in superb agreement (to
well within $\pm100\,$K) with the empirical relationships of GCC96 and AAM96b over a
range in $b-y$ corresponding to F- and G-type stars.  

\subsection{\Stromgren [Fe/H] Calibration for Dwarf Stars}
\label{subsec:dwarfmetalcomp}

\begin{figure} 
\plotone{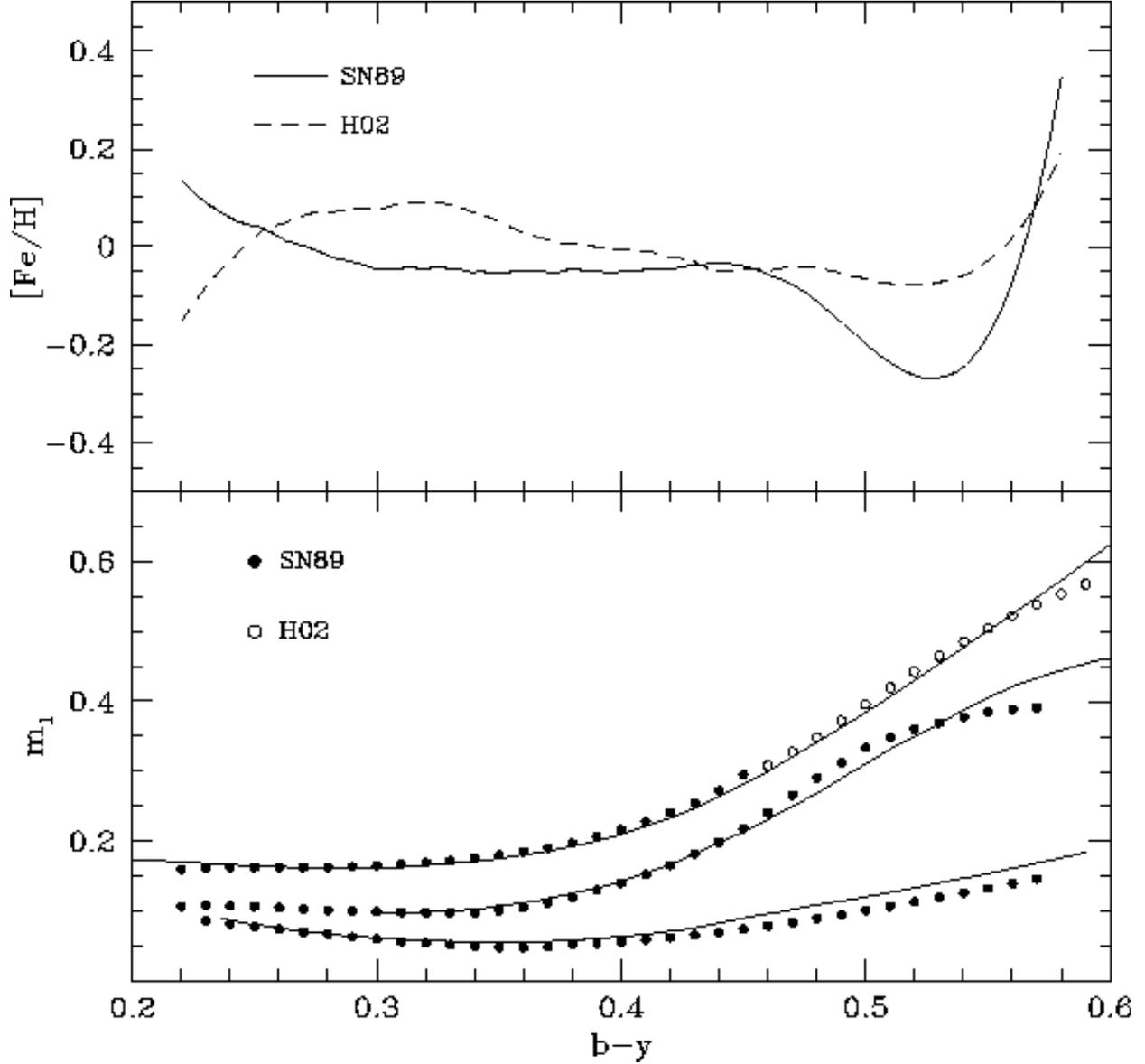} 
\caption{{\it Top panel}: The $\FeH$$-(b-y)$ relations computed from the 
\citet{SchusterNissen1989a} and \citet{Haywood2002} \Stromgren metallicity 
calibrations using the colors from our solar metallicity ZAMS model.  Note the large 
discrepancy in the SN89 calibration for $b-y\gtrsim0.47$.  {\it Bottom panel}: The 
$b-y$ and $m_1$ predictions from SN89 and H02 calibrations are compared with our ZAMS 
models for $\FeH=0.0$, $-1.0$, and $-2.0$.}
\label{fig:fehdwarf} 
\end{figure}

The \Stromgren $\FeH$ calibrations of \citet[hereafter SN89]{SchusterNissen1989a} have
served as an efficient way to determine the photometric metallicities of F- and G-type
dwarf stars, and they have been widely used in the past to investigate the metallicity
distributions of field stars in the Galactic disk and halo.  The SN89 relationships
were derived from a sample of Population~I and II stars having $\FeH$ estimates from
high-dispersion spectroscopy and employed the standard $uvby$ indices rather than
relying on the differential indices, $\delta$$m_1=m_1^{Hyades}-m_1^{star}$ and
$\delta$$c_1=c_1^{Hyades}-c_1^{star}$, used in previous \Stromgren metallicity
calibrations \citep[e.g., the relations of][]{Crawford1975, Olsen1984}.  To further
investigate the quality of our corrected transformations at $\FeH=0.0$, we have
employed the colors of our solar metallicity ZAMS model to derive $\FeH$ values as a
function of $b-y$ using the analytic formulae given by SN89.  As shown in the top
panel of Figure \ref{fig:fehdwarf}, the resultant relationship tends to hold constant
at a value of $\FeH\approx-0.05$ at $0.3\lesssim\,b-y\,\lesssim0.45$, but it deviates
towards lower metallicity values at $0.46\lesssim\,b-y\,\lesssim0.52$ and towards
higher values at $b-y\gtrsim0.53$.  (Note that we would obtain a horizontal line at
$\FeH=0.0$ if our respective calibrations were identical.)  From this evidence one
could conclude that our colors for cooler stars are in error.  However, given the
rather good agreement between our ZAMS model and the {\it Hipparcos} data (and
empirical standard relations) presented in Section \ref{subsec:Hipparcos}, we are
reluctant to make any changes to the transformations at $\FeH=0.0$ that would correct
the discrepancies in Figure \ref{fig:fehdwarf}.

We do note that a similar trend in the photometric $\FeH$ values for solar
neighborhood F- and G-type dwarf stars computed using the SN89 calibrations has also
been discovered recently by \citet{Twarog2002} and \citet[hereafter H02]{Haywood2002}.
Both of these studies employ a sample of stars from the Hyades as well as nearby {\it
Hipparcos} field stars to illustrate that their computed $\FeH$ values from SN89 are
systematically underestimated by 0.1 to 0.4~dex in the color range corresponding to
mid- to late-G dwarfs.  While the Twarog study argues that the $c_1$ index for cooler
dwarf stars is strongly affected by variations in $\FeH$, H02 concludes there is no
obvious additional dependence of the SN89 calibrations on $c_1$:  the latter rederives
the coefficients in the SN89 metallicity calibrations.  In the top panel of Figure
\ref{fig:fehdwarf} we also plot the relationship between $b-y$ and the $\FeH$ values
derived using the new H02 calibrations ({\it dashed line}).  Although we would argue
that the H02 relationship tends to slightly overestimate $\FeH$ values for stars with
$0.26\lesssim\,b-y\,\lesssim0.40$ and slightly underestimate them at
$0.40\lesssim\,b-y\,\lesssim0.56$, much of the discrepancy at $b-y\gtrsim0.47$ has
been removed.  In addition, the fact that the computed H02 $\FeH$ values are within
$\pm0.1$~dex of $\FeH=0.0$ across wide range in $b-y$ is quite reassuring since the
photometric metallicity estimates derived from these calibrations are expected to be
uncertain by at least this amount.

In the bottom panel of Figure \ref{fig:fehdwarf}, we compare the relationship between
$b-y$ and $m_1$ given by the SN89 calibrations ({\it filled circles}) to our ZAMS
predictions for $\FeH$ values of 0.0, $-1.0$, and $-2.0$.  Since the SN89 calibrations
are known to underestimate $\FeH$ for metallicities near solar and $b-y\geq0.450$ (as
discussed above), we have elected to plot the late-G dwarf calibration of H02 ({\it
open circles}, their Eq.~4) for $b-y$ colors redder than this value.  Clearly, the
agreement between our ZAMS models and the calibrations of SN89 and H02 is quite
satisfactory.

\subsection{\Stromgren [Fe/H] Calibration for Giant Stars}
\label{subsec:giantmetalcomp}

\begin{figure}
\plotone{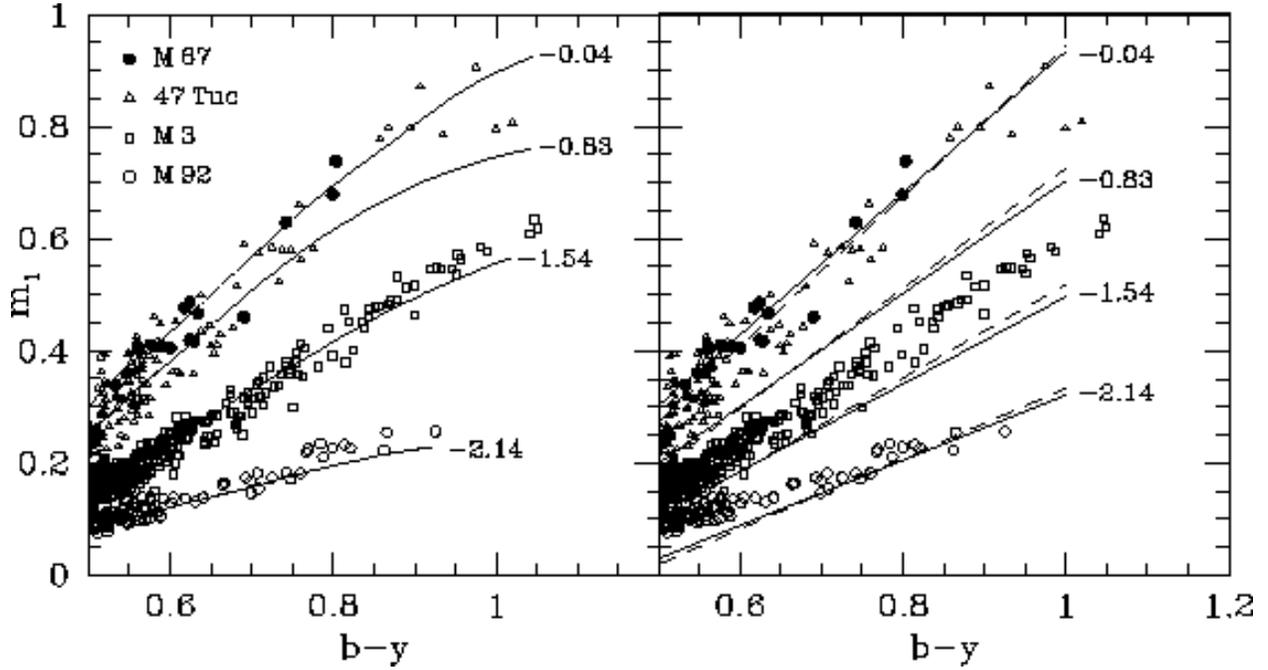}
\caption{The location of RGB stars from the clusters M$\,$92, M$\,$3, 47$\,$Tuc, and
M$\,$67 in ($b-y$,~$m_1$) space.  The {\it solid lines} in the left-hand panel denote 
the giant branch predictions from the same isochrones used to fit these clusters in 
the preceding sections.  The {\it solid} and {\it dashed lines} in the right panel are 
the two giant star metallicity calibrations of \citet{Hilker2000} plotted for the 
indicated $\FeH$ values.}
\label{fig:fehgiant} 
\end{figure}

In Figure \ref{fig:fehgiant} we plot on the ($b-y$,~$m_1$) plane the locations of RGB
stars from some of the clusters that have been considered in this study.  The cluster
photometry has been corrected for reddening using the same values (derived from the
reddening maps of \citeauthor{Schlegel1998}) denoted in the previous figures.  With
the exception of 47$\,$Tuc, the RGB stars from each cluster generally follow a fairly
tight and nearly linear relationship between their $b-y$ and $m_1$ colors.  The
increased scatter seen in the RGB of 47$\,$Tuc can likely be explained by the presence
of star-to-star differences in the amount of CN absorption.  Variations in the
strength of CN bands among stars in 47$\,$Tuc were first observed by
\citet{NorrisFreeman1979} and analyzed for abundances by \citet{Dickens1979}.  
Similar variations in main-sequence stars were first found by \citet{Hesser1978} and
have been analyzed by several authors including \citet{HesserBell1980}.  More recent
work has been carried out by several authors including \citet{Briley1997},
\citet{Cannon1998}, and \citet{Harbeck2003}.  In terms of the $uvby$ photometry
presented in Figure \ref{fig:fehgiant}, those RGB stars with strong or weak CN
features will manifest themselves as a scatter in the $m_1$ colors due to the location
of a prominent CN absorption band within the \Stromgren $v$ filter.  Therefore, stars
with abnormally large CN absorption will be scattered to large $m_1$ values while
those with relatively weak CN bands will lie at smaller $m_1$.

In the left-hand panel of Figure \ref{fig:fehgiant}, we overlay the RGB colors from
the same isochrones that were used in the previous sections to fit the cluster data.  
Upon inspection of this plot it appears that our RGB models do a good job in
reproducing not only the locations, but also the slopes of the observed giant branches
for M$\,$92, M$\,$3, and M$\,$67.  In the case of 47$\,$Tuc, however, it seems that
our adopted isochrone lies on the blue side of the RGB distribution in $m_1$,
especially for stars with $b-y\gtrsim0.8$.  As mentioned in the paragraph above, the
fairly large scatter in the cluster photometry can likely be associated with stars
having various amounts of CN absorption in their spectra.  While we have no data at
the moment that would help differentiate the 47$\,$Tuc stars in Figure
\ref{fig:fehgiant} according to their CN band strength, it suffices to say here that
our $\FeH=-0.83$ RGB model most likely matches the locus of stars with weak CN
absorption.

The right-hand panel of Figure \ref{fig:fehgiant} plots the same data for each cluster
but overlaid with lines of constant $\FeH$ predicted from the red-giant metallicity
calibrations of \citet{Hilker2000}.  In this diagram we see that the two calibrations
presented by Hilker, one based on 4 coefficients ({\it solid line}, their Eq.~1) and
the other using 5 coefficients ({\it dashed line}, their Eq.~3), tend to {\it
overestimate} the abundances of some globular clusters (by as much as 0.5~dex in
the case of M$\,$3 and 47$\,$Tuc).  The only case of good agreement with our results
is for M$\,$67 where both Hilker calibration lines fit the data nicely, and they
correspond well with our RGB predictions (left-hand panel).  In his derivation of the
red-giant calibrations, Hilker employed a sample of RGB stars both from globular
clusters (M$\,$22, M$\,$55, and $\omega\,$Cen) as well as from the field that had
\Stromgren photometry and spectroscopic metallicity estimates.  While they claim to
have placed their metallicities on the same scale as \citeauthor{ZinnWest1984}, Figure
\ref{fig:fehgiant} does not support this as, for example, M$\,$3 has $\FeH\approx-1.6$
on this scale.  It is possible that the reddening values Hilker adopted for the
globular clusters are in error:  an overestimate in E$(B-V)$ of 0.05~mag translates
approximately to a change in $\FeH$ by +0.25~dex, according to the Hilker
calibrations.  Therefore, the adoption of the correct reddening is a critical factor
in determining the location of the iso-metallicity lines based on the cluster
photometry.  (However, since Hilker also used a number of nearby field giant stars
with known metallicities, uncertainties in their reddening estimates of this magnitude
is hardly possible.)

\section{Summary \& Conclusions} 
\label{sec:conclusions}

In this investigation, we have produced an extensive set of color-temperature
relations for the \Stromgren $uvby$ system that covers a broad range in $\Teff$,
$\logg$, and $\FeH$.  To be specific, at each $\FeH$ value in the range from $-3.0$ to
$+0.5$, in steps of 0.5~dex, $b-y$, $m_1$, and $c_1$ colors are provided at
$-0.5\leq\logg\leq5.0$ and $3000\leq\Teff\leq6000\,$K (in a ``low-temperature" table),
as well as at $2.0\leq\logg\leq5.0$ and $6000<\Teff\leq40,000\,$K (in a
``high-temperature"  table).  The $v-b$ and $u-v$ colors are readily obtained from
$m_1=(v-b)-(b-y)$ and $c_1=(u-v)-(v-b)$.  By convention, \Stromgren $y$ and Johnson
$V$ magnitudes are similarly normalized, with the result that the bolometric
corrections to $y$ are nearly identical with those to $V$ and typically 
differ by $\lesssim\,$0.01$-$0.02~mag over most of parameter space.  As it is
generally the case that observed $y$ magnitudes are calibrated to be on the $V$
magnitude scale, the $BC_V$ values reported in Paper I have been included in our color
transformation tables in order for these tables and associated interpolation software
to be a self-contained package.

The main thrust of this study has been to determine what corrections to purely
synthetic color--$\Teff$ relations are needed to ensure that the latter do the best
possible job of satisfying empirical constraints.  The starting point for our analysis
consisted of a set of synthetic colors that were computed from MARCS model atmospheres
and SSG synthetic spectra at $\Teff\leq8000\,$K, to which we appended the predictions
for hotter stars as derived by \citet{Castelli1997} from Kurucz non-overshooting model
atmospheres.  There were no obvious problems with this composite grid (in conjunction
with modern isochrones) to reproduce \Stromgren observations for extremely metal-poor
turnoff and giant stars (those having $\FeH\lesssim-2.0$), as in the globular cluster
M$\,$92.  Indeed, a completely consistent interpretation of the cluster data was found
regardless of which color plane was examined, and that interpretation was virtually
identical to that obtained from a consideration of $BV$ data in Paper I.  
Discrepancies between theory and observations became evident for cool dwarfs with
$\FeH\approx-2.0$ via our comparisons to a sample of extremely metal-poor field stars.  
Moreover, as the metallicity increased, the predicted $uvby$ colors tended to be
systematically bluer than observed ones.  We assume that at least part of these
discrepancies, particularly in the \Stromgren $u$ and $v$ filters, can be attributed to
incomplete atomic and molecular line lists, not incorporating the Fe~{\sc I} continuous
opacity source \citep{Bell2001} in the SSG computations, and neglecting CN processing
in the atmospheres of our giant star models.  It is important to note, however, that
the neglect of such CN variations in giants (or the improved treatment of Fe~{\sc I}
continuous opacity) is not a concern for the {\it calibrated} color--$\Teff$ relations
that are presented in this paper because our transformations have been constrained to
reproduce the colors of {\it observed} stars.\footnote{It is precisely for this reason
that we have not put in the time and effort to do the best possible job of the
synthetic color transformations.  There is no doubt that the latter could be
significantly improved for cool, metal-rich stars (in particular) if we implemented a
better treatment of the bound-free opacity due to neutral iron, allowed for CN
processing in giants, etc., but the main results of this investigation do not depend on
the accuracy of the synthetic colors.  Even if we had started with different
model-atmosphere based transformations, we would still have obtained the same
empirically constrained color--$\Teff$ relations in the end.  While it is
desirable to determine the extent to which our results can be reproduced solely from
theory, this is outside the scope of the present project.  One concern that should be
kept in mind, however, is the possibility that the observational equipment employed by
some workers differs appreciably from that used to define the original $uvby$ system.  
This may account for the the wide variation in the colors sometimes found for the same
star by different observers.  In such cases, the calculated colors would not be
expected to match those observations very well.}

We first corrected the synthetic colors for $\FeH=0.0$ using a fairly large sample of
solar metallicity field stars having accurate $\Teff$ determinations from the infrared
flux method.  These adjustments tended to be rather large for cool stars, but they
were nonetheless necessary to place the colors on the same system as the photometric
data.  Important confirmation of the accuracy of the corrected transformations was
provided by fitting isochrones to the CMD of the open cluster M$\,$67 where models for
the observed metallicity, and whose $\Teff$ scale was precisely normalized to the Sun,
yielded a superb match to the observations.  [As cluster data played a only limited
role in the derivation of the color adjustments, and as the same interpretation of the
data was obtained here as in Paper I, which analyzed $BV(RI)_C$ observations, we
conclude that our transformations are accurate and fully consistent with those
reported in Paper I for the Johnson-Cousins photometric systems.]
 
We found that the necessary adjustments to the synthetic colors at $\Teff>5500\,$K or
$\logg<3.5$ varied nearly linearly with $\FeH$, whereas those outside this regime were
constrained based on a sample of cool field and cluster dwarf stars.  Indeed, upon
applying these corrections at $-1.5\leq\FeH\leq-0.5$ and $\FeH=+0.5$, it was possible
to achieve very good fits of isochrones to the CMDs of M$\,$3 and 47$\,$Tuc, which
have $\FeH\approx-1.55$ and $-0.75$, respectively, as well as to that of the Hyades
($\FeH=0.12$).  Once again, there was excellent consistency with the results reported
in Paper I concerning the $BV(RI)_C$ system.  Only at higher metal abundances did
inconsistencies arise as demonstrated by the fact that our fit of the same isochrones
used in Paper I to the CMD of the $\FeH\approx\,+0.4$ open cluster NGC$\,$6791
required a significant color offset to match the photometry if the same reddening used
in Paper I was adopted.  As we have been unable to understand this discrepancy, which
may require further observations to resolve it, our transformations at
$\FeH\gtrsim0.15$ (probably those in Paper I as well) should be used with caution.  
More observational constraints are needed to put the color--$T_{\rm eff}$ relations
for super-metal-rich stars on a firm foundation.

We have found good consistency between our $(b-y)$--$\Teff$ relation for dwarf stars
having near solar abundances and those derived by \citet{SaxnerHammarback1985},
Gratton et al.~\citeyearpar{Gratton1996}, and Alonso et al.~\citeyearpar{Alonso1996b}.  
Comparisons between the predicted and observed colors for the classic Population II
subdwarfs have also offered encouraging support for our results.  Moreover, the
dependence of $m_1$ on $b-y$ predicted from our color transformations agrees well with
the relations for solar-metallicity dwarfs derived by \citet{SchusterNissen1989a} at
$b-y\lesssim0.45$ and by \citet{Haywood2002} at redder colors, and with the relations
for giants as inferred from a sample RGB stars from a number of globular and open
clusters that encompass a wide range in $\FeH$.

It is our intention to update the transformations reported in this paper as new
constraints become available.  Even at this stage, however, we believe that our
results represent a considerable improvement over currently available color--$\Teff$
relations for the \Stromgren photometric system and that their use will lead to
important refinements in our our understanding of stars and stellar populations.  The
next paper in this series (Clem, VandenBerg, \& Stetson, in preparation) will present
color transformations for the Sloan $u^\prime g^\prime r^\prime i^\prime z^\prime$
system.

\acknowledgements 

We thank Angel Alonso for providing a machine readable version of Table 6 in Alonso
et~al.~(1999).  Erik Heyn Olsen is thanked for making his large database of $uvby$
photometry available to one of us (F.G.) for use in our analysis.  We would also
like to extend our appreciation to the anonymous referee whose helpful comments and
suggestions ultimately improved the quality of this paper.  This research has made
extensive use of the SIMBAD and Vizier databases operated at the Canadian
Astronomical Data Center by the Herzberg Institute of Astrophysics in Victoria,
British Columbia.  This work has been supported by an Operating Grant to D.A.V. from
the Natural Sciences and Engineering Research Council of Canada.  F.G. gratefully
acknowledges financial support from the Carlsberg Foundation.

\clearpage

\bibliographystyle{astron}
\bibliography{clem}

\begin{thebibliography}{}

\bibitem[Allende Prieto \& Lambert(2000)]{AllendePrietoLambert2000} 
Allende Prieto, C.~\& Lambert, D.~L.\ 2000, \aj, 119, 2445 

\bibitem[Alonso, Arribas, \& Mart{\'{\i}}nez-Roger(1996a)]{Alonso1996a} 
Alonso, A., Arribas, S., \& Mart{\'{\i}}nez-Roger, C.\ 1996a, \aaps, 117, 227~(AAM96)

\bibitem[Alonso, Arribas, \& Mart{\'{\i}}nez-Roger(1996b)]{Alonso1996b} 
Alonso, A., Arribas, S., \& Mart{\'{\i}}nez-Roger, C.\ 1996b, \aap, 313, 873~(AAM96b)

\bibitem[Alonso, Arribas, \& Mart{\'{\i}}nez-Roger(1999)]{Alonso1999a} 
Alonso, A., Arribas, S., \& Mart{\'{\i}}nez-Roger, C.\ 1999, \aaps, 139, 335~(AAM99)

\bibitem[Anders \& Grevesse(1989)]{AndersGrevesse1989} 
Anders, E.~\& Grevesse, N.\ 1989, \gca, 53, 197 

\bibitem[Anthony-Twarog(1987a)]{Anthony-Twarog1987a} 
Anthony-Twarog, B.~J.\ 1987a, \aj, 93, 647 

\bibitem[Anthony-Twarog(1987b)]{Anthony-Twarog1987b} 
Anthony-Twarog, B.~J.\ 1987b, \aj, 93, 1454 

\bibitem[Anthony-Twarog \& Twarog(1987)]{Anthony-TwarogTwarog1987} 
Anthony-Twarog, B.~J.~\& Twarog, B.~A.\ 1987, \aj, 94, 1222 

\bibitem[Anthony-Twarog \& Twarog(1994)]{Anthony-TwarogTwarog1994} 
Anthony-Twarog, B.~J.~\& Twarog, B.~A.\ 1994, \aj, 107, 1577 

\bibitem[Axer, Fuhrmann, \& Gehren(1994)]{Axer1994} 
Axer, M., Fuhrmann, K., \& Gehren, T.\ 1994, \aap, 291, 895 

\bibitem[Bell(1988)]{Bell1988} 
Bell, R.~A.\ 1988, \aj, 95, 1484 

\bibitem[Bell, Balachandran, \& Bautista(2001)]{Bell2001} 
Bell, R.~A., Balachandran, S.~C., \& Bautista, M.\ 2001, \apjl, 546, L65 

\bibitem[Bell \& Gustafsson(1989)]{BellGustafsson1989} 
Bell, R.~A.~\& Gustafsson, B.\ 1989, \mnras, 236, 653~(BG89)

\bibitem[Bell, Paltoglou, \& Tripicco(1994)]{Bell1994} 
Bell, R.~A., Paltoglou, G., \& Tripicco, M.~J.\ 1994, \mnras, 268, 771~(BPT94)

\bibitem[Bergbusch \& VandenBerg(2001)]{BergbuschVandenBerg2001} 
Bergbusch, P.~A.~\& VandenBerg, D.~A.\ 2001, \apj, 556, 322 

\bibitem[Blackwell \& Lynas-Gray(1994)]{BlackwellLynasGray1994} 
Blackwell, D.~E.~\& Lynas-Gray, A.~E.\ 1994, \aap, 282, 899 

\bibitem[Boesgaard \& Friel(1990)]{BoesgaardFriel1990} 
Boesgaard, A.~M.~\& Friel, E.~D.\ 1990, \apj, 351, 467 

\bibitem[Bonifacio, Caffau, \& Molaro(2000)]{Bonifacio2000} 
Bonifacio, P., Caffau, E., \& Molaro, P.\ 2000, \aaps, 145, 473 

\bibitem[Briley(1997)]{Briley1997} 
Briley, M.~M.\ 1997, \aj, 114, 1051 

\bibitem[Caldwell et al.(1993)]{Caldwell1993} 
Caldwell, J.~A.~R., Cousins, A.~W.~J., Ahlers, C.~C., van Wamelen, P., \& Maritz, E.~J.\ 1993, SAAO~Circ., 15, 1 

\bibitem[Cannon et al.(1998)]{Cannon1998} 
Cannon, R.~D., Croke, B.~F.~W., Bell, R.~A., Hesser, J.~E., \& Stathakis, R.~A.\ 1998, \mnras, 298, 601 

\bibitem[Carbon et al.(1987)]{Carbon1987} 
Carbon, D.~F., Barbuy, B., Kraft, R.~P., Friel, E.~D., \& Suntzeff, N.~B.\ 1987, \pasp, 99, 335 

\bibitem[Carney(1996)]{Carney1996} 
Carney, B.~W.\ 1996, \pasp, 108, 900 

\bibitem[Carretta \& Gratton(1997)]{CarrettaGratton1997} 
Carretta, E.~\& Gratton, R.~G.\ 1997, \aaps, 121, 95 

\bibitem[Carretta et al.(2000)]{Carretta2000} 
Carretta, E., Gratton, R.~G., Clementini, G., \& Fusi Pecci, F.\ 2000, \apj, 533, 215 

\bibitem[Castelli, Gratton, \& Kurucz(1997)]{Castelli1997} 
Castelli, F., Gratton, R.~G., \& Kurucz, R.~L.\ 1997, \aap, 318, 841~(CGK97)

\bibitem[Castelli \& Kurucz(1994)]{CastelliKurucz1994} 
Castelli, F.~\& Kurucz, R.~L.\ 1994, \aap, 281, 817 

\bibitem[Cayrel, Cayrel de Strobel, \& Campbell(1985)]{Cayrel1985} 
Cayrel, R., Cayrel de Strobel, G., \& Campbell, B.\ 1985, \aap, 146, 249 

\bibitem[Cayrel de Strobel, Soubiran, \& Ralite(2001)]{CayreldeStrobel2001} 
Cayrel de Strobel, G., Soubiran, C., \& Ralite, N.\ 2001, \aap, 373, 159 

\bibitem[Claret, Diaz-Cordoves, \& Gimenez(1995)]{Claret1995} 
Claret, A., Diaz-Cordoves, J., \& Gimenez, A.\ 1995, \aaps, 114, 247 

\bibitem[Clegg \& Bell(1973)]{CleggBell1973} 
Clegg, R.~E.~S.~\& Bell, R.~A.\ 1973, \mnras, 163, 13 

\bibitem[Clementini et al.(1999)]{Clementini1999} 
Clementini, G., Gratton, R.~G., Carretta, E., \& Sneden, C.\ 1999, \mnras, 302, 22 

\bibitem[Crawford(1975)]{Crawford1975} 
Crawford, D.~L.\ 1975, \aj, 80, 955 

\bibitem[Crawford \& Barnes(1969)]{CrawfordBarnes1969} 
Crawford, D.~L.~\& Barnes, J.~V.\ 1969, \aj, 74, 818 

\bibitem[Crawford \& Barnes(1970a)]{CrawfordBarnes1970a} 
Crawford, D.~L.~\& Barnes, J.~V.\ 1970a, \aj, 75, 946 

\bibitem[Crawford \& Barnes(1970b)]{CrawfordBarnes1970b} 
Crawford, D.~L.~\& Barnes, J.~V.\ 1970b, \aj, 75, 978 

\bibitem[Crawford \& Mandwewala(1976)]{CrawfordMandwewala1976} 
Crawford, D.~L.~\& Mandwewala, N.\ 1976, \pasp, 88, 917 

\bibitem[Crawford \& Perry(1966)]{CrawfordPerry1966} 
Crawford, D.~L.~\& Perry, C.~L.\ 1966, \aj, 71, 206 

\bibitem[Crawford \& Perry(1976)]{CrawfordPerry1976} 
Crawford, D.~L.~\& Perry, C.~L.\ 1976, \aj, 81, 419 

\bibitem[de Bruijne, Hoogerwerf, \& de Zeeuw(2001)]{deBruijne2001} 
de Bruijne, J.~H.~J., Hoogerwerf, R., \& de Zeeuw, P.~T.\ 2001, \aap, 367, 111 

\bibitem[Dickens, Bell, \& Gustafsson(1979)]{Dickens1979} 
Dickens, R.~J., Bell, R.~A., \& Gustafsson, B.\ 1979, \apj, 232, 428 

\bibitem[Dreiling \& Bell(1980)]{DreilingBell1980} 
Dreiling, L.~A.~\& Bell, R.~A.\ 1980, \apj, 241, 736 

\bibitem[Fuhrmann(1998)]{Fuhrmann1998} 
Fuhrmann, K.\ 1998, \aap, 338, 161 

\bibitem[Fulbright(2000)]{Fulbright2000} 
Fulbright, J.~P.\ 2000, \aj, 120, 1841 

\bibitem[Gratton, Carretta, \& Castelli(1996)]{Gratton1996} 
Gratton, R.~G., Carretta, E., \& Castelli, F.\ 1996, \aap, 314, 191~(GCC96)

\bibitem[Gratton et al.(2000)]{Gratton2000} 
Gratton, R.~G., Sneden, C., Carretta, E., \& Bragaglia, A.\ 2000, \aap, 354, 169 

\bibitem[Grevesse et al.(1990)]{Grevesse1990} 
Grevesse, N., Lambert, D.~L., Sauval, A.~J., van Dishoeck, E.~F., Farmer, C.~B., \& Norton, R.~H.\ 1990, \aap, 232, 225 

\bibitem[Grevesse et al.(1991)]{Grevesse1991} 
Grevesse, N., Lambert, D.~L., Sauval, A.~J., van Dishoek, E.~F., Farmer, C.~B., \& Norton, R.~H.\ 1991, \aap, 242, 488 

\bibitem[Grundahl(1999)]{Grundahl1999} 
Grundahl, F.\ 1999, ASP Conf.~Ser.~192: Spectrophotometric Dating of Stars and Galaxies, 223 

\bibitem[Grundahl et al.(2002a)]{Grundahl2002a} 
Grundahl, F., Briley, M., Nissen, P.~E., \& Feltzing, S.\ 2002a, \aap, 385, L14 

\bibitem[Grundahl, Stetson, \& Andersen(2002b)]{Grundahl2002b} 
Grundahl, F., Stetson, P.~B., \& Andersen, M.~I.\ 2002b, \aap, 395, 481 

\bibitem[Grundahl, VandenBerg, \& Andersen(1998)]{Grundahl1998} 
Grundahl, F., VandenBerg, D.~A., \& Andersen, M.~I.\ 1998, \apjl, 500, L179 

\bibitem[Grundahl et al.(2000a)]{Grundahl2000a} 
Grundahl, F., VandenBerg, D.~A., Bell, R.~A., Andersen, M.~I., \& Stetson, P.~B.\ 2000a, \aj, 120, 1884 

\bibitem[Grundahl et al.(2000b)]{Grundahl2000b} 
Grundahl, F., VandenBerg, D.~A., Stetson, P.~B., Andersen, M.~I., \& Briley, M.\ 2000b, The Galactic Halo : From Globular Cluster to Field Stars, 503 

\bibitem[Gustafsson et al.(1975)]{Gustafsson1975} 
Gustafsson, B., Bell, R.~A., Eriksson, K., \& Nordlund, A.\ 1975, \aap, 42, 407 

\bibitem[Harbeck, Smith, \& Grebel(2003)]{Harbeck2003} 
Harbeck, D., Smith, G.~H., \& Grebel, E.~K.\ 2003, \aj, 125, 197 

\bibitem[Hauck \& Mermilliod(1998)]{HauckMermilliod1998} 
Hauck, B.~\& Mermilliod, M.\ 1998, \aaps, 129, 431~(HM98)

\bibitem[Hayes \& Latham(1975)]{HayesLatham1975} 
Hayes, D.~S.~\& Latham, D.~W.\ 1975, \apj, 197, 593 

\bibitem[Haywood(2001)]{Haywood2001} 
Haywood, M.\ 2001, \mnras, 325, 1365 

\bibitem[Haywood(2002)]{Haywood2002} 
Haywood, M.\ 2002, \mnras, 337, 151~(H02)

\bibitem[Hesser(1978)]{Hesser1978} 
Hesser, J.~E.\ 1978, \apjl, 223, L117 

\bibitem[Hesser \& Bell(1980)]{HesserBell1980} 
Hesser, J.~E.~\& Bell, R.~A.\ 1980, \apjl, 238, L149 

\bibitem[Hesser et al.(1987)]{Hesser1987} 
Hesser, J.~E., Harris, W.~E., VandenBerg, D.~A., Allwright, J.~W.~B., Shott, P., \& Stetson, P.~B.\ 1987, \pasp, 99, 739 

\bibitem[Hilker(2000)]{Hilker2000} 
Hilker, M.\ 2000, \aap, 355, 994 

\bibitem[Hilker \& Richtler(2000)]{HilkerRichtler2000} 
Hilker, M.~\& Richtler, T.\ 2000, \aap, 362, 895 

\bibitem[Houdashelt et al.(2000a)]{Houdashelt2000a} 
Houdashelt, M.~L., Bell, R.~A., Sweigart, A.~V., \& Wing, R.~F.\ 2000a, \aj, 119, 1424

\bibitem[Houdashelt, Bell, \& Sweigart(2000b)]{Houdashelt2000b} 
Houdashelt, M.~L., Bell, R.~A., \& Sweigart, A.~V.\ 2000b, \aj, 119, 1448~(HBS2000)

\bibitem[Hughes \& Wallerstein(2000)]{HughesWallerstein2000} 
Hughes, J.~\& Wallerstein, G.\ 2000, \aj, 119, 1225 

\bibitem[Kraft \& Ivans(2003)]{KraftIvans2003} 
Kraft, R.~P.~\& Ivans, I.~I.\ 2003, \pasp, 115, 143 

\bibitem[Kraft et al.(1998)]{Kraft1998} 
Kraft, R.~P., Sneden, C., Smith, G.~H., Shetrone, M.~D., \& Fulbright, J.\ 1998, \aj, 115, 1500 

\bibitem[Kurucz(1979)]{Kurucz1979} 
Kurucz, R.~L.\ 1979, \apjs, 40, 1 

\bibitem[Kurucz(1993)]{Kurucz1993} 
Kurucz, R.\ 1993, ATLAS9 Stellar Atmosphere Programs and 2 km/s grid.~Kurucz CD-ROM No.~13.~ Cambridge, Mass.: Smithsonian Astrophysical Observatory, 1993., 13,  

\bibitem[Lester, Gray, \& Kurucz(1986)]{Lester1986} 
Lester, J.~B., Gray, R.~O., \& Kurucz, R.~L.\ 1986, \apjs, 61, 509 

\bibitem[Nissen(1988)]{Nissen1988} 
Nissen, P.~E.\ 1988, \aap, 199, 146 

\bibitem[Nissen, Twarog, \& Crawford(1987)]{Nissen1987} 
Nissen, P.~E., Twarog, B.~A., \& Crawford, D.~L.\ 1987, \aj, 93, 634 

\bibitem[Nordgren et al.(1999)]{Nordgren1999} 
Nordgren, T.~E.~et al.\ 1999, \aj, 118, 3032~(N99)

\bibitem[Nordgren et al.(2001)]{Nordgren2001} 
Nordgren, T.~E., Sudol, J.~J., \& Mozurkewich, D.\ 2001, \aj, 122, 2707~(N01)

\bibitem[Norris \& Freeman(1979)]{NorrisFreeman1979} 
Norris, J.~\& Freeman, K.~C.\ 1979, \apjl, 230, L179 

\bibitem[Olsen(1983)]{Olsen1983} 
Olsen, E.~H.\ 1983, \aaps, 54, 55 

\bibitem[Olsen(1984)]{Olsen1984} 
Olsen, E.~H.\ 1984, \aaps, 57, 443 

\bibitem[Olsen(1993)]{Olsen1993} 
Olsen, E.~H.\ 1993, \aaps, 102, 89 

\bibitem[Olsen(1994)]{Olsen1994} 
Olsen, E.~H.\ 1994, \aaps, 106, 257 

\bibitem[Olsen(1995)]{Olsen1995} 
Olsen, E.~H.\ 1995, \aap, 295, 710 

\bibitem[Peterson \& Green(1998)]{PetersonGreen1998} 
Peterson, R.~C.~\& Green, E.~M.\ 1998, \apjl, 502, L39 

\bibitem[Philip \& Egret(1980)]{PhilipEgret1980} 
Philip, A.~D.~\& Egret, D.\ 1980, \aaps, 40, 199 

\bibitem[Philip \& Egret(1983)]{PhilipEgret1983} 
Philip, A.~D.~\& Egret, D.\ 1983, \aap, 123, 39 

\bibitem[Relyea \& Kurucz(1978)]{RelyeaKurucz1978} 
Relyea, L.~J.~\& Kurucz, R.~L.\ 1978, \apjs, 37, 45 

\bibitem[Saxner \& Hamm\"{a}rback(1985)]{SaxnerHammarback1985} 
Saxner, M.~\& Hamm\"{a}rback, G.\ 1985, \aap, 151, 372~(SH85)

\bibitem[Schlegel et al.(1998)]{Schlegel1998} 
Schlegel, D.~J., Finkbeiner, D.~P., \& Davis, M.\ 1998, \apj, 500, 525 

\bibitem[Schuster \& Nissen(1988)]{SchusterNissen1988} 
Schuster, W.~J.~\& Nissen, P.~E.\ 1988, \aaps, 73, 225 

\bibitem[Schuster \& Nissen(1989a)]{SchusterNissen1989a} 
Schuster, W.~J.~\& Nissen, P.~E.\ 1989a, \aap, 221, 65~(SN89)

\bibitem[Schuster \& Nissen(1989b)]{SchusterNissen1989b} 
Schuster, W.~J.~\& Nissen, P.~E.\ 1989b, \aap, 222, 69 

\bibitem[Schuster, Parrao, \& Contreras Martinez(1993)]{Schuster1993} 
Schuster, W.~J., Parrao, L., \& Contreras Martinez, M.~E.\ 1993, \aaps, 97, 951 

\bibitem[Sekiguchi \& Fukugita(2000)]{SekiguchiFukugita2000} 
Sekiguchi, M.~\& Fukugita, M.\ 2000, \aj, 120, 1072 

\bibitem[Spiesman(1992)]{Spiesman1992} 
Spiesman, W.~J.\ 1992, \apjl, 397, L103 

\bibitem[Stetson, Bruntt, \& Grundahl(2003)]{Stetson2003} 
Stetson, P.~B., Bruntt, H., \& Grundahl, F.\ 2003, \pasp, 115, 413 

\bibitem[Stetson \& Harris(1988)]{StetsonHarris1988} 
Stetson, P.~B.~\& Harris, W.~E.\ 1988, \aj, 96, 909 

\bibitem[Str{\" o}mgren(1963)]{Stromgren1963} 
Str{\" o}mgren, B.\ 1963, \qjras, 4, 8 

\bibitem[Tomkin \& Lambert(1999)]{TomkinLambert1999} 
Tomkin, J.~\& Lambert, D.~L.\ 1999, \apj, 523, 234 

\bibitem[Twarog, Anthony-Twarog, \& Tanner(2002)]{Twarog2002} 
Twarog, B.~A., Anthony-Twarog, B.~J., \& Tanner, D.\ 2002, \aj, 123, 2715 

\bibitem[VandenBerg \& Bell(1985)]{VandenBergBell1985} 
Vandenberg, D.~A.~\& Bell, R.~A.\ 1985, \apjs, 58, 561~(VB85)

\bibitem[VandenBerg \& Clem(2003)]{VandenBergClem2003} 
VandenBerg, D.~A.~\& Clem, J.~L.\ 2003, \aj, 126, 778~(Paper I)

\bibitem[Zhao \& Magain(1990)]{ZhaoMagain1990} 
Zhao, G.~\& Magain, P.\ 1990, \aap, 238, 242 

\bibitem[Zinn \& West(1984)]{ZinnWest1984} 
Zinn, R.~\& West, M.~J.\ 1984, \apjs, 55, 45 

\end{thebibliography}

\clearpage

\begin{deluxetable}{llrrrrrr}
\tabletypesize{\footnotesize}
\tablecaption{Comparison of $\Teff$ and $\Fbol$ Estimates}
\tablewidth{0pt}
\tablehead{\colhead{HR \#} &
	   \colhead{Spec. Type} &
	   \multispan{2}{$\hfil$ ~AAM96/AAM99 $\hfil$} &
	   \multispan{2}{$\hfil$ ~~~~BG89 $\hfil$} &
	   \multispan{2}{$\hfil$ ~~~SH85 $\hfil$} \\
	        & &  $\Fbol$~ &  $\Teff$~  & $\Fbol$~  & $\Teff$~ & $\Fbol$~ & $\Teff$~}
\tablenotetext{~}{Note. --- $\Fbol$ has units of 10$^{-8}$~ergs~cm$^{-2}$~s$^{-1}$.}
\startdata
219.....	&   G0 V      &   113.80  &  5817  &   119.30  &  5839   &     -  &   -    \\
434.....        &   K4 III    &    63.26  &  4046  &    68.40  &  4046   &     -  &   -    \\
458.....	&   F8 V      &    60.34  &  6155  &      -    &   -     &   61.7 &  6154  \\
464.....	&   K3 III    &   156.50  &  4359  &   167.30  &  4425   &     -  &   -    \\
483.....	&   G1.5 V    &    27.88  &  5874  &      -    &   -     &   29.0 &  5856  \\
617.....	&   K2 III    &   625.20  &  4490  &   652.00  &  4499   &     -  &   -    \\
937.....	&   G0 V      &    63.74  &  5996  &      -    &   -     &   65.7 &  5994  \\
1084...		&   K2 V      &   100.30  &  5076  &   109.30  &  5156   &     -  &   -    \\
1101...		&   F9 IV-V   &    51.86  &  5998  &      -    &   -     &   52.0 &  5963  \\
1325...		&   K1 V      &    53.11  &  5040  &    56.09  &  5114   &     -  &   -    \\
1457...		&   K5 III    &  3247.00  &  3866  &  3422.00  &  3943   &     -  &   -    \\
1543...		&   F6 V      &   137.40  &  6482  &      -    &   -     &  139.0 &  6373  \\
1729...         &   G1.5 IV-V &    35.09  &  5847  &      -    &   -     &   36.5 &  5819  \\
1907...         &   K0 IIIb   &    81.93  &  4693  &    86.55  &  4719   &     -  &   -    \\
1983... 	&   F7 V      &    95.74  &  6260  &      -    &   -     &   97.6 &  6259  \\
2085...         &   F1 V      &    83.01  &  7013  &      -    &   -     &  128.0 &  8144  \\
2852...		&   F0 V      &    55.17  &  7020  &      -    &   -     &   55.1 &  6957  \\
2943...		&   F5 IV-V   &  1844.00  &  6579  &      -    &   -     & 1860.0 &  6601  \\
2990...		&   K0 IIIb   &  1140.00  &  4854  &  1236.00  &  4896   &     -  &   -    \\
3323...		&   G5 III    &   135.50  &  5136  &   139.70  &  5176   &     -  &   -    \\
4247...         &   K0 III    &   105.10  &  4643  &   112.40  &  4692   &     -  &   -    \\
4496...         &   G8 V      &    20.99  &  5342  &    23.19  &  5552   &     -  &   -    \\
4518... 	&   K0.5 IIIb &   135.50  &  4348  &   144.00  &  4421   &     -  &   -    \\
4540...         &   F9 V      &    94.19  &  6095  &      -    &   -     &  100.0 &  6147  \\
4785...	        &   G0 V      &    53.19  &  5867  &    54.09  &  5861   &   54.3 &  5842  \\
4883...		&   G0 IIIp   &    28.69  &  5589  &    30.50  &  5761   &     -  &   -    \\
4932...  	&   G8 III    &   225.10  &  5043  &   236.20  &  5052   &     -  &   -    \\
4983... 	&   F9.5 V    &    52.41  &  5964  &    55.09  &  6024   &     -  &   -    \\
5340... 	&   K1.5 III  &  4830.00  &  4233  &  5159.00  &  4321   &     -  &   -    \\
5429...		&   K3 III    &   163.80  &  4271  &   175.30  &  4303   &     -  &   -    \\
5447...		&   F2 V      &    43.28  &  6707  &      -    &   -     &   42.5 &  6696  \\
5634...	        &   F5 V      &    27.55  &  6571  &      -    &   -     &   28.0 &  6616  \\
5681... 	&   G8 III    &   134.20  &  4798  &   142.30  &  4832   &     -  &   -    \\
5868...         &   G0 V      &    45.14  &  5897  &      -    &   -     &   46.7 &  5940  \\
5914...		&   F8 Ve     &    39.00  &  5774  &      -    &   -     &   41.7 &  5802  \\
5933...		&   F6 IV     &    75.90  &  6233  &      -    &   -     &   78.4 &  6246  \\
6220...		&   G7.5 IIIb &   126.10  &  4942  &   131.90  &  4913   &     -  &   -    \\
6603...		&   K2 III    &   295.20  &  4533  &   310.80  &  4603   &     -  &   -    \\
6705...		&   K5 III    &   805.80  &  3934  &   890.60  &  3955   &     -  &   -    \\
6869...         &   K0 III-IV &   165.80  &  4835  &   177.50  &  4949   &     -  &   -    \\
7429...		&   K3 IIIb   &    65.39  &  4473  &    69.71  &  4456   &     -  &   -    \\
7462...         &   K0 V      &    40.12  &  5227  &    42.49  &  5253   &     -  &   -    \\
7503...		&   G1.5 Vb   &    11.15  &  5763  &    11.33  &  5826   &     -  &   -    \\
7504... 	&   G3 V      &     8.96  &  5767  &     9.31  &  5664   &     -  &   -    \\
7615...		&   K0 III    &    93.60  &  4796  &    96.90  &  4887   &     -  &   -    \\
7957...	        &   K0 IV     &   136.60  &  4908  &   147.90  &  4996   &     -  &   -    \\
8085...		&   K5 V      &    37.15  &  4323  &    39.43  &  4463   &     -  &   -    \\
8086...		&   K7 V      &    22.20  &  3865  &    25.33  &  4252   &     -  &   -    \\
8255...		&   K1 III    &    39.63  &  4609  &    43.30  &  4715   &     -  &   -    \\
8832...		&   K3 V      &    20.44  &  4785  &    22.42  &  4896   &     -  &   -    \\
8905...		&   F8 IV     &    46.12  &  5954  &      -    &   -     &   47.5 &  5897  \\
\enddata	
\label{tab:comparisons}
\end{deluxetable}

\begin{deluxetable}{lcccr} 
\tabletypesize{\footnotesize} 
\tablecaption{Comparison of Stellar Angular Diameters}
\tablewidth{0pt}
\tablehead{\colhead{HR \#} &
	   \colhead{$\theta_{BG89}$} &
	   \colhead{$\theta_{AAM99}$} &
	   \colhead{$\theta_{N99/N01}$} &
	   \colhead{$\pi$ (mas)}  \\ } 
\tablenotetext{~}{Note. --- The errors quoted in columns 2 and 
3 are derived assuming a 5\% and $\pm100\,$K uncertainty in the $\Fbol$ and $\Teff$ 
estimates presented by BG89 and AAM99.}
\startdata
163.....        &          -         &   1.72 $\pm$ 0.08  &   1.77 $\pm$ 0.08     &    19.34 $\pm$ 0.76  \\
165.....        &          -         &   4.20 $\pm$ 0.22  &   4.24 $\pm$ 0.06     &    32.19 $\pm$ 0.68  \\
168.....        &          -         &   5.66 $\pm$ 0.28  &   5.65 $\pm$ 0.08     &    14.27 $\pm$ 0.57  \\
464.....        &   3.62 $\pm$ 0.19  &   3.61 $\pm$ 0.19  &   3.76 $\pm$ 0.07     &    18.76 $\pm$ 0.74  \\
489.....        &          -         &   3.00 $\pm$ 0.16  &   2.81 $\pm$ 0.03     &     8.86 $\pm$ 0.77  \\
617.....        &   6.91 $\pm$ 0.35  &   6.80 $\pm$ 0.35  &   6.94 $\pm$ 0.08     &    49.48 $\pm$ 0.99  \\
824.....        &   1.99 $\pm$ 0.10  &          -         &   1.88 $\pm$ 0.11     &    18.06 $\pm$ 0.84  \\
1409...         &   2.64 $\pm$ 0.13  &          -         &   2.41 $\pm$ 0.11     &    21.04 $\pm$ 0.82  \\
2943...         &   5.42 $\pm$ 0.21  &   5.43 $\pm$ 0.21  &   5.43 $\pm$ 0.07     &   285.93 $\pm$ 0.88  \\
2990...         &   8.04 $\pm$ 0.39  &   7.88 $\pm$ 0.38  &   7.95 $\pm$ 0.09     &    96.74 $\pm$ 0.87  \\
3249...         &   5.17 $\pm$ 0.29  &          -         &   5.13 $\pm$ 0.04     &    11.23 $\pm$ 0.97  \\
3547...         &          -         &   3.24 $\pm$ 0.16  &   3.29 $\pm$ 0.08     &    21.64 $\pm$ 0.99  \\
3705...         &          -         &   7.33 $\pm$ 0.42  &   7.50 $\pm$ 0.09     &    14.69 $\pm$ 0.81  \\
3873...         &   2.72 $\pm$ 0.12  &          -         &   2.70 $\pm$ 0.10     &    13.01 $\pm$ 0.88  \\
3980...         &   3.41 $\pm$ 0.19  &          -         &   3.33 $\pm$ 0.04     &    11.89 $\pm$ 0.72  \\
4247...         &   2.64 $\pm$ 0.13  &   2.61 $\pm$ 0.13  &   2.54 $\pm$ 0.03     &    33.40 $\pm$ 0.78  \\
4301...         &   6.79 $\pm$ 0.34  &          -         &   6.91 $\pm$ 0.08     &    26.38 $\pm$ 0.53  \\
4335...         &   4.18 $\pm$ 0.21  &          -         &   4.08 $\pm$ 0.07     &    22.21 $\pm$ 0.68  \\
4432...         &          -         &   3.23 $\pm$ 0.18  &   3.21 $\pm$ 0.03     &     5.40 $\pm$ 0.99  \\
4518...         &   3.36 $\pm$ 0.17  &   3.37 $\pm$ 0.18  &   3.23 $\pm$ 0.02     &    16.64 $\pm$ 0.60  \\
4932...         &   3.30 $\pm$ 0.15  &   3.23 $\pm$ 0.15  &   3.23 $\pm$ 0.05     &    31.90 $\pm$ 0.87  \\
5253...         &   2.26 $\pm$ 0.09  &          -         &   2.28 $\pm$ 0.07     &    13.01 $\pm$ 0.63  \\
5602...         &   2.61 $\pm$ 0.12  &          -         &   2.48 $\pm$ 0.08     &    14.91 $\pm$ 0.57  \\
5681...         &   2.80 $\pm$ 0.14  &   2.76 $\pm$ 0.13  &   2.76 $\pm$ 0.03     &    27.94 $\pm$ 0.61  \\
5854...         &   4.96 $\pm$ 0.25  &          -         &   4.83 $\pm$ 0.09     &    44.54 $\pm$ 0.71  \\
6220...         &   2.61 $\pm$ 0.12  &   2.52 $\pm$ 0.12  &   2.50 $\pm$ 0.08     &    29.11 $\pm$ 0.52  \\
6418...         &   5.52 $\pm$ 0.30  &          -         &   5.26 $\pm$ 0.06     &     8.89 $\pm$ 0.52  \\
7314...         &   2.40 $\pm$ 0.12  &          -         &   2.23 $\pm$ 0.09     &     4.24 $\pm$ 0.49  \\
7328...         &          -         &   2.22 $\pm$ 0.11  &   2.07 $\pm$ 0.07     &    26.48 $\pm$ 0.49  \\
7602...         &   2.16 $\pm$ 0.10  &          -         &   2.18 $\pm$ 0.09     &    72.95 $\pm$ 0.83  \\
7957...         &   2.67 $\pm$ 0.13  &   2.66 $\pm$ 0.13  &   2.65 $\pm$ 0.04     &    69.73 $\pm$ 0.49  \\
8632...         &          -         &   2.62 $\pm$ 0.14  &   2.63 $\pm$ 0.05     &    10.81 $\pm$ 0.56  \\
8684...         &   2.47 $\pm$ 0.12  &          -         &   2.53 $\pm$ 0.09     &    27.95 $\pm$ 0.77  \\
8923...         &          -         &   1.53 $\pm$ 0.07  &   1.61 $\pm$ 0.17     &    18.34 $\pm$ 0.74  \\
8961...         &          -         &   2.73 $\pm$ 0.14  &   2.66 $\pm$ 0.08     &    38.74 $\pm$ 0.68  \\
\enddata
\label{tab:angdiameters}
\end{deluxetable}

\begin{deluxetable}{rcccccccccr}
\tabletypesize{\footnotesize}
\tablecaption{Field Star Sample}
\tablewidth{0pt}
\tablehead{\colhead{Hip. ID}   &
           \colhead{$\Teff$}   &
           \colhead{$\logg$}   &
           \colhead{$\FeH$}    &
           \colhead{V}         &
           \colhead{$(B-V)$}   &
           \colhead{E$(B-V)$}  &
           \colhead{$(b-y)$}   &
           \colhead{$m_1$}     &
           \colhead{$c_1$}     &
           \colhead{Source\tablenotemark{a}\tablenotetext{a}{Source of $\Teff$, $\logg$, $\FeH$, and $\EBV$ estimates: (1)=HBS2000, (2)=AAM96, and (3)=AAM99.}}    \\ }
\tablenotetext{~}{Note. --- The complete version of this table is in the electronic edition of the Journal.  The printed edition contains only a sample.}
\startdata
    80 & 5859 & 4.24 & -0.59 &   8.40 & 0.566 & 0.000 & 0.372 & 0.143 & 0.309 & 2 \\
   434 & 5351 & 2.66 & -1.45 &   9.04 & 0.692 & 0.008 & 0.434 & 0.092 & 0.474 & 3 \\
   484 & 4968 & 2.60 & -1.10 &   9.66 & 0.787 & 0.000 & 0.513 & 0.155 & 0.349 & 3 \\
   910 & 6148 & 4.11 & -0.35 &   4.89 & 0.487 & 0.003 & 0.328 & 0.130 & 0.405 & 1 \\
  1298 & 5265 & 2.60 & -1.10 &   9.58 & 0.710 & 0.015 & 0.460 & 0.129 & 0.464 & 3 \\
  1301 & 5784 & 4.15 & -0.85 &   9.74 & 0.569 & 0.000 & 0.383 & 0.135 & 0.284 & 2 \\
\enddata
\label{tab:starlist}
\end{deluxetable}

\begin{deluxetable}{lcccccc} 
\tabletypesize{\footnotesize}
\tablecaption{Coefficients for the Calibrations Between Synthetic and Observed Color}
\tablewidth{0pt}
\tablehead{\colhead{Color}    &
           \colhead{A}        &
           \colhead{B}        &
           \colhead{C}        &
           \colhead{N}        &
           \colhead{$\sigma$} &
           \colhead{Range}    \\ }
\tablenotetext{~}{Note. --- Calibrations take the form y$\,$=$\,$A$\,$+$\,$B$x\,$+$\,$C$x^2$, where $x$ is the synthetic color 
and $y$ is the observed color}
\startdata
$(b-y)$....         &                      &                     &                     &     &       &                           \\
~~~dwarfs           & ~0.003$\,\pm\,$0.006 & 1.077$\,\pm\,$0.011 &   -                 & 102 & 0.020 & $0.15\leq\,b-y\,\leq0.78$ \\
~~~giants           & ~0.004$\,\pm\,$0.007 & 1.084$\,\pm\,$0.012 &   -                 &  61 & 0.025 & $0.15\leq\,b-y\,\leq0.82$ \\ 
~~~dwarfs \& giants & -0.001$\,\pm\,$0.005 & 1.080$\,\pm\,$0.009 &   -                 & 163 & 0.023 & $0.15\leq\,b-y\,\leq0.82$ \\ 
$m_1$.....          &                      &                     &                     &     &       &                           \\
~~~dwarfs           & ~0.028$\,\pm\,$0.006 & 0.346$\,\pm\,$0.071 & 3.573$\,\pm\,$0.173 & 102 & 0.047 & $0.10\leq\,m_1\,\leq0.40$ \\
~~~giants           & -0.127$\,\pm\,$0.007 & 2.083$\,\pm\,$0.026 &   -                 &  61 & 0.044 & $0.10\leq\,m_1\,\leq0.45$ \\
\enddata
\label{tab:calibrations}
\end{deluxetable}

\begin{deluxetable}{lcccc}
\tabletypesize{\footnotesize}
\tablecaption{Various $\Teff$ Estimates for Selected Population~II Subdwarfs} 
\tablewidth{0pt}
\tablehead{\colhead{HD \#} &
           \colhead{$T_{\rm{eff,IRFM}}$\tablenotemark{a}
\tablenotetext{a}{Derived from the IRFM by AAM96.}} &
           \colhead{$T_{\rm{eff,CT}}$\tablenotemark{b}
\tablenotetext{b}{Derived from color-temperature relations by Gratton et~al. \citeyear{Gratton1996, Gratton2000} 
and/or Clementini et~al. \citeyear{Clementini1999}.}} &
           \colhead{$T_{\rm{eff,Balmer}}$\tablenotemark{c}
\tablenotetext{c}{Derived from fitting Balmer line profiles by Axer et~al. \citeyear{Axer1994} 
and/or Fuhrmann \citeyear{Fuhrmann1998}.}} &
           \colhead{$T_{\rm{eff,UV}}$\tablenotemark{d}
\tablenotetext{d}{Derived from UV-flux distributions by Allende Prieto \& Lambert \citeyear{AllendePrietoLambert2000}.}} \\ }
\startdata
19445.....  &  6050  &      6054     &    6040    &    6065   \\
25329.....  &  4842  &      4845     &     -      &    4870   \\
64090.....  &  5441  &      5506     &    5499    &    5456   \\
74000.....  &  6224  &      6275     &    6211    &     -     \\
84937.....  &  6330  &      6344     &    6353    &    6389   \\
103095...   &  5029  &      5097     &    5110    &    5069   \\
132475...   &  5788  &      5758     &     -      &     -     \\
134439...   &  4974  &      5120     &     -      &    5110   \\
134440...   &  4746  &      4879     &     -      &     -     \\
140283...   &  5691  &      5763     &    5814    &     -     \\
188510...   &  5564  &      5628     &    5500    &    5597   \\
201891...   &  5909  &      5974     &    5797    &    5929   \\ 
\enddata
\label{tab:subdwarftemp}
\end{deluxetable}

\begin{deluxetable}{lcccccccccccc}
\tabletypesize{\footnotesize}
\tablecaption{Fundamental Parameters and Photometry for Selected Population II Subdwarfs.}
\tablewidth{0pt}
\tablehead{\colhead{HD \#}           &
           \colhead{$\logg$\tablenotemark{a}\tablenotetext{a}{Mean value from Cayrel~de~Strobel et~al.~\citeyear{CayreldeStrobel2001}.}}  &
           \colhead{$\logg$\tablenotemark{b}\tablenotetext{b}{Derived from isochrones using {\it Hipparcos} parallax}} &
           \colhead{$\FeH$\tablenotemark{a}} &
           \colhead{$\FeH$\tablenotemark{c}\tablenotetext{c}{From Schuster \& Nissen \citeyear{SchusterNissen1989b}.}} &
           \colhead{$\EBV$\tablenotemark{d}\tablenotetext{d}{From Carretta et~al. \citeyear{Carretta2000}.}} &
	   \colhead{$(B-V)_o$\tablenotemark{d}}       &
           \colhead{$(b-y)_o$}       &
           \colhead{$m_o$}           &
           \colhead{$c_o$}           \\ }
\startdata                              
19445.....  & 4.38 & 4.44 & -1.99 & -1.92 & 0.002 & 0.458 & 0.351 & 0.052 & 0.203 \\   
25329.....  & 4.66 & 4.68 & -1.68 & -1.63 & 0.000 & 0.864 & 0.533 & 0.307 & 0.131 \\
64090.....  & 4.49 & 4.58 & -1.65 & -1.69 & 0.000 & 0.614 & 0.428 & 0.110 & 0.126 \\
74000.....  & 4.18 & 4.08 & -2.01 & -1.69 & 0.000 & 0.431 & 0.311 & 0.067 & 0.295 \\
84937.....  & 3.97 & 4.05 & -2.09 & -2.14 & 0.009 & 0.382 & 0.296 & 0.058 & 0.352 \\
103095...   & 4.62 & 4.61 & -1.30 & -1.33 & 0.000 & 0.752 & 0.484 & 0.222 & 0.155 \\
132475...   & 3.85 & 3.84 & -1.52 & -1.32 & 0.037 & 0.522 & 0.373 & 0.072 & 0.279 \\
134439...   & 4.63 & 4.63 & -1.41 & -1.33 & 0.005 & 0.767 & 0.480 & 0.225 & 0.164 \\
134440...   & 4.59 & 4.62 & -1.43 & -1.24 & 0.005 & 0.845 & 0.520 & 0.298 & 0.172 \\
140283...   & 3.50 & 3.69 & -2.40 & -2.49 & 0.024 & 0.463 & 0.362 & 0.039 & 0.280 \\
188510...   & 4.54 & 4.57 & -1.53 & -1.57 & 0.001 & 0.598 & 0.415 & 0.100 & 0.163 \\
201891...   & 4.33 & 4.29 & -0.97 & -1.08 & 0.003 & 0.514 & 0.360 & 0.095 & 0.260 \\  
\enddata
\label{tab:subdwarfparam}
\end{deluxetable}

\begin{deluxetable}{lrrrrcl}
\tabletypesize{\footnotesize}
\tablecaption{The Computed $uvby$ (and $B-V$) Photometry for Population~II Subdwarfs}
\tablewidth{0pt}
\tablehead{\colhead{HD \#}           &
           \colhead{$(b-y)_o$}       &
           \colhead{$m_o$}           &
           \colhead{$c_o$}           &
           \colhead{$(B-V)_o$}       &
	   \colhead{}                &
           \colhead{Notes}          \\ }
\tablenotetext{~}{Note. --- Numbers in parentheses are model parameters ($\Teff$/$\logg$/$\FeH$).}
\startdata
\bf 19445..... &  \bf 0.351 &  \bf 0.052 &  \bf 0.203 &  \bf 0.458 &  & Intrinsic colors                \\
               &      0.345 &      0.051 &      0.212 &      0.450 &  & Model A = (6050/4.38/-1.99)     \\
               & $\mp$0.014 & $\pm$0.002 & $\pm$0.018 & $\mp$0.020 &  & Model A: $\Teff\pm100\,$K       \\   
               & $\pm$0.005 & $\pm$0.001 & $\mp$0.030 & $\pm$0.006 &  & Model A: $\logg\pm0.25\,$dex    \\
               & $\mp$0.003 & $\pm$0.008 & $\pm$0.010 & $\pm$0.008 &  & Model A: $\FeH\pm0.25\,$dex     \\
\bf 25329..... &  \bf 0.533 &  \bf 0.307 &  \bf 0.131 &  \bf 0.864 &  & Intrinsic colors                \\
               &      0.538 &      0.250 &      0.122 &      0.844 &  & Model A = (4842/4.66/-1.68)     \\
               &      0.540 &      0.298 &      0.124 &      0.844 &  & Model B = Model A w/ [C/Fe]=[N/Fe]=+0.4   \\
               &      0.540 &      0.297 &      0.100 &      0.844 &  & Model C = Model A w/ [C/Fe]=+0.4          \\
               &      0.538 &      0.251 &      0.148 &      0.844 &  & Model D = Model A w/ [N/Fe]=+0.4          \\
\bf 64090..... &  \bf 0.428 &  \bf 0.110 &  \bf 0.126 &  \bf 0.614 &  & Intrinsic colors                \\
               &      0.431 &      0.119 &      0.154 &      0.625 &  & Model A = (5441/4.49/-1.65)     \\
\bf 74000..... &  \bf 0.311 &  \bf 0.067 &  \bf 0.295 &  \bf 0.431 &  & Intrinsic colors                \\
               &      0.321 &      0.052 &      0.283 &      0.412 &  & Model A = (6224/4.18/-2.01)     \\
\bf 84937..... &  \bf 0.296 &  \bf 0.058 &  \bf 0.352 &  \bf 0.382 &  & Intrinsic colors                \\
               &      0.306 &      0.054 &      0.355 &      0.387 &  & Model A = (6330/3.97/-2.09)     \\
\bf 103095...  &  \bf 0.484 &  \bf 0.222 &  \bf 0.155 &  \bf 0.752 &  & Intrinsic colors                \\
               &      0.493 &      0.256 &      0.149 &      0.780 &  & Model A = (5029/4.62/-1.30)     \\
               & $\mp$0.016 & $\mp$0.025 & $\mp$0.003 & $\mp$0.035 &  & Model A: $\Teff\pm100\,$K       \\
               & $\pm$0.001 & $\pm$0.007 & $\mp$0.011 & $\pm$0.009 &  & Model A: $\logg\pm0.25\,$dex    \\
               & $\mp$0.007 & $\pm$0.037 & $\mp$0.003 & $\pm$0.004 &  & Model A: $\FeH\pm0.25\,$dex     \\
\bf 132475...  &  \bf 0.373 &  \bf 0.072 &  \bf 0.279 &  \bf 0.522 &  & Intrinsic colors                \\
               &      0.367 &      0.081 &      0.257 &      0.513 &  & Model A = (5788/3.85/-1.52)     \\
\bf 134439...  &  \bf 0.480 &  \bf 0.225 &  \bf 0.164 &  \bf 0.767 &  & Intrinsic colors                \\
               &      0.505 &      0.257 &      0.140 &      0.798 &  & Model A = (4974/4.63/-1.41)     \\
               &      0.480 &      0.217 &      0.145 &      0.748 &  & Model B = (5120/4.63/-1.41)     \\
               &      0.479 &      0.227 &      0.153 &      0.749 &  & Model C = (5120/4.63/-1.33)     \\
\bf 134440...  &  \bf 0.520 &  \bf 0.298 &  \bf 0.172 &  \bf 0.845 &  & Intrinsic colors                \\
               &      0.550 &      0.322 &      0.131 &      0.881 &  & Model A = (4746/4.59/-1.43)     \\
               &      0.523 &      0.279 &      0.137 &      0.830 &  & Model B = (4879/4.59/-1.43)     \\
               &      0.519 &      0.308 &      0.152 &      0.833 &  & Model C = (4879/4.59/-1.24)     \\
\bf 140283...  &  \bf 0.362 &  \bf 0.039 &  \bf 0.280 &  \bf 0.463 &  & Intrinsic colors                \\
               &      0.384 &      0.039 &      0.266 &      0.493 &  & Model A = (5691/3.50/-2.40)     \\
               &      0.365 &      0.038 &      0.294 &      0.467 &  & Model B = (5814/3.50/-2.40)     \\
               & $\mp$0.015 & $\pm$0.001 & $\pm$0.025 & $\mp$0.021 &  & Model B: $\Teff\pm100\,$K       \\
               & $\pm$0.003 & $\pm$0.001 & $\mp$0.046 & $\pm$0.006 &  & Model B: $\logg\pm0.25\,$dex    \\
               & $\mp$0.001 & $\pm$0.007 & $\pm$0.007 & $\pm$0.005 &  & Model B: $\FeH\pm0.25\,$dex     \\
\bf 188510...  &  \bf 0.415 &  \bf 0.100 &  \bf 0.163 &  \bf 0.598 &  & Intrinsic colors                \\
               &      0.412 &      0.115 &      0.162 &      0.593 &  & Model A = (5564/4.54/-1.53)     \\
\bf 201891...  &  \bf 0.360 &  \bf 0.095 &  \bf 0.260 &  \bf 0.514 &  & Intrinsic colors                \\
               &      0.360 &      0.109 &      0.249 &      0.523 &  & Model A = (5909/4.33/-0.97)     \\
               & $\mp$0.013 & $\mp$0.006 & $\pm$0.016 & $\mp$0.025 &  & Model A: $\Teff\pm100\,$K       \\
               & $\pm$0.002 & $\pm$0.005 & $\mp$0.032 & $\pm$0.008 &  & Model A: $\logg\pm0.25\,$dex    \\
               & $\pm$0.005 & $\pm$0.016 & $\pm$0.021 & $\pm$0.015 &  & Model A: $\FeH\pm0.25\,$dex     \\
\enddata
\label{tab:subdwarfcalc}
\end{deluxetable} 



\clearpage

\clearpage

\clearpage

\clearpage

\clearpage

\clearpage

\clearpage

\clearpage

\clearpage
               
\clearpage

\clearpage

\clearpage

\clearpage

\clearpage

\clearpage

\clearpage

\clearpage

\clearpage

\clearpage

\clearpage

\clearpage

\clearpage

\clearpage

\clearpage

\clearpage

\clearpage

\end{document}